\documentclass[twocolumn,tighten, twocolappendix]{aastex631}
\turnoffeditone
\usepackage{physics,bm}	% Advanced maths commands
\usepackage[mathscr]{euscript}
\usepackage{xcolor}
\usepackage{chemformula} %to write chemical formulas with  \let\ce\ch

\received{March 18, 2024}
\accepted{May 12, 2024}

\shorttitle{Interstellar Polarization Survey IV}
\shortauthors{Angarita et al.}
\begin{document}

\tighten

\title{Interstellar Polarization Survey. IV. \\ Characterizing the magnetic field strength and turbulent dispersion using optical starlight polarization in the diffuse interstellar medium}

\correspondingauthor{Y. Angarita}
\email{y.angarita@astro.ru.nl}

\author[0000-0001-5016-5645]{Y. Angarita}
\affiliation{Department of Astrophysics/IMAPP, Radboud University, PO Box 9010, 6500 GL Nijmegen, The Netherlands}

\author[0000-0003-0400-8846]{M.J.F. Versteeg}
\affiliation{Department of Astrophysics/IMAPP, Radboud University, PO Box 9010, 6500 GL Nijmegen, The Netherlands}

\author[0000-0002-5288-312X]{M. Haverkorn}
\affiliation{Department of Astrophysics/IMAPP, Radboud University, PO Box 9010, 6500 GL Nijmegen, The Netherlands}

\author[0000-0002-5501-232X]{A. Marchal}
\affiliation{Research School of Astronomy and Astrophysics, The Australian National University, Canberra, ACT 2611, Australia}

\author[0000-0002-9459-043X]{C.V. Rodrigues}
\affiliation{Divis\~ao de Astrof\'isica, Instituto Nacional de Pesquisas Espaciais (INPE/MCTI), Av. dos Astronautas, 1758, S\~ao Jos\'e dos Campos, SP, Brazil}

\author[0000-0002-1580-0583]{A.M. Magalh\~aes}
\affiliation{IAG, Universidade de S\~ao Paulo, Brazil}

\author[0000-0001-6880-4468]{R. Santos-Lima}
\affiliation{IAG, Universidade de S\~ao Paulo, Brazil}

\author[0000-0001-6099-9539]{Koji S. Kawabata}
\affiliation{Hiroshima Astrophysical Science Center, Hiroshima University, Kagamiyama, Higashi-Hiroshima, Hiroshima, 739-8526, Japan}

\begin{abstract}
    Angular dispersion functions are typically used to estimate the fluctuations in polarization angle around the mean magnetic field orientation in dense regions, such as molecular clouds. The technique provides accurate turbulent to regular magnetic field ratios, $\langle B_t^2\rangle^{1/2}/B_{pos}$, which are often underestimated by the classic Davis-Chandrasekhar-Fermi method. We assess the technique’s suitability to characterize the turbulent and regular plane-of-sky magnetic field in low-density structures of the nearby interstellar medium (ISM), particularly when the turbulence outer scale, $\delta$, is smaller than the smallest scale observed, $\ell_{min}$. We use optical polarization maps of three intermediate-latitude fields ($|b| \gtrsim 7\fdg5$) with dimensions of $0\fdg3 \times 0\fdg3$, sourced from the Interstellar Polarization Survey--General ISM (IPS-GI) catalog. We decomposed the \ion{H}{1} emission detected by the Galactic All-Sky Survey (GASS) within our fields to estimate the multiphase ISM properties associated with the structure coupled to the magnetic field. We produced maps of the plane-of-sky magnetic field strength ($B_{pos}$), mass density ($\rho$), and turbulent velocity dispersion ($\sigma_{v,turb}$). In the regions with well-defined structures at $d<400$~pc, the average $B_{pos}$ ranges from ${\sim}3~\mu$G to ${\sim}9~\mu$G, depending on the method and physical properties. In the region where structures extend up to $1,000$~pc, $B_{pos}$ varies from ${\sim}1~\mu$G to ${\sim}3~\mu$G.
    The results agree with previous estimations in the local, diffuse ISM. Finally, optical starlight polarization and thermal dust polarization at 353 GHz consistently reveal a highly regular plane-of-sky magnetic field orientation unfazed by diffuse dust structures observed at $12~\micron$.
\end{abstract}


%...... Begin Section ......................................................
\section{Introduction} \label{sec:intro}

    The Galactic magnetic field (GMF) interacts with dust, gas, and cosmic rays, influencing the interstellar medium (ISM) from the diffuse medium to molecular clouds and star-formation regions \citep[e.g., see][and references therein]{Crutcher_2012, Haverkorn_2015, Vazquez_Semadeni_2015}. Starlight polarization produced by the dichroic extinction of background stars reveals the orientation of the Galactic magnetic field (GMF) projected on the plane of the sky \citep{Hiltner_1949, Davis_Greenstein_1951}. Moreover, linear polarization observations are useful for estimating the average plane-of-sky magnetic field strength with the Davis-Chandrasekhar-Fermi method \citep{Davis_1951, Chandrasekhar&Fermi_1953} or any analogous approach. The DCF method estimates the magnetic field strength based on the dispersion of the polarization angle under the assumption that this dispersion measures the turbulence in the medium. In cases where this assumption is not true, a structure function (SF) or angular dispersion function (ADF) can be used to estimate the dispersion around the large-scale field and, subsequently, the magnetic field strength.
    
    The second-order SF of the polarization angle, $\langle \Delta\theta^2(\ell) \rangle$, measures the behavior of the dispersion about the mean magnetic field direction as a function of angular or spatial distance between two measurements. The SF method was first explored by \cite{Serkowski_1958, Serkowski_1962} and then by \cite{Kobulnicky_1994} to study the polarizing properties of the ISM in open clusters. Later, \citet{Falceta_Gonzalves_2008} used the SFs along with simulations of polarization maps to characterize the turbulent cascade and the magnetic field. \cite{Hildebrand_2009} defined the ADF, $\langle \Delta\theta^2(\ell) \rangle^{1/2}$, to measure the turbulent and large-scale contribution to the angular dispersion in molecular clouds. The technique (hereafter referred to as ADF-DCF) finds the magnetic field strength by avoiding inaccurate estimates of the field dispersion, $\sigma_{\theta}$, based on simple Gaussian fits to the polarization angle distribution. However, the ADF-DCF method is valid only at scale lengths larger than the turbulent correlation length, $\delta$, or the observed resolution scale.

    The ADF-DCF method and its variations \citep{Houde_2009} have been extensively used with starlight and submillimeter polarization maps of discrete and localized dense regions, e.g.,~nearby molecular clouds \citep[e.g., see][and references therein]{Hildebrand_2009, Franco_2010, Soler_2016, Wang_JCMT_BISTRO_2019}, whose physical properties are well known. In the diffuse ISM, on the other hand, starlight polarization observations have much lower spatial density \citep[e.g., see][]{Heiles_2000, Panopoulou_catalog_2023}{}{} and there is little knowledge about the physical properties of diffuse structures (i.e.,~precise measurements of velocity dispersion and densities). Fortunately, optical polarimetry data from the Interstellar Polarization Survey (IPS) (\citealt{Magalhaes_2005} and Magalh\~aes et al. 2024, in preparation) in the diffuse ISM \citep[IPS-GI,][]{Versteeg_2023} offers a great opportunity to study the properties of the interstellar magnetic fields in diffuse dust close to the Galactic thin disk. 
    
    The IPS-GI database \citep{Versteeg_2023} has demonstrated the potential of optical polarization maps, combined with precise \textit{Gaia}-EDR3  \citep{Gaia_Collaboration_2021b} distance, optical extinction \citep{Anders_2022}, and atomic and molecular line emission to study the magnetized ISM (e.g.,~see \citealt{Angarita_2023} and \citealt{Versteeg_2024}). In this paper, we use the IPS-GI optical polarization measurements at intermediate-latitude ($|b|>7\fdg5$) regions to test the suitability of the classic DCF, the ADF-DCF, and other DCF-like techniques to study the magnetic field properties in the diffuse medium and at small angular scales ($<0\fdg3$). We report the plane-of-sky magnetic field and ISM properties of the nearby ($d<1,000$~pc) diffuse polarizing clouds deduced from observations. In Section~\ref{sec:Observ_data}, we present the polarization data and ancillary data needed to estimate the properties of the interstellar structures. We explain the DCF, the ADF-DCF, and other methods to estimate the magnetic field strength in Section~\ref{sec:MF_cal}. We describe the polarimetric properties of the selected regions and calculate their physical properties using atomic hydrogen emission and three-dimensional (3D) dust maps in Section~\ref{sec:selec_fields}. We present our results on the turbulent dispersion and magnetic field strength in Section~\ref{sec:results}. We discuss the turbulent scales observed in our sample and the magnetic field strength obtained from different methods, compare our results with other estimations, and outline the characteristics of the nearby magnetic field in Section \ref{sec:discu}. Finally, we summarize our work in Section \ref{sec:summary}.

%...... Begin Section ......................................................
\section{Observations and data} \label{sec:Observ_data}

    In this section, we present the optical linear polarization observations and ancillary data used in our research. 

    %...... Begin Sub-section ......................................................
    \subsection{Optical polarization data}  \label{subsec:obs_IPS-GI}

        This research uses data from the Interstellar Polarization Survey (IPS) (\citealt{Magalhaes_2005} and Magalh\~aes et al. 2024, in preparation) in the general ISM, which was reduced and presented in detail by \cite{Versteeg_2023}. The stellar catalog (IPS-GI) consists of 38 fields of view, each measuring $0\fdg3\times0\fdg3$, sparsely distributed near and within the Galactic thin disk in the Southern sky. The photometric and polarimetric parameters were obtained with the reduction pipeline SOLVEPOL \citep{Ramirez_2017}. We use the quality filters presented in \cite{Versteeg_2023} and \cite{Angarita_2023}, including a strong constraint in the signal-to-noise of $P/\sigma_P \geq 5$, which ensures the measurements are not significantly affected by the Ricean bias, as discussed by \cite{Ramirez_2017} and \cite{Versteeg_2023}. The low instrumental polarization, $0.07\%$, below the median fractional polarization error measured, does not need to be corrected. The final catalog consists of ${\sim}10,500$ stars with high-quality optical polarization measurements, distance, optical extinction, and among other parameters from \cite{Anders_2022} and \cite{Gaia_Collaboration_2021b}.         

    %...... Begin Sub-section ......................................................
    \subsection{\texorpdfstring{\ion{H}{1}}{HI} map}  \label{subsec:obs_HI_map}
    
        The Galactic All-Sky Survey (GASS, \citealt{McClure_Griffiths_2009}) has atomic hydrogen (\ion{H}{1}) spectral line emission data for the entire Southern sky. The survey has a resolution of $16\arcmin$, comparable with the size of the IPS-GI field-of-view. Therefore, we used position-position-velocity (PPV) data cubes\footnote{\url{https://www.atnf.csiro.au/research/GASS/Data.html}} of neutral atomic hydrogen within an area of $0\fdg7\times0\fdg7$, centered on the IPS-GI fields' coordinates (Section~\ref{sec:selec_fields}).  The velocity range covers from $-400$~km~s$^{-1}$ to $500$~km~s$^{-1}$ with $1$~km~s$^{-1}$ of spectral resolution. The average \textit{rms} noise in the \ion{H}{1} map is $0.057$~K.
         
    %...... Begin Sub-section ......................................................
    \subsection{3D Dust maps of the local Galaxy}  \label{subsec:obs_dust_maps}

        High-resolution 3D dust maps give us accurate insights into the dust morphology and properties along our lines of sight in the local Galactic neighborhood. We considered three 3D maps: the \cite{Vergely_2022} dust map, with $10$~pc resolution covering roughly $1.5$~kpc and sampling $5$~pc; the \cite{Leike_2020} map, with radii ranging from about $370$~pc to $590$~pc and resolution of $1$~pc; and the \cite{Edenhofer_dustmap_2023} map, spanning $1.25$~kpc with resolutions between $0.4-7$~pc. We compare the three dust maps with various resolutions to look for consistencies between the morphology of line-of-sight features in each field studied. We primarily use the \cite{Vergely_2022} 3D dust map, retrieved through the G-Tomo\footnote{\url{https://explore-platform.eu/sdas}} platform, to estimate the total column density along the line of sight. We queried \cite{Leike_2020} and \cite{Edenhofer_dustmap_2023} high-resolution 3D dust maps through the Python \textit{Dustmaps} module\footnote{\url{https://dustmaps.readthedocs.io/en/latest/index.html}} \citep{Green_dustmaps_2018}. We use \cite{Leike_2020} for the volume density and spatial scale estimations in nearby clouds. Additionally, we use \cite{Edenhofer_dustmap_2023} for comparison with the other maps. We obtained the extinction density, the cumulative extinction profiles with distance, and their respective errors using the central coordinates of the selected IPS-GI fields (see Section~\ref{sec:selec_fields}).

%...... Begin Section ......................................................
\section{Magnetic field strength calculation} \label{sec:MF_cal}

    The total magnetic field is assumed to have a regular but not constant component, $\bm{B}_{pos}$, and a turbulent component, $\bm{B}_{t}$, in the plane of the sky. Polarization measurements combined with interstellar gas properties allow us to estimate the regular and turbulent plane-of-sky magnetic field strength through the DCF method and its modifications under certain considerations. We described the methods and their assumptions in the following sections.
    
    %...... Begin Section ......................................................
    \subsection{The DCF method} \label{subsec:MF_cal_DCF}
         
        The DCF method estimates the average plane-of-sky magnetic field strength from polarization angle deviations around a mean magnetic field due to small, non-thermal turbulent gas motions \citep{Davis_1951, Chandrasekhar&Fermi_1953}. The regular plane-of-sky magnetic field strength is estimated by the DCF method as
        \begin{equation}
        \label{eq:DCF}
            B_{pos}^{DCF} \simeq f \sqrt{4\pi\rho} \frac{\sigma_{v}}{\sigma_{\theta}}   ~,
        \end{equation}
        where $\sigma_{v}$ and $\rho$ are the velocity dispersion and mass density of the magnetized medium, respectively, $f$ is the fudge factor that corrects the magnetic field strength estimation, and $\sigma_{\theta}$ is the dispersion of the polarization angle. To obtain Equation~(\ref{eq:DCF}), the method assumes equipartition between the gas turbulent kinetic energy and magnetic energy, isotropic turbulence (i.e.,~$\sigma_{v_{pos}} = \sqrt{2}\sigma_{v_{los}}$), and a small random magnetic field component (i.e.,~$B_{t} \ll B_{pos}$). Furthermore, the DCF method assumes that the field dispersion is represented entirely by the deviations in polarization angles at the observed scales (i.e.,~$B_{t}/B_{pos} \approx \sigma_{\theta}$), and this dispersion is solely due to gas turbulent motion composed of Alfv\'en modes, i.e.,~incompressible transverse magnetohydrodynamic (MHD) waves \citep[e.g., see][and references therein for more details]{Crutcher_2012, Beck_2013, Jones_bookChap_2015, Andersson_2015, Liu_2021}. 
        
        It is well-known that interstellar turbulence is anisotropic. Indeed, the theory by \cite{Goldreich_Sridhar_1995} predicts larger anisotropies for smaller scales, with turbulent eddies aligning with the local magnetic field. However, as the size of turbulent cells decreases, more cells contribute to the line-of-sight integration, each with its own local magnetic field orientation. Therefore, we assume that the anisotropies average out on these scales, allowing us to use the DCF method for isotropic turbulence. Studies on the DCF method based on synthetic maps created from numerical simulations, which include the natural anisotropies in the MHD turbulence below the injection scale, have demonstrated that under these assumptions, the DCF method for isotropic turbulence provides reasonable results \citep{Cho_Yoo_2016}.       

    %...... Begin Section ......................................................
    \subsection{The \texorpdfstring{\cite{Skalidis_Tassis_2021}}{Skalidis \& Tassis (2021)} method} \label{subsec:MF_cal_ST-DCF}

        The \cite{Skalidis_Tassis_2021} method (hereinafter ST-DCF) is a modification of the classic DCF method that calculates the plane-of-sky magnetic field strength as
        \begin{equation}
        \label{eq:ST_DCF}
            B_{pos}^{ST-DCF} \simeq \sqrt{2\pi\rho} \frac{\sigma_{v}}{\sqrt{\sigma_{\theta}}}   ~.
        \end{equation}
        Contrary to the classic DCF method, the ST-DCF assumes that the fluctuating magnetic energy is dominated by the cross-term (i.e.,~$\bm{B_{t} \cdot B_{pos}} \neq 0$), which represents the compressible modes. Although this assumption may not be valid in all cases, e.g.,~in dense regions where self-gravity is important \citep{Liu_2021, Liu_2022}, in the case of the diffuse ISM, ST-DCF may provide reliable estimates of the regular plane-of-sky magnetic field strength. The method demonstrates a mean relative deviation of $17\%$ from the true value in 3D compressible MHD simulations, with its accuracy independent of the turbulence properties.

    %...... Begin Section ......................................................
    \subsection{Angular dispersion function} \label{subsec:MF_cal_ADF}
    
        The ADF is defined as the squared root of the second-order SF of the polarization angle \citep{Hildebrand_2009, Poidevin_2010},
        \begin{equation}
        \label{eq:ADF}
            \langle\Delta\theta^2(\ell)\rangle^{1/2} =  \biggl[ \frac{1}{N(\ell)}\sum^{N(\ell)}_{i=1} [\theta(x)-\theta(x+\ell)]^2 \biggr]^{1/2},
        \end{equation}
        where $\theta(x)-\theta(x+\ell)$ is the difference in polarization angle between pairs of stars with angular separation $\ell$, and $N$ is the number of pairs per angular separation bin, which has $N(N-1)/2$ distinct combinations \citep[see][]{Falceta_Gonzalves_2008}. The $\Delta\theta$ magnitude is constrained in the range [0, 90]$^{\circ}$, but in practice, $\Delta \theta$ is confined to small angles to ensure that the small-angle approximation in the DCF method is valid and to avoid the $n\pi$ ambiguity.
    
        As described by \cite{Hildebrand_2009}, if the observed scales $\ell$ are much smaller than the maximum scale of variation in $\bm{B}_{pos}$, this variation increases approximately linearly with scale length and slope $m$. If these scales, in addition, are larger than the maximum scale of turbulence $\delta$, the turbulence contributes a constant value denoted by $b$. Since the two contributions are statistically independent, we can add them in quadrature with the contribution from the measurement uncertainties, $\sigma_{M,k}$ in the $k-$th scale-length bin, to find the total observed SF of polarization angle for small $\Delta\theta(\ell)$: 
        \begin{equation}
        \label{eq:SF_tot_fit}
             \langle \Delta\theta^2(\ell)\rangle_{obs,k} \simeq b^2 + m^2\ell_k^2 + \sigma_{M,k}^2 ~,
        \end{equation}
        where $\sigma_{M,k} = \langle \sigma^2_{\Delta\theta}\rangle_k^{1/2}$ is obtained from the propagation of the errors as $\sigma_{\Delta\theta} = (\sigma_{\theta_{\bm{x}}}^2 + \sigma_{\theta_{\bm{x+\ell}}}^2)^{1/2}$. 
        
        \cite{Houde_2009} presented a similar approach as in Equation~(\ref{eq:SF_tot_fit}), departing from the same basic assumptions as \cite{Hildebrand_2009} and assuming a Gaussian form for the autocorrelation function for the turbulence. They expanded the study of the dispersion by including the contributions from the signal integration through the thickness of the clouds and over the telescope beam. The method applies to starlight polarization by setting the angular resolution term as $W = 0$. This approach is valuable for determining the turbulent correlation length $\delta$ when the SF or the ADF drops at scales nearing $\ell=0$, possibly suggesting that $\delta$ is being resolved (Section~\ref{subsec:MF_cal_ADF_interpre}), e.g.,~as in \cite{Franco_2010} and \cite{Wang_JCMT_BISTRO_2019}. Nevertheless, as reported in Section~\ref{subsec:Resul_ADF}, this is not the case for our IPS-GI regions in the diffuse ISM. Therefore, we use the ADF \citep{Hildebrand_2009, Poidevin_2010} corrected for the measurement errors $\sigma_M$, as
        \begin{equation}
            \label{eq:ADF_corr}
            \langle\Delta\theta^2(\ell)\rangle^{1/2} = (\langle\Delta\theta_{obs}^2(\ell)\rangle - \sigma_M^2)^{1/2} ~,
        \end{equation}
        for straightforward comparison with the polarization angle dispersion and subsequently fit Equation~(\ref{eq:SF_tot_fit}) to the scales at which the ADF increases approximately linearly.
    
    %...... Begin Sub-section ......................................................
        \subsubsection{Interpretation of the ADF} \label{subsec:MF_cal_ADF_interpre}
    
            A qualitative analysis of the ADF profile as a function of the angular scale length can give us an idea of the properties of the magnetic field and the ISM in the observed region:
            \begin{enumerate}
                \item Flat ADF profiles mean that the fluctuations in polarization angle are scale-independent over the scales probed. This can indicate random noise or that fluctuations in polarization angle exist only on scales smaller than the smallest scale probed by the observations $\ell_{min}$ \citep{Sun_Han_SF_2004}.
                \item The ADFs may exhibit multiple bumps or wiggles at different scales due to the contributions of more than one discrete structure that differ in polarization properties \citep{Sun_Han_SF_2004} or due to irregular sampling of sources \citep{Soler_2016}. In such a case, care must be taken in interpreting each slope or bump. 
                \item Most commonly, ADFs increase with scale $\ell$, indicating that the amount of fluctuations grows with scale. This is generally interpreted in either of two ways:
                \begin{enumerate}
                    \item The increasing ADF reflects purely turbulence, which shows a power-law behavior in log-log space \citep[e.g.,][]{Falceta_Gonzalves_2008}. If the ADF flattens out, the scale of this turn-over can be interpreted as the turbulence correlation length $\delta$ \citep{Haverkorn_2008}. 
                    \item If the measured scales are larger than the turbulence correlation length, i.e.,~$\delta<\ell$, the rising ADF is caused by gradients and other large-scale variations in the regular magnetic field. In such a case, the slope represents variations in the field due to a gradual change in polarization angle across the field of view. 
                \end{enumerate}
                A change in the slope of an ADF as a function of scale may indicate detection of the turbulent outer scale \citep{Houde_2009}, may reflect properties of the turbulence \citep{Falceta_Gonzalves_2008}, or suggest various sources of structure \citep{Haverkorn_2004}.
            \end{enumerate}
        
            As discussed in Section~\ref{subsec:selec_fields_SF_PA_gradient}, the case \textit{3.b} may apply to our data. In this case, characteristics of the ADF give a better estimation of polarization angle fluctuations than its dispersion, $\sigma_{\theta}$ \citep[e.g.,~see][]{Liu_2021}. Therefore, the ADFs with DCF analysis combination gives an estimate of the magnetic field in the region \citep{Hildebrand_2009, Houde_2009}. We call this modified DCF method as ADF-DCF hereinafter. We describe the generalities of the technique and its difference with DCF in the following section.
            %
            %%% --- Begin Figure -----------------------------------------------------------
            \begin{figure*}[ht!]
                \epsscale{1.18}
                \plotone{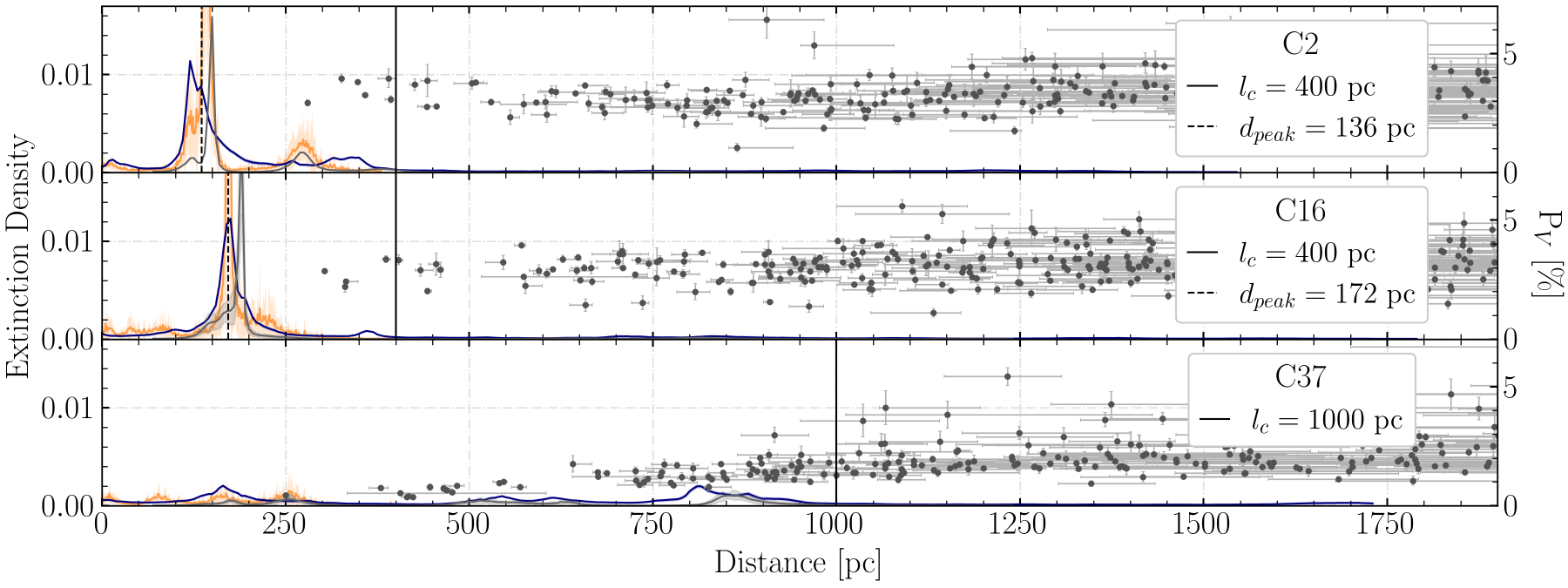}
                \caption{Extinction density as a function of the distance retrieved from \cite{Vergely_2022} at $10$~pc resolution in units of mag~pc$^{-1}$ (the blue solid line), \cite{Leike_2020} at $1$~pc resolution in units of e-foldings~pc$^{-1}$ (the orange solid line), and \cite{Edenhofer_dustmap_2023} with resolution between $0.4-7$~pc in the units of the parameter $E$ from \cite{Zhang_2023} (the gray solid line) in direction of \textit{C2} (top), \textit{C16} (middle), and \textit{C37} (bottom). The gray dots show the IPS-GI optical polarization as a function of the distance. The vertical black dashed lines show the distance to the dust peaks taken from \cite{Zucker_2021}. The vertical black solid lines are the total dust depth assumed, $l_c$.
                \label{fig:Av_dens_vs_dist}}
            \end{figure*}
            %%%    

        %...... Begin Sub-section ......................................................
        \subsubsection{Implementing ADFs in the DCF method: ADF-DCF} \label{subsec:MF_cal_ADF_H09_B0} 
        
            We can now apply the DCF method, which traditionally is used by equating the turbulent to regular magnetic field ratio $\langle B_{t}^2\rangle^{1/2}/B_{pos}$ to the dispersion in polarization angle $\sigma_{\theta}$. This assumes that the polarization angle dispersion is caused solely by turbulence, which is not completely valid in the case \textit{3.b} discussed above. Considering the turbulent contribution in Equation~(\ref{eq:SF_tot_fit}), \cite{Hildebrand_2009} showed that, in this case, the turbulent to regular magnetic field ratio can be described in terms of the turbulent parameter $b$ as
            \begin{equation}
            \label{eq:Bt_B0_ratio}
                \frac{\langle B_{t}^2\rangle^{1/2}}{B_{pos}} = \frac{b}{\sqrt{2-b^2}} ~.
            \end{equation} 
            
            Equation~(\ref{eq:Bt_B0_ratio}) can then be used to define the magnetic field strength in terms of the gas velocity dispersion, the density, and the turbulent parameter as
            \begin{equation}
            \label{eq:ADF_DCF}
                B_{pos} \simeq \sqrt{8\pi\rho} \frac{\sigma_{v}}{b}   ~~~~~~ \textrm{for}~~~ B_{t} \ll B_{pos} ~.
            \end{equation}
            If the observed scale is smaller than the turbulent scale (i.e.,~$\delta > \ell_{max}$), we can no longer use Equation~(\ref{eq:ADF_DCF}) to calculate the magnetic field strength. Instead, we can implement the classic DCF method (Equation~(\ref{eq:DCF}), Section~\ref{subsec:MF_cal_DCF}) or the \cite{Skalidis_Tassis_2021} method (Equation~(\ref{eq:ST_DCF}), Section~\ref{subsec:MF_cal_ST-DCF}), assuming that the field of view is larger or equal to the outer scale of fluctuations so that the angle dispersion is a good measure of the magnetic field fluctuations in the DCF approximation.

%...... Begin Section ......................................................
\section{IPS-GI field selection and properties} \label{sec:selec_fields}

    In the following sections, we describe the properties of the IPS-GI fields selected for this work, along with the reasons for their selection. We also highlight the impact of extreme values, subsampling, and small fields of view on the ADF's performance.
    \\

    %...... Begin Section ......................................................
    \subsection{IPS-GI field selection criteria} \label{subsec:selec_fields_criteria}
    
        The ADF probes the plane-of-sky magnetic field fluctuations by comparing the polarization angles between different lines of sight as a function of the physical distance between them. As we observe angular distances, the 2D ADF defined in Equation~(\ref{eq:ADF}) can only be applied when we can map angular to physical distances. This is only the case if the dust distribution is similar for all stars, i.e.,\ we probe the same dust in front of all stars in the sample. 

        In that context, we focus on the IPS-GI fields where starlight passes through a single, clearly defined, diffuse polarizing dust structure. The IPS-GI intermediate Galactic latitude fields, i.e.,~\textit{C2},  \textit{C16}, and perhaps \textit{C37}, at $|b| \ga 7\fdg5$, are good candidates for this study. \cite{Angarita_2023} found that these fields have, on average, low optical dust extinction ($A_V < 1$ mag). Additionally, \cite{Angarita_2023} demonstrated that optical starlight polarization in \textit{C2}, \textit{C16}, and \textit{C37} originates in the nearby ISM, in or just behind the Local Bubble (LB) wall. This is evident, especially in \textit{C2} and \textit{C16}, as most stars behind the dust structures exhibit a constant degree of polarization regardless of distance.
        
        Figure~\ref{fig:Av_dens_vs_dist} shows the extinction density profile with distance in the three fields. We show the models from \cite{Vergely_2022}, \cite{Leike_2020}, and \cite{Edenhofer_dustmap_2023} (Section~\ref{subsec:obs_dust_maps}) to assess the consistency of the dust structures at different resolutions. We see dust structures at $d<400$~pc in \textit{C2} and \textit{C16}, and $d<1,000$~pc  in \textit{C37}. Optical polarization measurements (the gray dots) are mostly behind these dust structures, confirming the origin of the polarization measured in the foreground. Therefore, we can assume that all lines of sight cross the polarizing dust at comparable distances, and the angular scales between stars correspond to actual physical scales within the cloud (Figure~\ref{fig:Av_dens_vs_dist}). The assumption holds less for \textit{C37}. Unfortunately, there are only a handful of data points at $d<1,000$~pc in \textit{C37} (Figure~\ref{fig:Av_dens_vs_dist}), so we can not disentangle the polarization properties of the individual structures. However, the low density of the dust, the constant optical polarization degree beyond $1,000$~pc, and the highly regular magnetic field in \textit{C37} \citep[see][for more details]{Angarita_2023} make it feasible to assume uniform polarizing properties in all dust structures along a pathlength of $1,000$~pc.
            
        %...... Begin Sub-section ......................................................
        \subsubsection{Polarization angle distribution in intermediate-latitude IPS-GI fields} \label{subsec:selec_fields_SF_PA_gradient} 

            As explained in Section~\ref{subsec:MF_cal_ADF_interpre}, the interpretation of the ADF depends on the relative observed scales concerning the turbulence correlation length. Interpretation of ADFs is different whether observed fluctuations in polarization angle are part of a turbulent cascade or large-scale gradients in the cloud. The IPS-GI field-of-view is exceedingly small (${\sim}1$~pc at the distance of the dust peak of Figure~\ref{fig:Av_dens_vs_dist}) but may be comparable to the turbulent outer scales in cold clouds (see Section~\ref{subsec:Disc_turb_scales}). The large-scale distribution of polarization angle may give a clue as to the question of whether observed fluctuations in polarization angle are due to turbulence or gradients. 
            
\begin{figure}[ht!]
                \epsscale{1.16}
                \plottwo{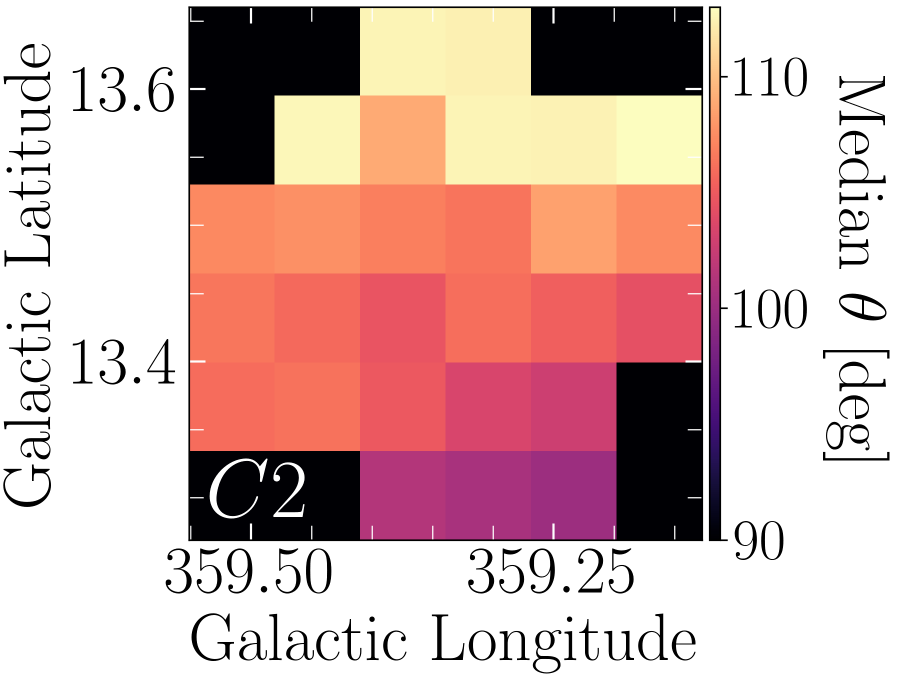}{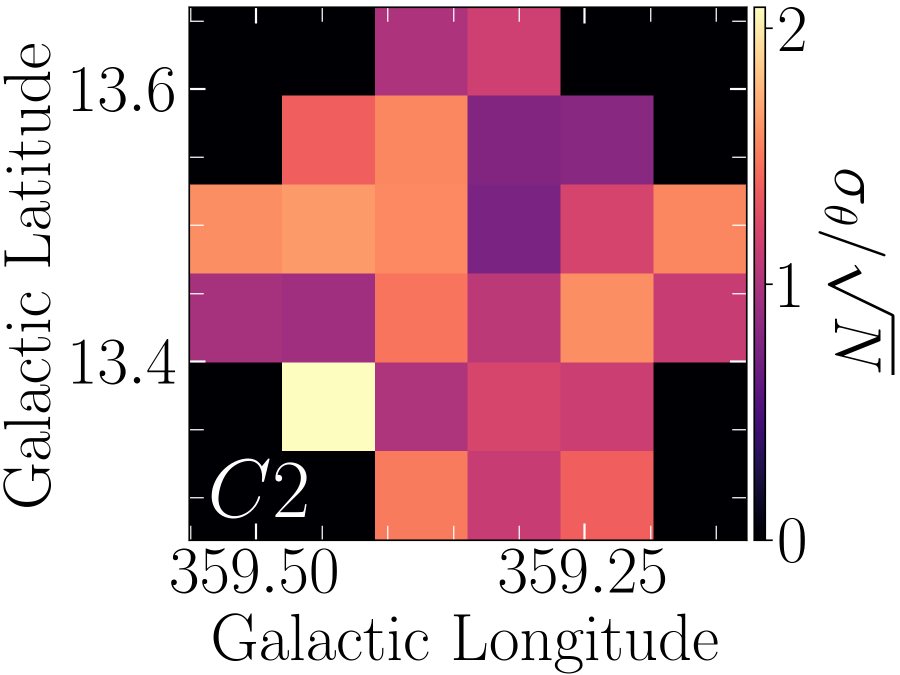}
                \epsscale{1.16}
                \plottwo{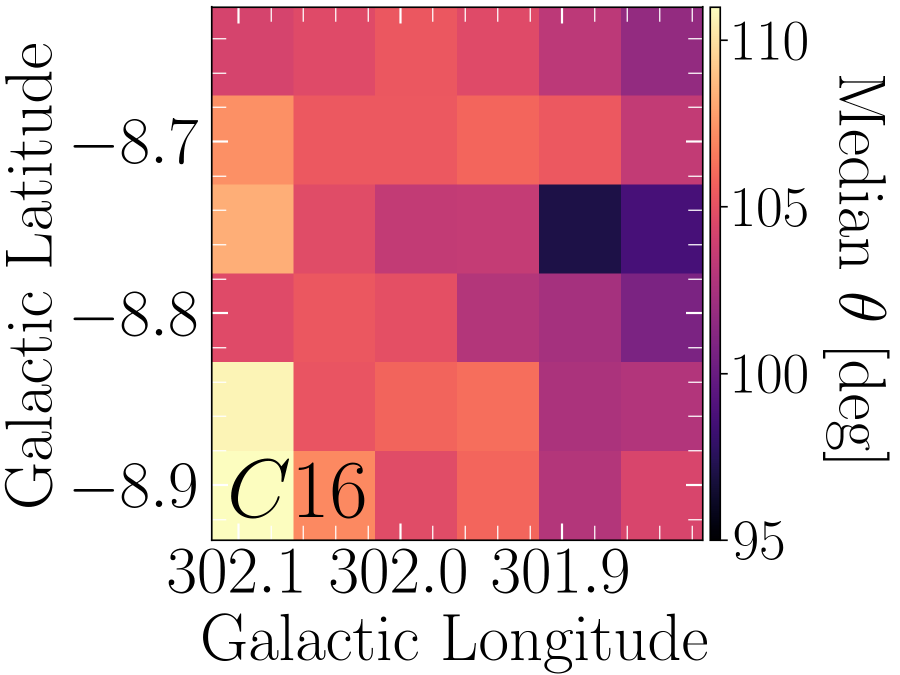}{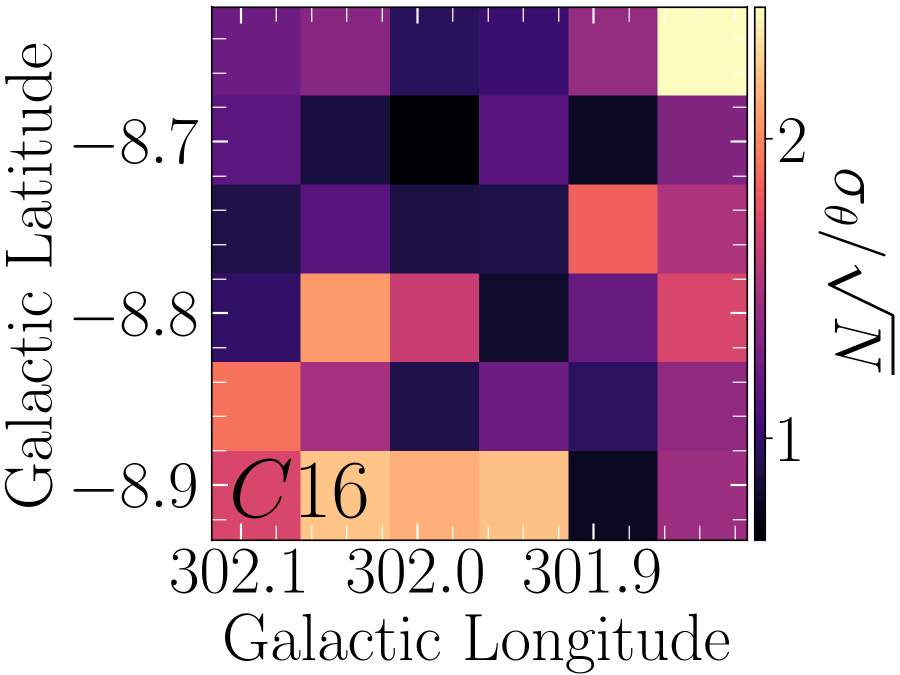}
                \epsscale{1.16}
                \plottwo{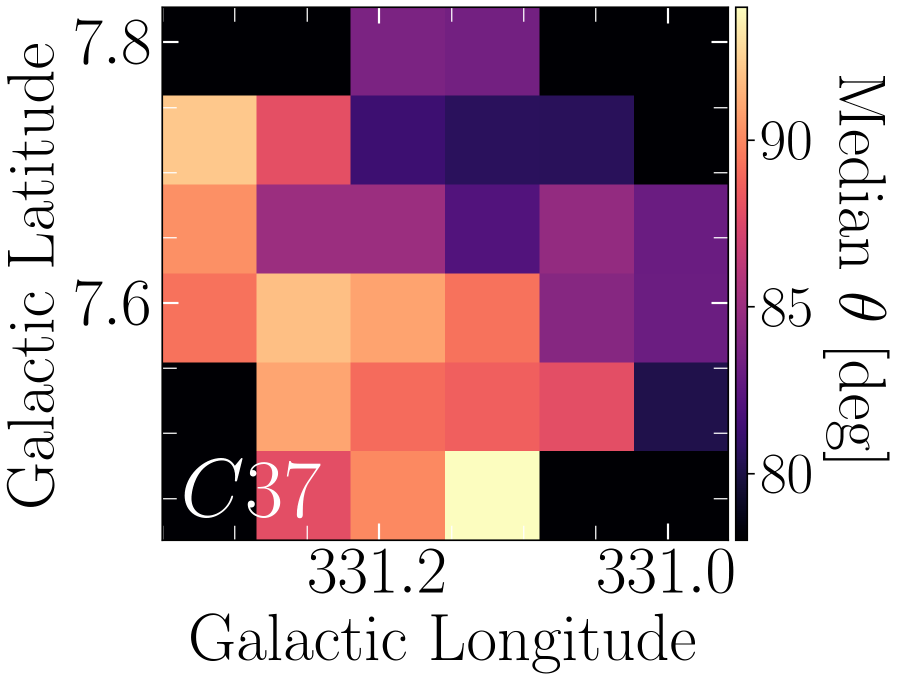}{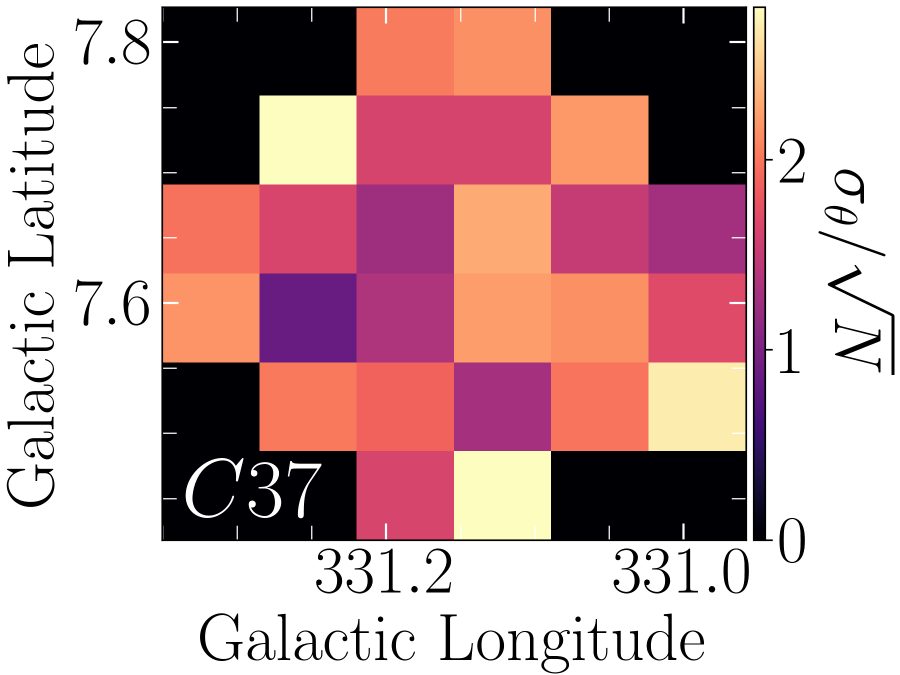}
                \caption{Distribution of the median polarization angle (left column) and the standard error (right column) in the field-of-view of \textit{C2} (top row), \textit{C16} (middle row), and \textit{C37} (bottom row).
                \label{fig:heatmap_PA_gradient}}
            \end{figure}

            In Figure~\ref{fig:heatmap_PA_gradient} (left), we see a gradual change in polarization angle across the fields of view. The smooth change in orientation by approximately $10^{\circ}$ suggests that the dispersion is likely caused by large-scale fluctuations from a structure larger than the field of view, but partially inside the field (i.e.,~$\delta<\ell$), rather than pure turbulence (i.e.,~$\delta>\ell_{max}$). Although it may be impossible to make the distinction definitively, the gradual change in the median $\theta$ is statistically significant according to the standard error in the right column of Figure~\ref{fig:heatmap_PA_gradient}. Interestingly, the direction of the change in the magnetic field orientation coincides with the direction of the \ion{H}{1} emission gradient in both the observations and the models, lending some more credence to the idea that the angle gradient across the field is part of a large-scale structure (see Section~\ref{subsec:selec_fields_ISM_proper} for details in \ion{H}{1} observations and modeling, see also the Figures in Appendix~\ref{sec:Append_HI_decomp} for comparison).
            %
            %%% --- Begin Figure -----------------------------------------------------------
            %%% -------------------------------------------------------------------------- 
            %    
            %
            %%% --- Begin Figure -----------------------------------------------------------
            \begin{figure}[t!]
                \epsscale{1.18}
                \plotone{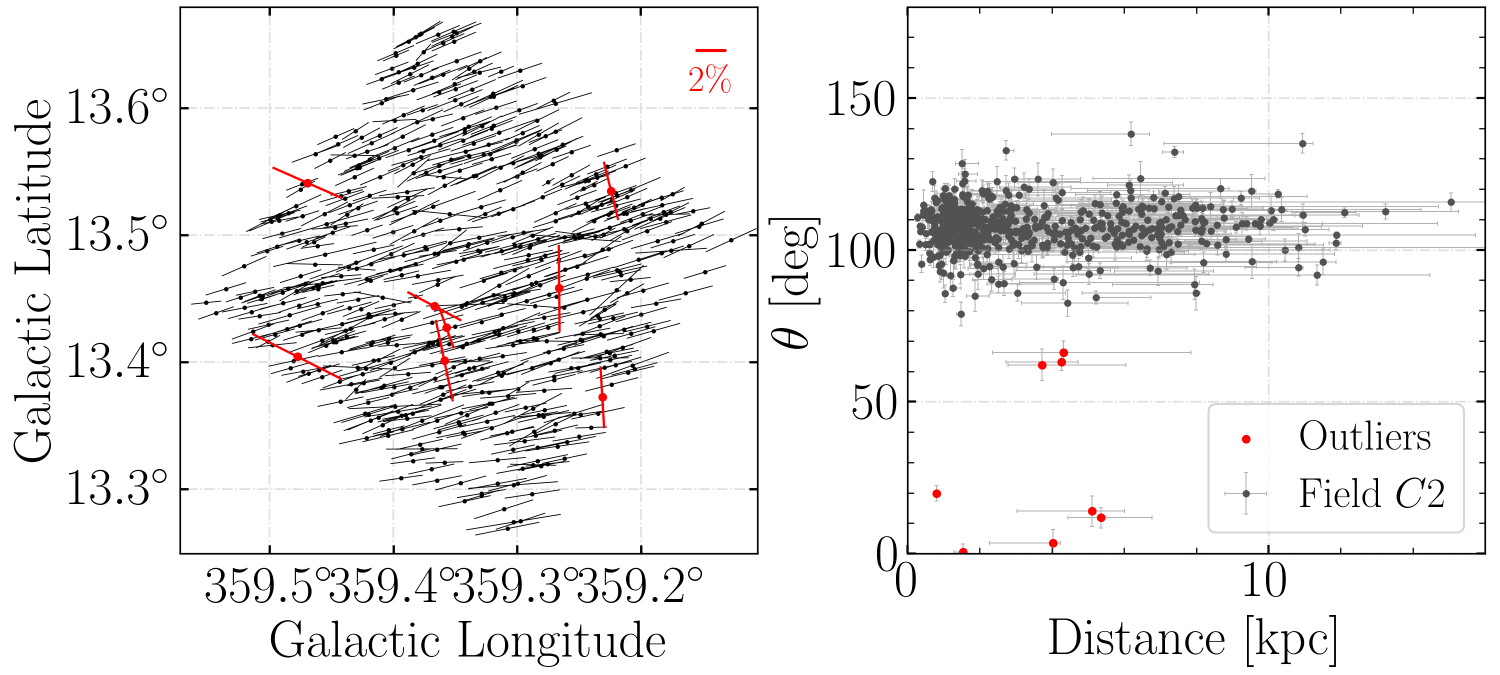}
                \epsscale{1.18}
                \plotone{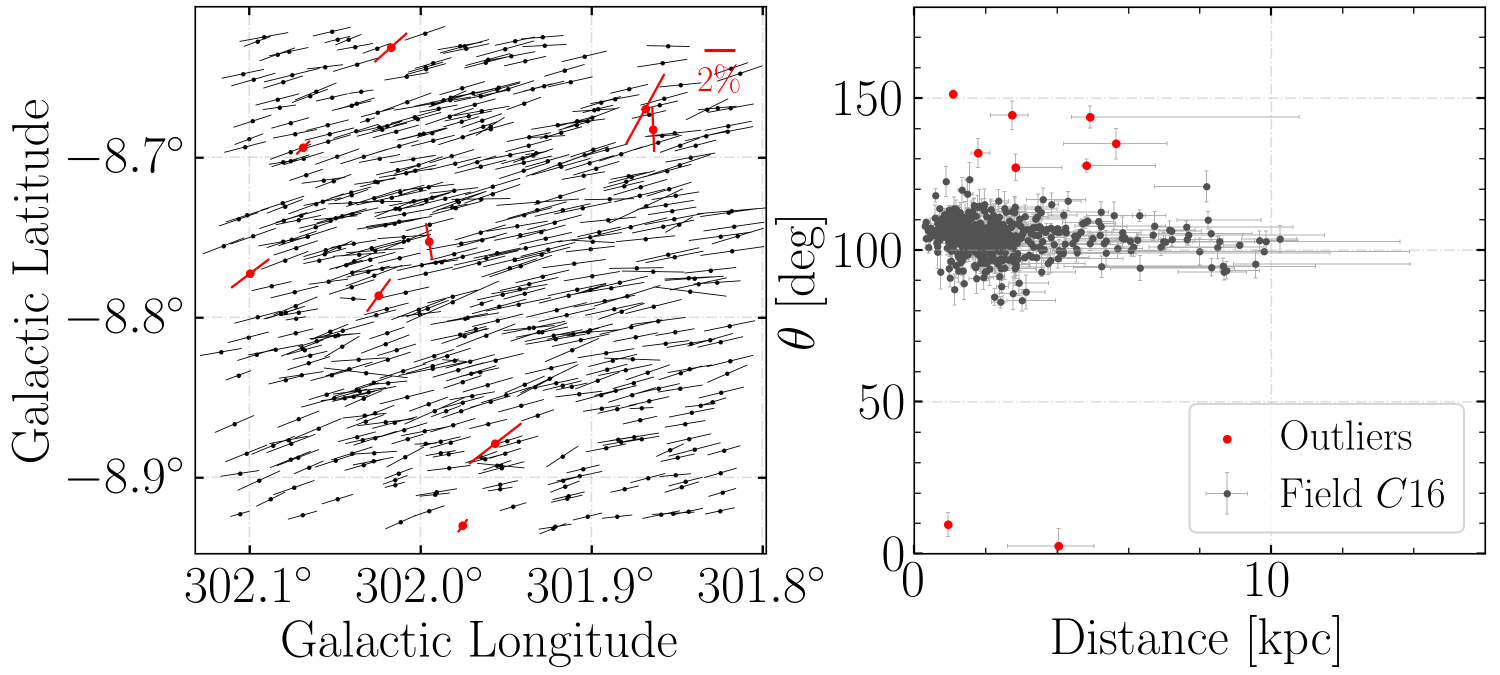}
                \epsscale{1.18}
                \plotone{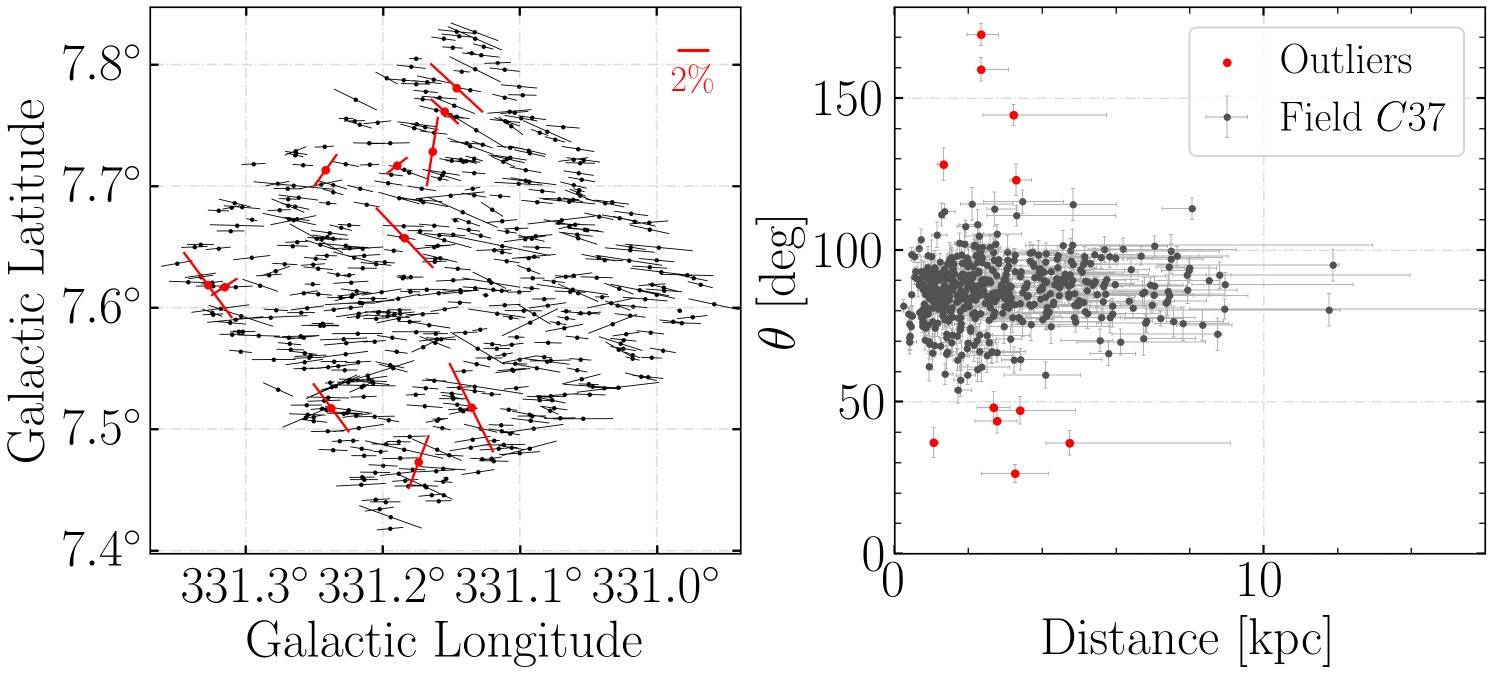}
                \caption{Optical polarization vectors map (left column) and polarization angle as a function of distance (right column) of fields \textit{C2} (top row), \textit{C16} (middle row), and \textit{C37} (bottom row). The red vectors represent unreliable polarization angle measurements above $3\sigma$ from the mean distribution and are excluded from the calculation of the ADF.
                \label{fig:vectors_map_outliers}}
            \end{figure}
            %%% --------------------------------------------------------------------------
            %
            %%% --- Begin Figure -----------------------------------------------------------
            \begin{figure*}[ht!]
                \epsscale{0.376}
                \plotone{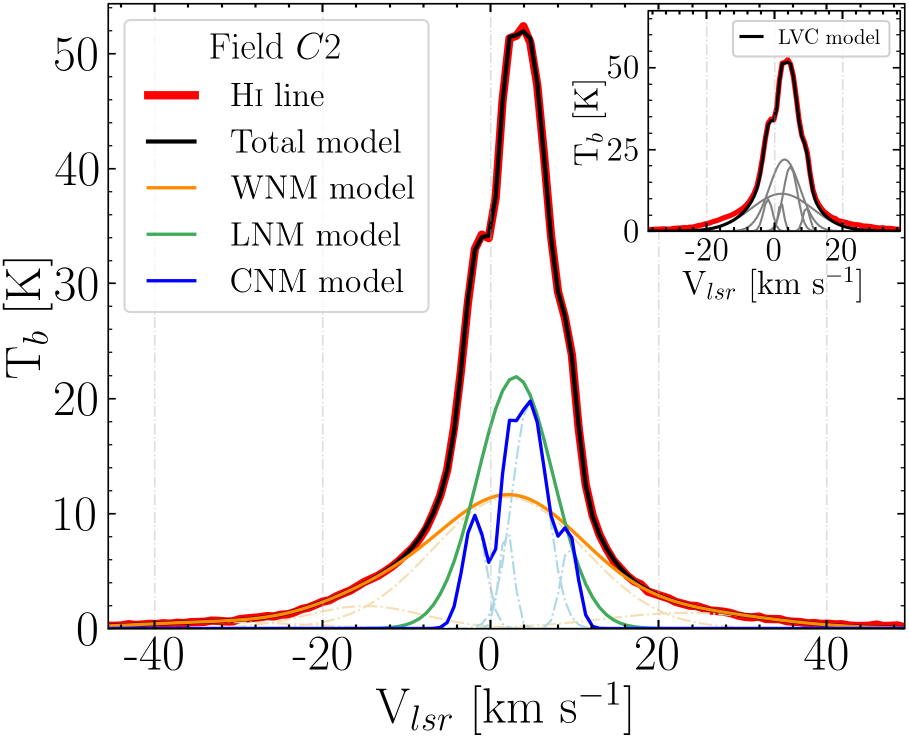}
                \epsscale{0.376}
                \plotone{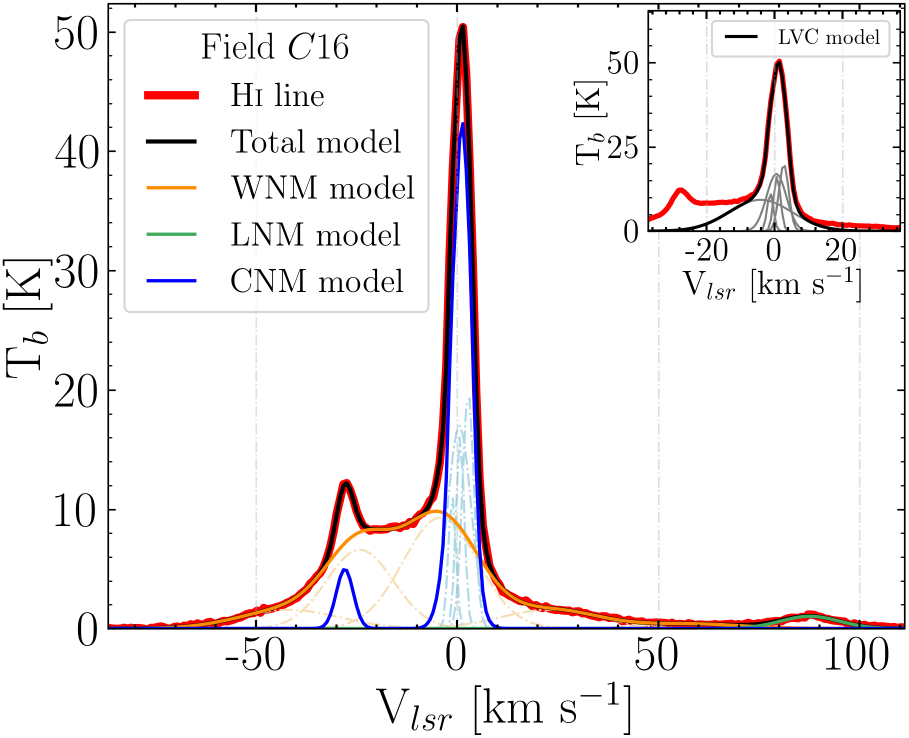}
                \epsscale{0.376}
                \plotone{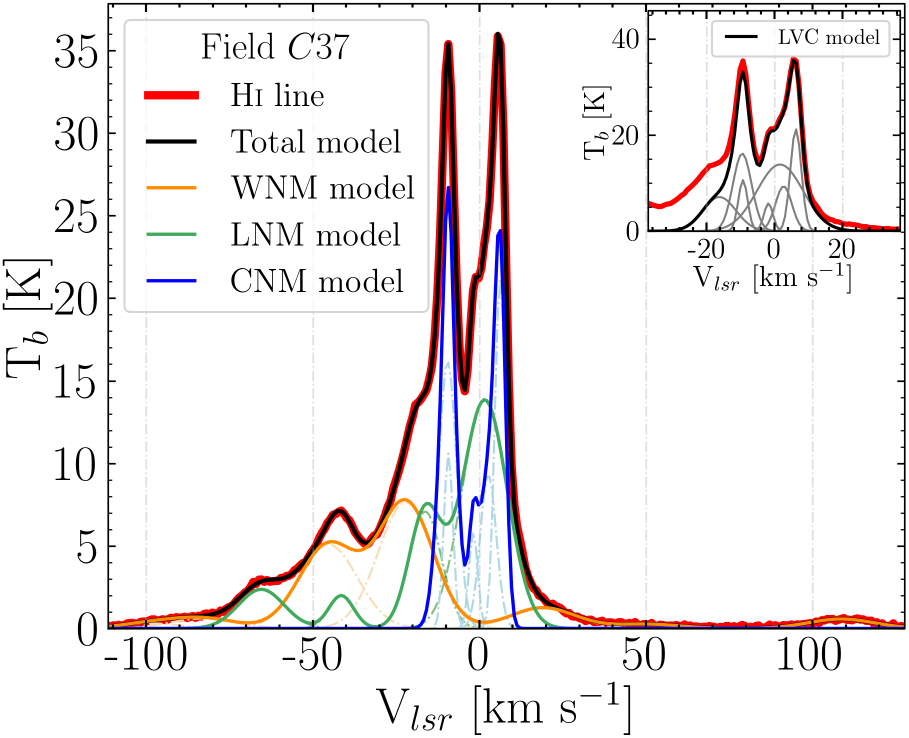}
                \caption{\ion{H}{1} emission line of GASS (red solid line) compared to ROHSA's total model (black solid line) and the CNM (blue solid line), LNM (green solid line), and WNM (orange solid line) models in the center of the map. The insets in each panel present the model built with the components of the low-velocity clouds (LVCs, $|V_{lsr}| < 20$~km~s$^{-1}$) from the ROHSA decomposition.
                \label{fig:ROHSA_Gauss_comp}}
            \end{figure*}
            %%% -------------------------------------------------------------------------  
            %
            %%% --- Begin Figure -----------------------------------------------------------
            \begin{figure*}[ht!]
                \epsscale{0.376}
                \plotone{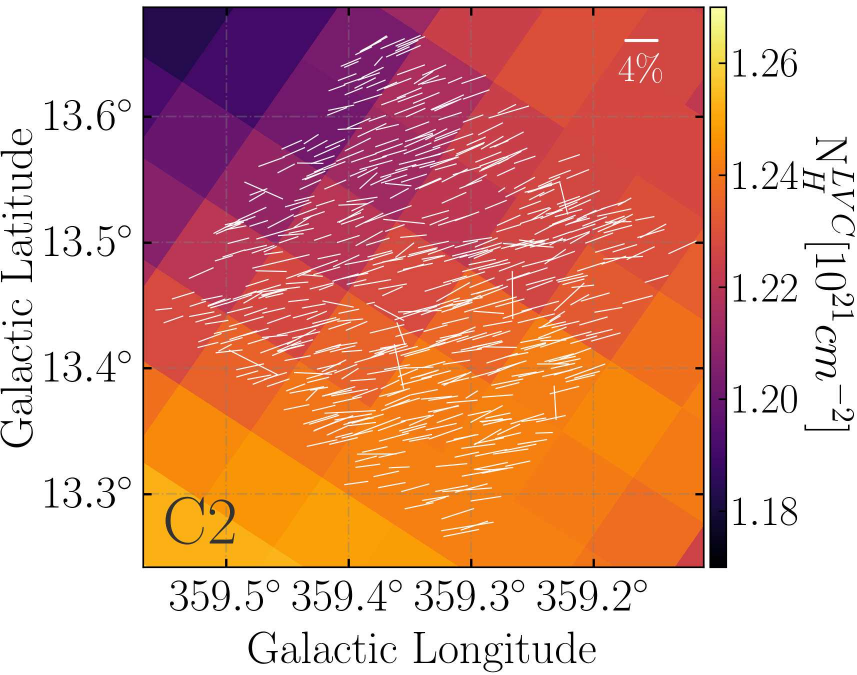}
                \epsscale{0.376}
                \plotone{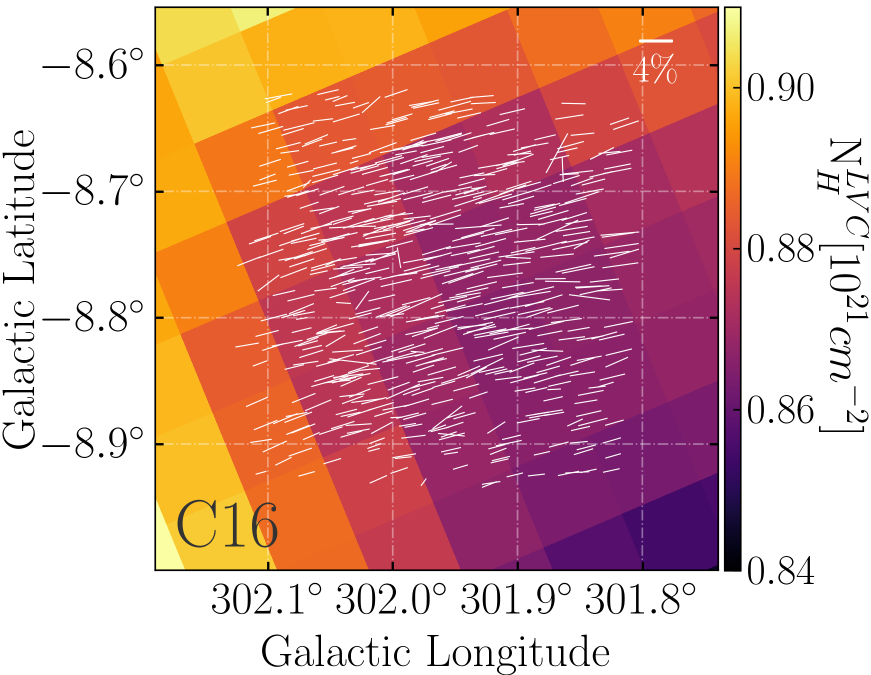}
                \epsscale{0.376}
                \plotone{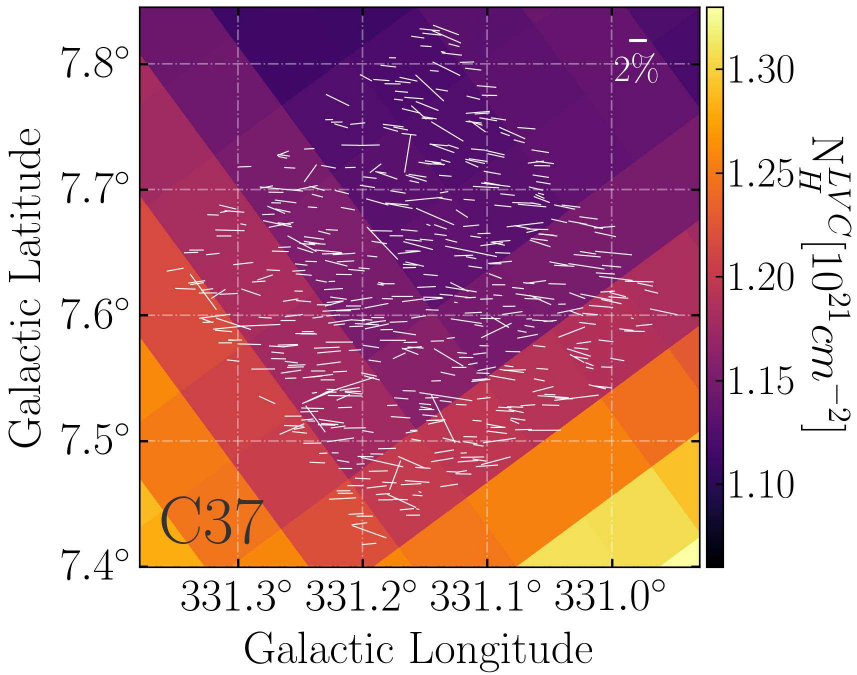}
                \caption{Gas column density of the LVC model obtained from ROHSA decomposition in field \textit{C2} (left), \textit{C16} (middle), and \textit{C37} (right). The white pseudo-vectors represent the IPS-GI optical polarization measurements for comparison.    
                \label{fig:NH_GASS_Model_Planck}}
            \end{figure*}
            %%% --------------------------------------------------------------------------

        %...... Begin Sub-section ......................................................
        \subsubsection{Extreme values and subsampling issues} \label{subsec:selec_fields_SF_outlier_subsam}          
            
            The ADFs are a powerful tool to measure fluctuations in the ISM. Still, they are very sensitive to extreme values and non-uniform distributions of the measurements \citep[e.g.,~see][]{Sun_Han_SF_2004, Soler_2016}. Firstly, to account for outlier issues, we removed data points with polarization angles exceeding three times the standard deviation, as they showed no significant correlation with structures along the lines of sight (the red polarization vectors and data points in Figure~\ref{fig:vectors_map_outliers}). The atypical polarization values are not correlated with either distance or sky position and are likely due to circumstellar matter causing intrinsic polarization in these stars. Additionally, it can also be caused by observational issues, for instance, in crowded regions where the superposition of different sources can produce mismatches between the ordinary and extraordinary images created by the Savart prism \citep{Magalhaes_2005, Versteeg_2023}. Secondly, we do not trust angular scales above $2/3$ (i.e.,~${\sim}0\fdg2$) of the observed field width since subsampling leads to large uncertainties. Thirdly, polarization angle dispersion enhancements or ``bumps'' may happen at a certain range of scales, leading to more than one slope in the ADF profile. Given that all fields studied present a gradient in the median polarization angle (Figure~\ref{fig:heatmap_PA_gradient}), we attribute the bumps to irregular sampling of the sources \citep{Soler_2016}.

    %...... Begin Section ......................................................
    \subsection{ISM properties in IPS-GI fields} \label{subsec:selec_fields_ISM_proper}
    
        To characterize the magnetic field strength, we need density $\rho$ and velocity dispersion $\sigma_v$ measurements of the ISM structure coupled to the magnetic field. The emission and absorption of atomic and molecular lines such as CO, \ion{K}{1}, \ion{Ca}{2}, OH, CN, and \ion{H}{1}, among others, have been extensively used to estimate the properties of the diffuse and dense clouds \citep[see e.g.][and references therein]{Heiles_Troland_2005, Andersson_Potter_CoalSack_2005, Medan_Andersson_2019}. Although the intermediate-latitude IPS-GI fields are often outside the coverage of the existing atomic and molecular line observations, we have full-sky \ion{H}{1} emission observations in the Southern sky (Section~\ref{subsec:obs_HI_map}).
    
        We use \ion{H}{1} PPV cubes from GASS to find the gas properties that better correlate with the polarizing dust structure in our lines of sight. To this end, we perform a Gaussian decomposition using the Regularized Optimization tool for Hyper-Spectra Analysis, ROHSA \citep{Marchal_ROHSA_2019}, which provides comprehensive information on the multiphase interstellar gas properties. This approach should be used with caution since not all neutral \ion{H}{1} gas might be associated with the polarizing structures, even though magnetohydrodynamic (MHD) turbulence permeates all different ISM phases \citep[e.g., see][]{Beresnyak_2019}. We obtained models for the cold, lukewarm, and warm neutral medium (CNM, LNM, and WNM, respectively), as well as the total atomic hydrogen emission in a region approximately four times larger than the IPS-GI fields. An example of the models centered on each IPS-GI field is presented in Figure~\ref{fig:ROHSA_Gauss_comp}.  The three fields present a total emission peaking principally at low velocities ($|V_{lsr}|{\sim}1.6$~km~s$^{-1}$), denoting that most of the gas is local. Furthermore, the local \ion{H}{1} emission peak comprises at least two ISM phases, with the CNM always being present (Figure~\ref{fig:ROHSA_Gauss_comp}).       
        Details of the Gaussian decomposition performed on each IPS-GI field can be found in Appendix~\ref{sec:Append_HI_decomp}. The following sections describe our $\rho$ and $\sigma_{v}$ calculations. All the measured properties are summarized in Table~\ref{tab:fields_proper}. 

        %
        % %%% --- Begin Table -----------------------------------------------------------
        \begin{deluxetable*}{llccc}
            \tablecaption{Properties of the intermediate-latitude IPS-GI fields. \label{tab:fields_proper}}
            \tablehead{
            \colhead{Parameter} & \colhead{Convention/units} & \colhead{\textit{C2}} & \colhead{\textit{C16}} & \colhead{\textit{C37}} 
            } 
            \startdata
            \textit{Total dust depth}                                           & $l_c$ (pc) & 400 & 400 & 1,000 \\[0.15cm]
            \textit{Cloud effective thickness}                                  & $l_{eff}$ (pc) & ${\sim}15$ & ${\sim}10$ & ${\sim}127$ \tablenotemark{*} \\[0.15cm]
            \textit{Cloud distance}                                             & $d_{peak}$ (pc) & $136_{-21}^{+14}$ & $172_{-1}^{+1}$ & 500 \tablenotemark{\dag} \\[0.15cm]
            \textit{Smallest scale observed}                                    & $\ell_{ips}$ (pc) & $0.024_{-0.004}^{+0.002}$ & $0.030_{-0.0002}^{+0.0002}$ & 0.087\tablenotemark{\dag} \\[0.15cm]
            \textit{Largest scale observed}                                     & $\ell_{FoV}$ (pc) & 0.712 & 0.901 & 2.618\tablenotemark{\dag} \\[0.15cm]
            \textit{Polarization angle dispersion corrected}                    & $\sigma_{\theta,corr}$ (deg) & 5.3 & 3.9 & 7.5 \\[0.15cm]
            \textit{Turbulent parameter}                                        & $b$ (deg) & $8.64\pm0.05$ & $6.53\pm0.03$ & $11.46\pm0.08$ \\[0.15cm]
            \textit{Turbulent to regular magnetic field ratio $^a$}             & $\rm{\langle B_t^2 \rangle^{1/2}/B_{pos}}$ & 0.11 & 0.08 & 0.14 \\[0.15cm]
            \textit{Total hydrogen column density associated to the dust $^b$}  & $N_\mathrm{H}^{A_V}$ (10$^{21}$~cm$^{-2}$) & $1.33\pm0.019$ & $1.08\pm0.004$ & $1.11\pm0.009$  \\[0.15cm]
            \textit{Average GASS atomic hydrogen column density $^c$}           & $\langle N_\mathrm{H}^{GASS} \rangle$ (10$^{21}$~cm$^{-2}$) & 1.38 & 1.42 & 1.90 \\[0.15cm]
            \textit{Average LVC atomic hydrogen column density}                 & $\langle N_\mathrm{H}^{LVC} \rangle$ (10$^{21}$~cm$^{-2}$) & 1.23 & 0.87 & 1.18 \\[0.15cm]
            \textit{Total volume density}                                       & $n_\mathrm{H}$ (cm$^{-3}$) & ${\sim}27$ & ${\sim}27$ & ${\sim}3$ \tablenotemark{*} \\[0.15cm]
            \textit{Average LVC mass density}                                   & $\langle \rho^{LVC} \rangle$ (10$^{-23}$~g~cm$^{-3}$) & 4.52 & 4.52 & 0.50 \\[0.15cm]
            \textit{Average total velocity dispersion}                          & $\langle \sigma_{T_b}\rangle$ (km~s$^{-1}$) & 3.52 & 1.23 & 7.04 \\[0.15cm]
            \textit{Average turbulent velocity dispersion}                      & $\langle \sigma_{v,turb}\rangle$ (km~s$^{-1}$) & 3.37 & 1.14 & 6.67 \\[0.15cm]
            \textit{Average turbulent velocity dispersion scaled $^d$}          & $\langle \sigma_{v,turb}\rangle_{ips}$ (km~s$^{-1}$) & 0.395 & 0.163 & 0.588\tablenotemark{\dag}  \\[0.15cm]
            \textit{}                                                           & $\langle \sigma_{v,turb}\rangle_{FoV}$ (km~s$^{-1}$) & 1.23 & 0.506 & 1.83\tablenotemark{\dag}  \\[0.15cm]
            \textit{}                                                           & $\langle \sigma_{v,turb}\rangle_{\ell_{1pc}}$ (km~s$^{-1}$) & 1.38 & 0.524 & 1.33\tablenotemark{\dag}  \\
            \enddata
            \tablecomments{\textit{(a)} The uncertainties propagated from the turbulent parameter errors are of the order of $10^{-3}$ or below. \textit{(b)} Integrated up to the cloud depth $l_{c}$, the uncertainty is propagated from the error of \citealt{Vergely_2022} models. \textit{(c)} The uncertainties propagated from the observed spectral \textit{rms} noise are in the order of $10^{17}$~cm$^{-2}$. \textit{(d)} Turbulent velocity dispersion scaled to the minimum scale probed, $\ell_{ips}$, the scale of the field of view, $\ell_{FoV}$, and $1$~pc scale, $\ell_{1pc}$.} 
            \tablenotetext{*}{An approximated average peak value rather than a maximum value.} 
            \tablenotetext{\dag}{Halfway to the total dust pathlength, $l_c = 1,000$~pc. Assumed as an average between the distances to all the dust peaks observed in Figure~\ref{fig:Av_dens_vs_dist}, bottom.} 
        \end{deluxetable*}

        %%% --------------------------------------------------------------------------

        %...... Begin Sub-section ......................................................
        \subsubsection{Hydrogen column density} \label{subsec:sel_fields_N_HI}
    
            We estimate the total column density of GASS \ion{H}{1} emission in the IPS-GI regions by integrating the brightness temperature over the entire range of velocity channels observed, 
            \begin{equation}
             \label{eq:NH_integr}
                \frac{N_{\mathrm{H}}^{GASS}}{\mathrm{(cm^{-2})}} = \frac{1.82\times10^{18}}{\mathrm{(cm^{-2} (K~km~s^{-1})^{-1})}} \int{\frac{T_b}{\mathrm{(K)}}\frac{dv}{\mathrm{(km~s}^{-1})}} ~,
            \end{equation} 
            where $dv$ is the velocity channel width equal to \mbox{0.824~km~s$^{-1}$} (see the maps in the Appendix~\ref{sec:Append_HI_decomp}, Figure~\ref{fig:ROHSA_models}). Equation~(\ref{eq:NH_integr}) is only valid under the assumption of optically thin emission \citep{Dickey_1990}\footnote{According to \cite{Dickey_1990}, the optically thin emission assumption is possible when the brightness temperature is about few tens.}, which is likely to be the case of the diffuse intermediate-latitude IPS-GI fields with volume densities below $30$~cm$^{-3}$ within a cloud with thickness $l_{eff}$ (see Table~\ref{tab:fields_proper} and Section~\ref{subsec:selec_fields_nH}).
    
            The total atomic hydrogen column density observed, $N_{\mathrm{H}}^{GASS}$, is integrated along the entire Galaxy. However, we need to evaluate the column density only in the region where the polarizing dust screen exists (see Figure~\ref{fig:Av_dens_vs_dist}), i.e.,~very nearby gas ($d<400$~pc in fields \textit{C2} and \textit{C16}, and $d<1,000$~pc in \textit{C37}). ROHSA decomposition of \ion{H}{1} spectra accommodates for this by distinguishing low-, intermediate-, and high-velocity clouds (LVCs, IVCs, and HVCs, respectively). We expect the gas associated with the polarizing dust structure to be part of an LVC (with $|V_{lsr}|<20$~km~s$^{-1}$, \citealt{Wakker_2001}). Using the LVC model of the \ion{H}{1} decomposition, consisting of all the Gaussian components at $|V_{lsr}|<20$~km~s$^{-1}$ (see the insets of Figure~\ref{fig:ROHSA_Gauss_comp}, also see Appendix~\ref{sec:Append_HI_decomp}), and Equation~(\ref{eq:NH_integr}), we calculated the column density map of the nearby gas ($N_\mathrm{H}^{LVC}$, Figure~\ref{fig:NH_GASS_Model_Planck}). 
            
            Alternatively, we can find an approximation to the total hydrogen column density (i.e.,~$N_{\ch{HI}} + 2N_{\ch{H2}}$) associated only with the dust, $N_\mathrm{H}^{A_V}$, by using the column density to reddening ratio, $N_{\mathrm{H}}/$E$(\bv)$, from \cite{bohlin_1978}, the total to selective extinction ratio of A$_V/$E$(\bv) = 3.1$ \citep{Savage_1979, Fitzpatrick_2004}, and the extinction density profiles from \cite{Vergely_2022} (Figure~\ref{fig:Av_dens_vs_dist}). The gas number density in the three fields is calculated as 
            \begin{equation}
             \label{eq:nH_av}
                \frac{N_{\mathrm{H}}^{A_V}}{\mathrm{(cm^{-2})}} = \frac{5.8\times10^{21}}{\mathrm{(cm^{-2} mag^{-1})}} \int_{l_{c}}{\frac{a_V(l)/3.1}{\mathrm{(mag~cm^{-1})}} \frac{dl}{\mathrm{(cm)}}} ~,
            \end{equation}
            where $a_V(l)$ is the extinction density (i.e.,~extinction per distance unit, also known as differential extinction), $l_{c}$ is the depth of the cloud assumed from the extinction density profiles (Figure~\ref{fig:Av_dens_vs_dist}, also see Table~\ref{tab:fields_proper}), and $dl = 5$~pc is the distance step.
            
            The $N_\mathrm{H}^{A_V}$, $\langle N_\mathrm{H}^{GASS} \rangle$, and $\langle N_\mathrm{H}^{LVC} \rangle$ values per IPS-GI region are shown in Table~\ref{tab:fields_proper}. The $\langle N_\mathrm{H}^{LVC} \rangle$ values account for roughly $61\%$ to $96\%$ of the $\langle N_\mathrm{H}^{GASS} \rangle$, indicating that a substantial fraction of the neutral gas resides in the local ISM. Furthermore, $\langle N_\mathrm{H}^{LVC} \rangle$ is very consistent with the corresponding $N_\mathrm{H}^{A_V}$ values calculated in the assumed total depth of the dust, $l_{c}$. The good agreement suggests a connection between the modeled atomic hydrogen column density and the dust structure within $l_{c}$. We, therefore, use $N_\mathrm{H}^{LVC}$ maps presented in Figure \ref{fig:NH_GASS_Model_Planck} to calculate the density of the polarizing cloud.

        %...... Begin Sub-section ......................................................
        \subsubsection{Volume density} \label{subsec:selec_fields_nH}

            The volume density maps based on the 3D dust map of \cite{Leike_2020} were used by \cite{Zucker_2021} to construct the average radial density profiles of local molecular clouds (i.e.,~local structures mostly within and beyond the LB wall, $d{\sim}100-400$~pc). The radial density profiles were fitted with a two-Gaussian function, leading to the characterization of an inner and outer layer with different standard deviations (referred to as ``widths'') and peak densities for each molecular cloud considered. \cite{Zucker_2021} found inner widths between $2.5-4.9$~pc, with peak densities of $25-52$~cm$^{-3}$, meanwhile, the outer widths are between $8-18$~pc and densities of $5-15$~cm$^{-3}$. The 3D volume densities reported by \cite{Zucker_2021} are likely lower limits due to the systematic uncertainties propagated from the 3D dust map.

            Following the basic idea from \cite{Zucker_2021}, we used the extinction density profiles from \cite{Vergely_2022} and \cite{Leike_2020} (see Figure~\ref{fig:Av_dens_vs_dist}) to estimate the peak volume density of the local dust structure found along our lines of sight. We followed the same conventions as in Section~\ref{subsec:sel_fields_N_HI} to convert the extinction density of \cite{Vergely_2022} into volume density considering distance steps of $5$~pc. Meanwhile, we used the conversion $n_\mathrm{H} = 880$~cm$^{-3} \times S_x$ to convert the $G$-band extinction density of \cite{Leike_2020} into volume density; see more details in \cite{Zucker_2021}.

\begin{figure}[ht!]
                \epsscale{2.125}
                \plottwo{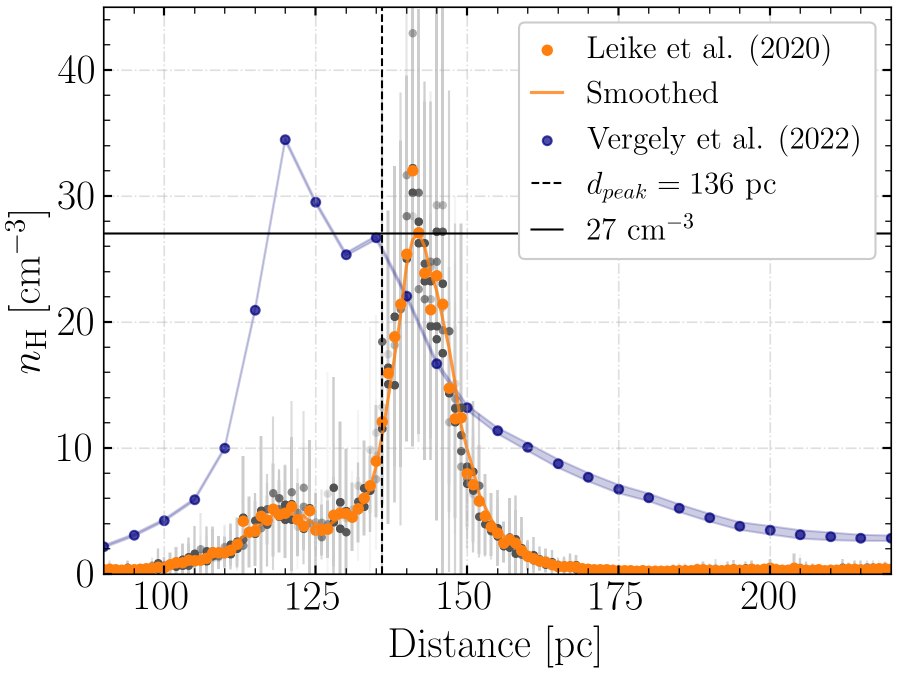}{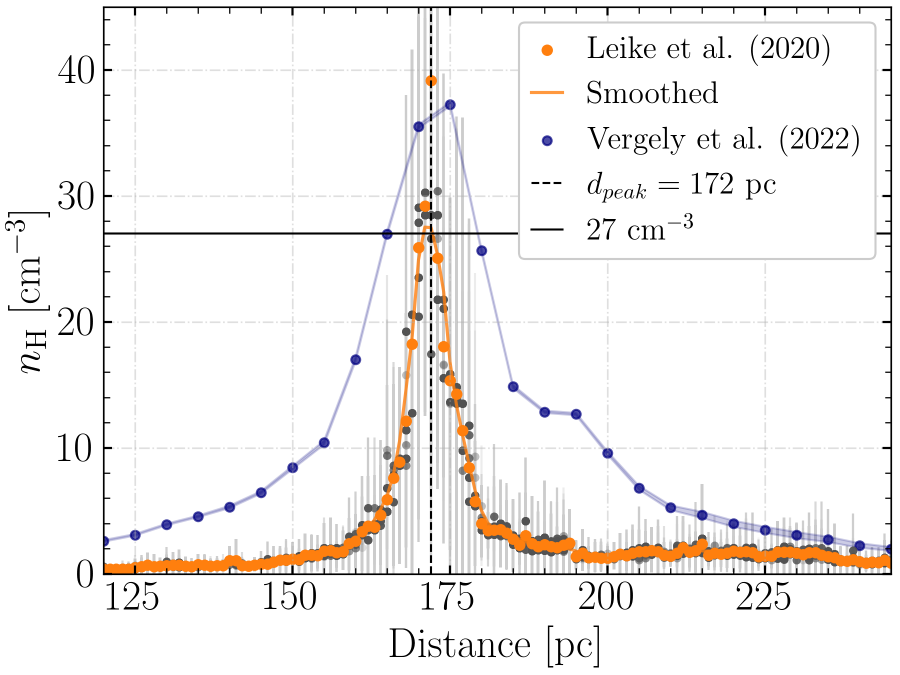}
                \epsscale{1.0625}
                \plotone{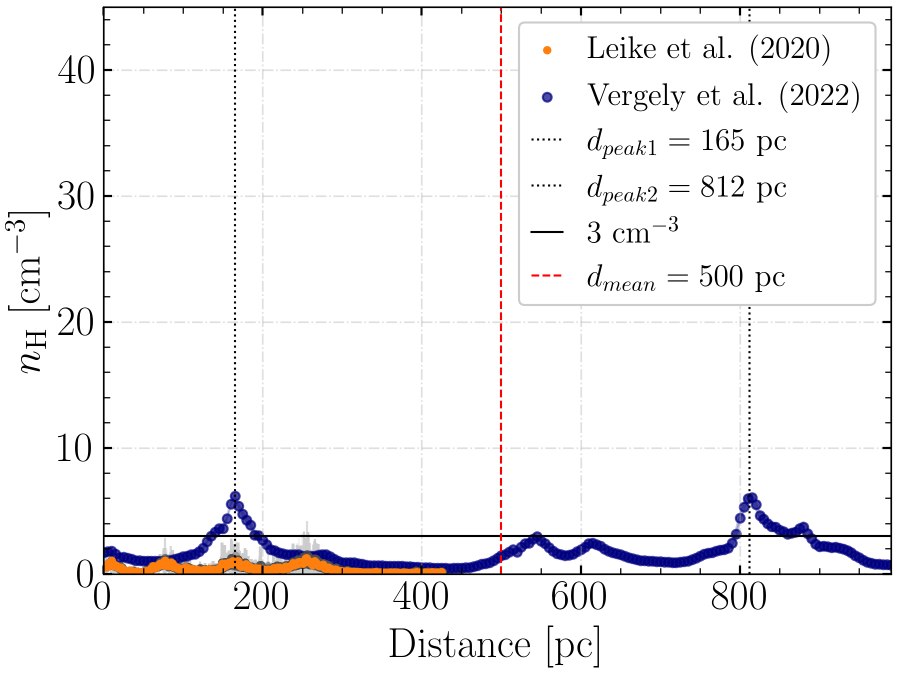}
                \caption{Volume density as a function of the distance obtained from \cite{Leike_2020} (the gray and orange dots), and \cite{Vergely_2022} (the blue dots) 3D dust maps in \textit{C2} (top), \textit{C16} (middle), and \textit{C37} (bottom). The orange solid line is the mean smoothed curve, with kernel 4. The gray bars are the standard deviation. The blue shaded curve is the error from \cite{Vergely_2022}. The horizontal black line marks the approximated peak volume density. The vertical black dashed line is the distance to the cloud in \textit{C2} and \textit{C16}, which is assumed from \cite{Zucker_2021}. In the case of \textit{C37}, the vertical red dashed line is the mean distance of the entire dust distribution.
                \label{fig:nH}}
            \end{figure}

            Figure~\ref{fig:nH} presents the total volume density profiles of fields \textit{C2}, \textit{C16}, and \textit{C37}. Considering the resolution limitations of the two dust maps compared (i.e.,~$2$~pc\footnote{Despite the claimed map resolution of $1$~pc, \cite{Leike_2020} consider that only structures at spatial scales greater than $2$~pc should be considered reliable; see also discussion from \cite{Zucker_2021}.} for \cite{Leike_2020} and $10$~pc for \cite{Vergely_2022}), we approximate the total volume density to $27$~cm$^{-3}$ in fields \textit{C2} and \textit{C16}. This is approximately the average peak value between the main dust structures observed in both dust maps compared. On the other hand, \textit{C37} presents a series of diffuse structures along a pathlength of $1,000$~pc depth. As it is mentioned in Section~\ref{subsec:selec_fields_criteria}, we have to assume a single structure with average properties, i.e.,~a $n_\mathrm{H} {\sim}3$~cm$^{-3}$ (see the black horizontal line in Figure~\ref{fig:nH}, bottom). 
            %
            %%% --- Begin Figure -----------------------------------------------------------
            %%% --------------------------------------------------------------------------

            Comparing our observations with the 3D models of \cite{Zucker_2021}, our field \textit{C2} lies approximately within the outer envelope modeled for Ophiuchus molecular cloud\footnote{\url{https://faun.rc.fas.harvard.edu/czucker/Paper_Figures/3D_Cloud_Topologies/ophiuchus_topology/ophiuchus.html}}. Meanwhile, \textit{C16} is within the inner layer of Musca molecular cloud\footnote{\url{https://faun.rc.fas.harvard.edu/czucker/Paper_Figures/3D_Cloud_Topologies/musca_topology/musca.html}}. This implies, based on \cite{Zucker_2021}'s findings, that the volume densities should be around the amplitudes found for the outer envelope of Ophiuchus: $12.6^{+1.1}_{-1.1}$~cm$^{-3}$, and the inner component of Musca: $31.8^{+1.5}_{-1.5}$~cm$^{-3}$. The latter is consistent with the value obtained for \textit{C16} from Figure~\ref{fig:nH}. However, it is more uncertain since the Musca molecular cloud is mostly unresolved by \cite{Leike_2020} 3D dust map at small radial distances (i.e.,~$< 2$~pc, see explanation in \citealt{Zucker_2021}). On the other hand, the profile of \textit{C2} in Figure~\ref{fig:nH} demonstrates that our field of view may be closer to the amplitude of the inner component of Ophiuchus, which is $n_\mathrm{H} = 31.1^{+1.6}_{-1.5}$~cm$^{-3}$.
            
            The $n_\mathrm{H}$ values obtained from 3D dust models include all the phases and molecular hydrogen (i.e.,~$n_\mathrm{H} = n_{\ch{HI}} + 2n_{\ch{H2}}$). However, we showed that the atomic hydrogen column density in the nearby ISM of the IPS-GI regions dominates in a large portion compared to that of the \ch{H2} (see the comparison between $N_\mathrm{H}^{A_V}$ and $\langle N_\mathrm{H}^{LVC} \rangle$ values, Section~\ref{subsec:sel_fields_N_HI}, and Table~\ref{tab:fields_proper}). Therefore, we expect the total volume density to have a contribution primarily from the neutral atomic medium.
         
        %...... Begin Sub-section ......................................................
        \subsubsection{Cloud effective thickness} \label{subsec:selec_fields_cloud_thickness}

            The cloud effective thickness can be obtained through the ratio between the average value of the LVC column density map (Section~\ref{subsec:sel_fields_N_HI}) and the approximated volume density (Section~\ref{subsec:selec_fields_nH}) as follows,
            \begin{equation} 
                \label{eq:l_eff}
                l_{eff} ~\mathrm{(pc)} \simeq \frac{\langle N_\mathrm{H}^{LVC} \rangle}{n_\mathrm{H}}  ~.
            \end{equation}
            Using Equation~(\ref{eq:l_eff}) and the values of Table~\ref{tab:fields_proper}, we found an effective thickness of approximately $15$~pc, $10$~pc, and $127$~pc for \textit{C2}, \textit{C16}, and \textit{C37}, respectively. The first two values are consistent with the low widths modeled by \cite{Zucker_2021} in the local clouds (i.e.,~Ophiuchus and Musca molecular cloud). Since the IPS-GI pathlengths observed are actual segments that cross the 3D structures modeled by \cite{Zucker_2021}, we expect the effective thickness to be around two times the widths modeled.

            The values obtained with Equation~(\ref{eq:l_eff}) are within the widths of the highest dust peaks found along the line of sight in the extinction density profiles of \cite{Vergely_2022}, \cite{Leike_2020}, and \cite{Edenhofer_dustmap_2023} (Figure~\ref{fig:Av_dens_vs_dist} and Figure~\ref{fig:nH}). Higher resolution 3D dust maps (e.g.,~\citealt{Leike_2020} and \citealt{Edenhofer_dustmap_2023}) are expected to show smaller spatial scales in the ISM structures than lower resolution maps (e.g.,~\citealt{Vergely_2022}). Therefore, the \cite{Leike_2020} 3D dust map yields lower $l_{eff}$ values through \cite{Zucker_2021} method than those that could be estimated with \cite{Vergely_2022} or \cite{Edenhofer_dustmap_2023} 3D dust maps.
    
        %...... Begin Sub-section ......................................................
        \subsubsection{Mass density} \label{subsec:sel_fields_density}
            
            The local \ion{H}{1} emission in our fields comprises almost all the interstellar gas phases (see Figure~\ref{fig:ROHSA_Gauss_comp}, also see Appendix~\ref{sec:Append_HI_decomp}), and the dust in the nearby polarizing screen (Figure~\ref{fig:Av_dens_vs_dist}) is consistent with the gas represented by the LVC model (Section~\ref{subsec:sel_fields_N_HI}). Therefore, we calculate the mass density map of the local clouds using the LVC column density map and the constant cloud thickness as follows,
            \begin{equation} 
                \label{eq:rho}
                \rho^{LVC} ~\mathrm{(g~cm^{-3})} \approx \frac{N_\mathrm{H}^{LVC}}{l_{eff}} m_\mathrm{H}  ~,
            \end{equation}
            where $m_\mathrm{H}$ is the mass of the hydrogen atom. Figure~\ref{fig:dens_&_turb_disp} presents the mass density maps per IPS-GI field. The average values per IPS-GI field are shown in Table~\ref{tab:fields_proper}.

        %...... Begin Sub-section ......................................................
        \subsubsection{Turbulent velocity dispersion} \label{subsec:sel_fields_sigma_v}
    
            At high and intermediate Galactic latitudes, the cold medium is often associated with dust structures. For instance, \cite{Lenz_2017} found that, at low column densities ($N_{\mathrm{H}}< 4\times10^{20}$~cm$^{-2}$), \ion{H}{1} is in good agreement with dust maps. \cite{Hensley_2022} showed that the CNM fraction strongly correlates with the dust fraction in PAHs. Furthermore, the cold medium also traces the magnetic field properties. For instance, the \ion{H}{1} small-scale structures, which are preferentially cold gas, are oriented predominantly with the magnetic field  \citep{Clark_2015, Clark_2019, Clark_Hensley_2019}. Moreover, \cite{Lei&Clark_2023} found a strong correlation between the CNM mass fraction and thermal dust polarization, implying that the polarization is created in the densest cold regions. A very short pathlength characterizes this cold medium. On the other hand, the WNM seems to contribute very little (about $4\%$) to the polarization due to depolarization in the warm medium \citep[see][for more details]{Lei&Clark_2023}. Nevertheless, the internal dynamics of the CNM are related to the dynamics in the WNM according to numerical simulations of the local multiphase ISM \citep{Saury_2014}. 
            %
            %%% --- Begin Figure -----------------------------------------------------------
            \begin{figure}[t!]
                \epsscale{2.4}
                \plottwo{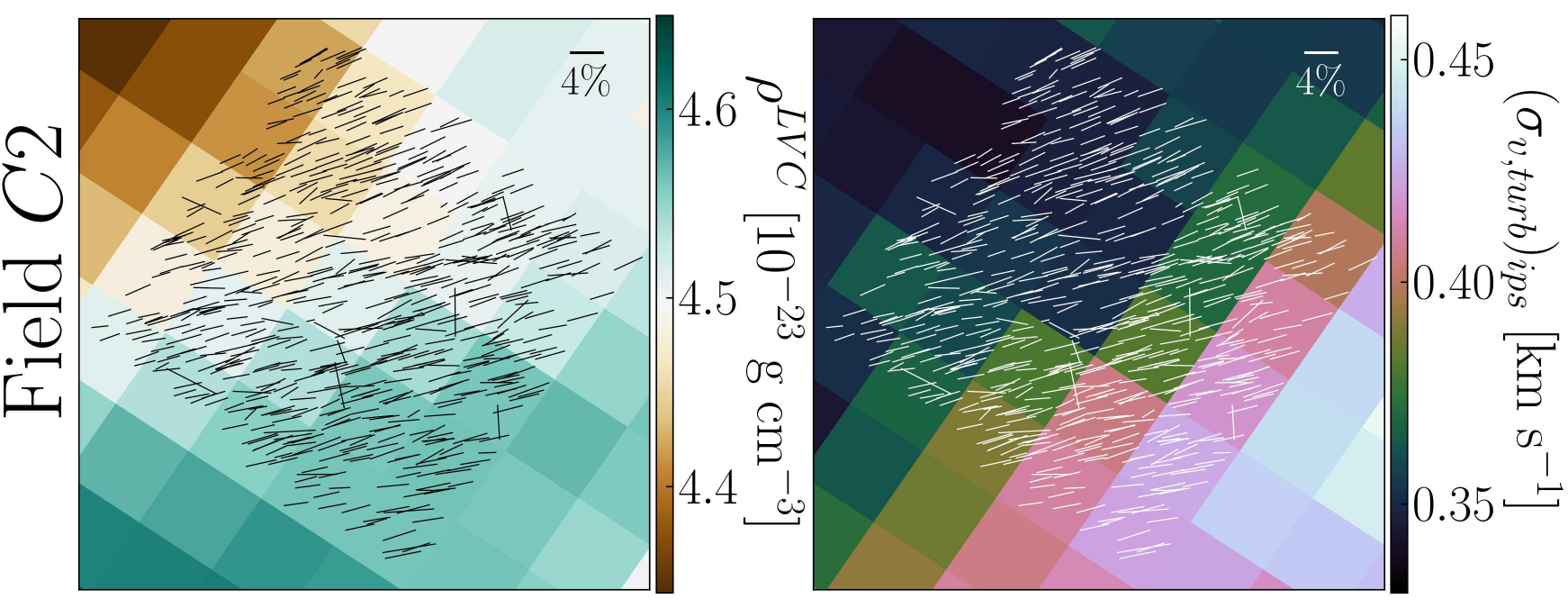}{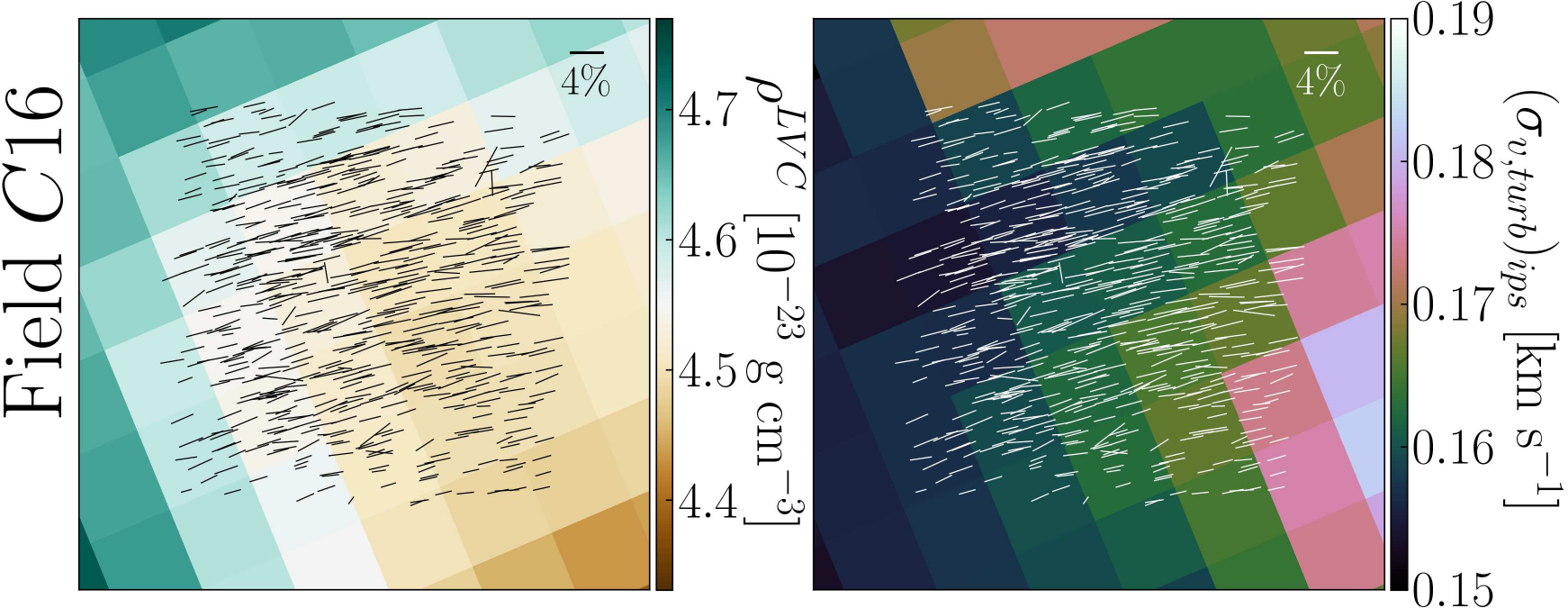}
                \epsscale{1.2}
                \plotone{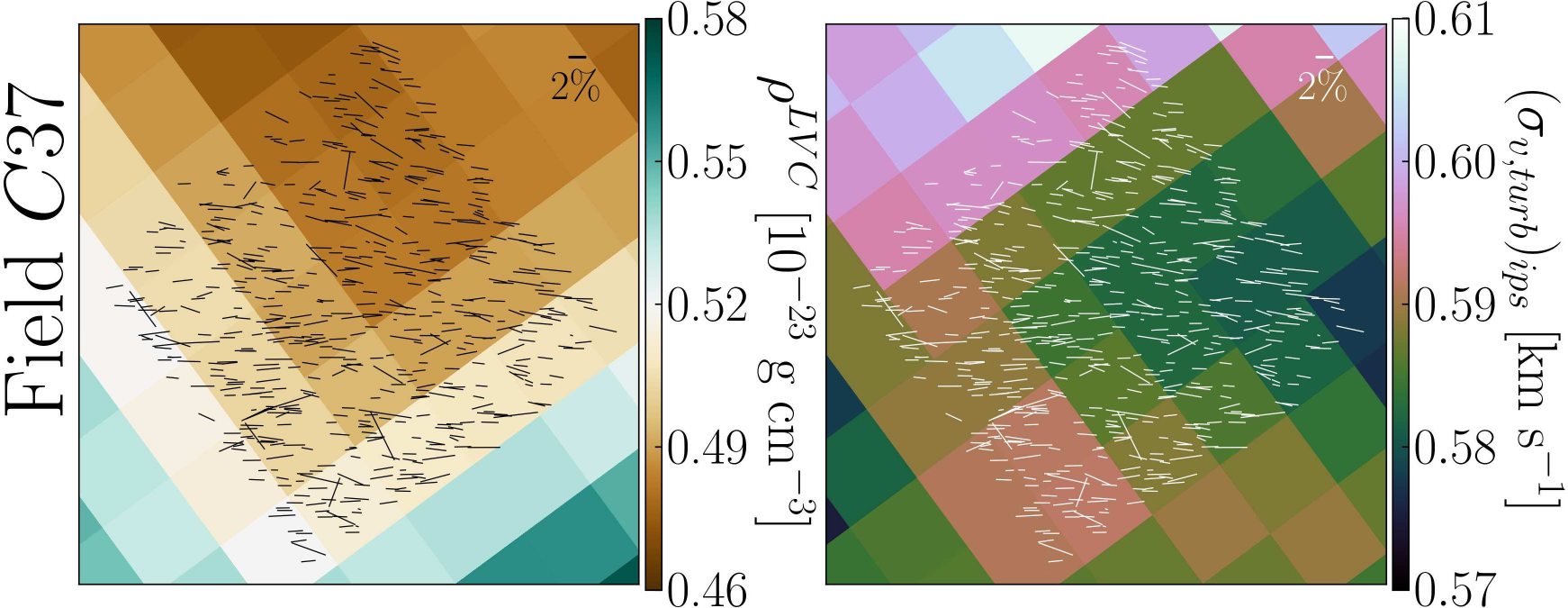}
                \caption{Mass density (left column) and scaled turbulent velocity dispersion (right column) 2D maps of fields \textit{C2} (upper row), \textit{C16} (middle row), and \textit{C37} (lower row). The black (left) and white (right) pseudo-vectors represent the IPS-GI optical polarization measurements for comparison.
                \label{fig:dens_&_turb_disp}}
            \end{figure}
            %%% --------------------------------------------------------------------------
            
            In this vein, following the approach from \cite{Marchal_2021}, we will treat the cold structures (CNM clouds) as test particles embedded within the warm medium (WNM). This ensures that the velocity dispersion among the cold clouds reflects the turbulent velocity field of the entire multiphase medium. Therefore, we calculate the velocity dispersion of the nearby cloud as the second-moment map of the \ion{H}{1} emission using only the CNM components of the LVC model obtained in the \ion{H}{1} Gaussian decomposition (Appendix~\ref{sec:Append_HI_decomp}) as follows,
            \begin{equation}
                \label{eq:vel_disp_M2}
                \frac{\sigma_{T_b}^2}{(\mathrm{km~s}^{-1})^2} = \frac{\sum_{i=0}^{N_{gauss}}{w_i (\mu_i-M_1)^2}}{\sum_{i=0}^{N_{gauss}}{w_i}} ~,
            \end{equation}
            where $N_{gauss}$ is the number of Gaussian components, $w_i = a_i\sigma_{v,i}$ is the weight, $a_i$, $\mu_i$, and $\sigma_{v,i}$ are the amplitude, central velocity, and velocity dispersion, respectively, of the $i$th component, and $M_1$ is the first-moment map of the \ion{H}{1} emission defined as
            \begin{equation}
                \label{eq:vel_M1}
                \frac{M_1}{(\mathrm{km~s}^{-1})} = \frac{\sum_{i=0}^{N_{gauss}}{w_i \mu_i}}{\sum_{i=0}^{N_{gauss}}{w_i}} ~.
            \end{equation}
        
            The velocity dispersion measured from \ion{H}{1} emission comprises a thermal and a turbulent contribution \citep[e.g., see][]{Marchal_2021} such that,
            \begin{equation}
                \label{eq:total_vel_disp}
                \sigma_{T_b}^2 = \sigma_{v, therm}^2 + \sigma_{v, turb}^2 ~.
            \end{equation}
            Approximating the turbulent sonic Mach number, $\mathcal{M}_s$, as the turbulent to thermal velocity dispersion ratio, 
            \begin{equation}
                \label{eq:mach_num}
                \mathcal{M}_s \approx \frac{\sigma_{v, turb}}{\sigma_{v, therm}} ~,
            \end{equation}
            we can use Equations~(\ref{eq:total_vel_disp}) and (\ref{eq:mach_num}) to calculate the CNM turbulent velocity dispersion map as follows,
            \begin{equation}
                \label{eq:turb_vel_disp}
                \sigma_{v, turb}^2 = \left( \frac{\mathcal{M}_s^2}{1 + \mathcal{M}_s^2}\right) \sigma_{T_b}^2  ~.
            \end{equation}        
            The turbulent Mach number for diffuse, cold neutral gas is typically around $3$, but with wide variations between $1-7$, according to \cite{Heiles_Troland_2003} and \cite{Murray_2015}. We, therefore, used the median Mach number, $\mathcal{M}_s=2.86$, from \cite{Murray_2015} (see their Figure~5), which is also in agreement with the typical value from \cite{Heiles_Troland_2003}. The final maps of the turbulent dispersion per field are presented in Figure~\ref{fig:dens_&_turb_disp}, and the average values are shown in Table~\ref{tab:fields_proper}. 
            %
            %%% --- Begin Figure -----------------------------------------------------------
            \begin{figure*}[ht!]
                \epsscale{0.377}
                \plotone{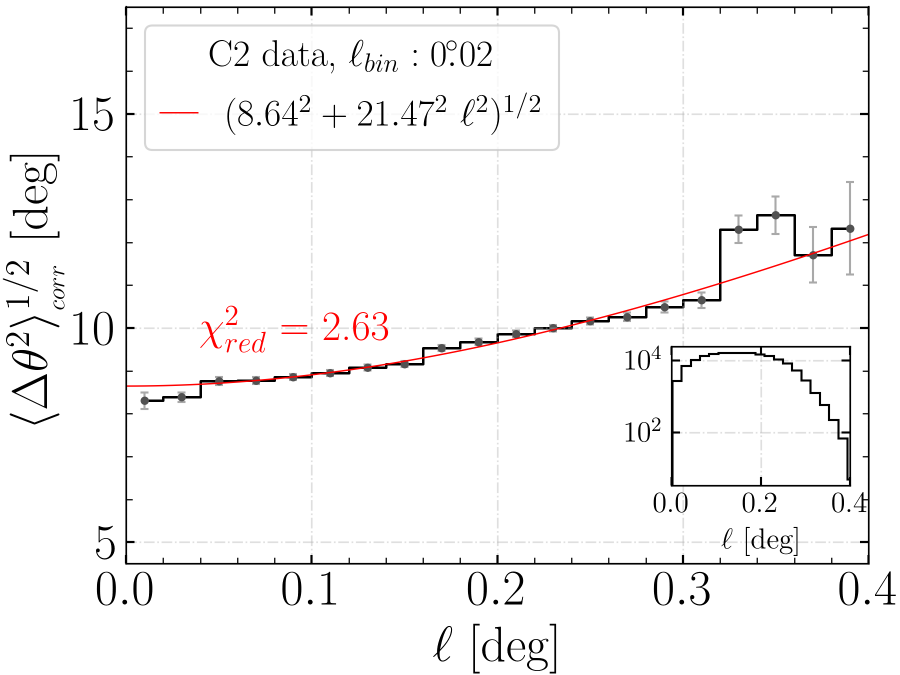}
                \epsscale{0.377}
                \plotone{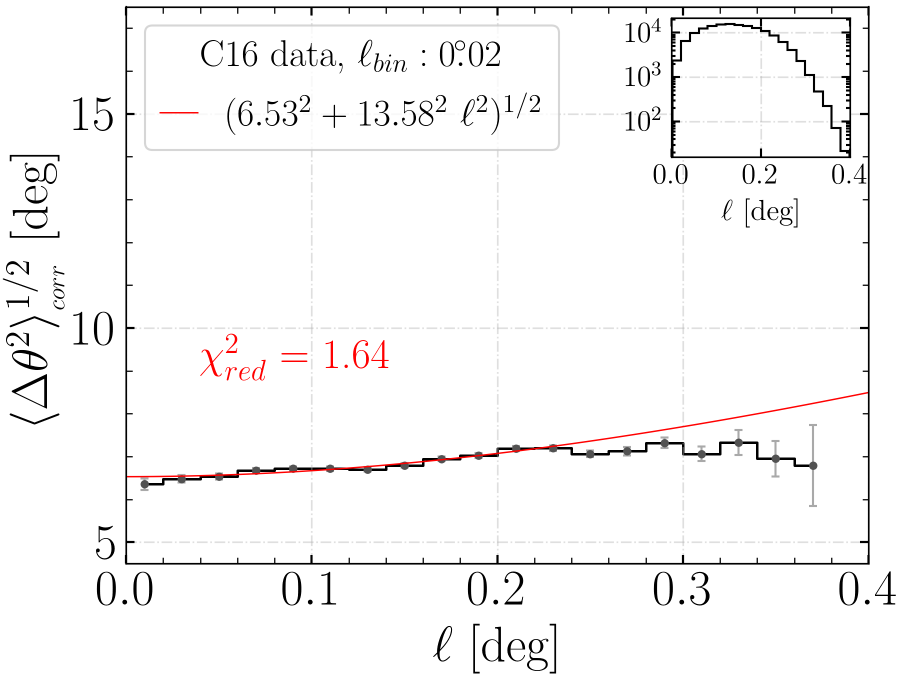}
                \epsscale{0.377}
                \plotone{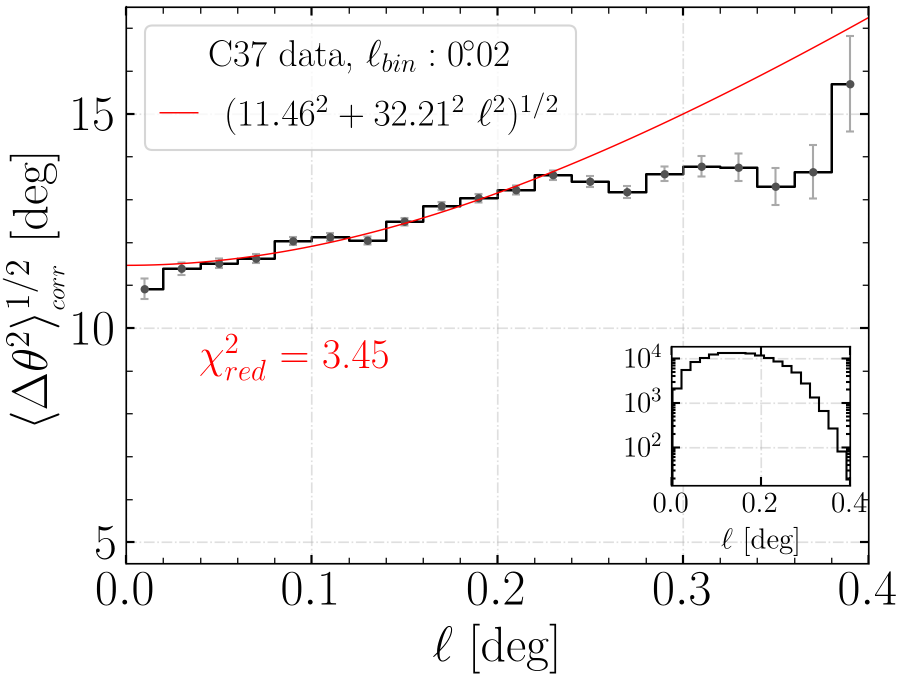}
                \caption{Dispersion of the polarization angle corrected by the measurement errors as a function of the scale length in field \textit{C2} (left), \textit{C16} (middle), and \textit{C37} (right). The red solid line is the best fit to the scales at which we see a linear increase in dispersion. The insets show a log-scale histogram depicting the number of sampled star pairs per angular scale bin.
                \label{fig:ADF}}
            \end{figure*}
            %%% --------------------------------------------------------------------------

            The scale at which the CNM turbulent velocity dispersion was measured in the PPV GASS cubes (i.e.,~along the line of sight) is larger than the largest spatial scale probed by the IPS-GI observations on the plane of the sky. In Sections~\ref{subsec:MF_cal_ADF_interpre} and \ref{subsec:selec_fields_SF_PA_gradient}, we argued that the presence of significant large-scale gradients in polarization angle across the IPS fields, corresponding to the regular magnetic field component, provides evidence for a scenario involving small-scale turbulence within the medium. Therefore, we can scale down the gas velocity dispersion $\sigma_{v, turb}$ to the scales of the IPS-GI fields as
            \begin{equation}
                \label{eq:IPS_CNM_turb_scale}
                (\sigma_{v,turb})_{ips} \approx \sigma_{v,turb} \left ( \frac{\ell_{ips}}{l_{eff}}\right )^{q} ~,
            \end{equation}
            where the minimum angular scale probed by the IPS-GI fields, $\ell_{ips} = 0\fdg01$, is assumed to be the upper limit of the turbulent correlation length (see Section~\ref{subsec:Resul_ADF}), corresponding to $0.024-0.030$~pc at a distance of $136-172$~pc in fields \textit{C2} and \textit{C16}, respectively (Table~\ref{tab:fields_proper}). The distances to the clouds result from the median value of the models in \cite{Zucker_2021} (see also $d_{peak}$ values in Figure~\ref{fig:nH}). In the same fields, the scale of the cold neutral medium is assumed to be $l_{eff}{\sim}10-15$~pc, i.e.,~the approximated effective thickness of the cloud (Section~\ref{subsec:selec_fields_cloud_thickness}). Let us recall that the cold medium in diffuse ($N_\mathrm{H}<10^{21}$~cm$^{-2}$) high-latitude ($|b|>30\degr$) regions is expected to have very short pathlengths and produce most of the polarization with high magnetic alignment according to \cite{Lei&Clark_2023}. 
            
            Finally, the exponent $q$ in Equation~(\ref{eq:IPS_CNM_turb_scale}) depends on the structure in the medium and is in principle unknown. For a Kolmogorov-like spectrum \citep{Kolmogorov_1941, Frisch_1995}, and also in the case of anisotropic MHD Alfv\'enic turbulence in the strong regime \citep{Goldreich_Sridhar_1995}, an exponent $q = 1/3$ is expected, although slightly different values are observed in \ion{H}{1} clouds \citep{Larson_1979} or predicted in models for the scales below the inertial range \citep{Boldyrev_2002}. We adopt \cite{Wolfire_2003} here, who propose that $q = 1/3$ is the appropriate value for the smaller length scales in the CNM. To probe our parameters in the scaling law of the turbulent velocity field, we calculated the mean velocity dispersion at $1$~pc scale, $\langle \sigma_{v,turb} \rangle_{\ell_{1pc}}$ (Table~\ref{tab:fields_proper}), finding values between ${\sim}0.5-1.3$~km~s$^{-1}$ consistent with the average values found in cold regions of the ISM, $0.4-1.2$~km~s$^{-1}$ \citep{Wolfire_2003, Miville-Deschenes_2017}.

            The scaling factor of the turbulent velocity dispersion in Equation~(\ref{eq:IPS_CNM_turb_scale}) is more difficult to determine in \textit{C37} due to the complex distribution of dust structures within a long pathlength ($l_c=1,000$~pc, Figure~\ref{fig:Av_dens_vs_dist}). However, to obtain an approximated estimation of the ISM properties in \textit{C37}, we assume a distance to the cloud roughly at $500$~pc, halfway to the total pathlength of the dust observed in Figure~\ref{fig:nH}, bottom. This leads to observed scales ($\ell_{ips}$) around $0.087$~pc. Nevertheless, we acknowledge the large uncertainties due to the approximations. 
           
%...... Begin Section ......................................................
\section{Results} \label{sec:results}

    In the following sections, we present our results of the ADFs and the magnetic field strength of the intermediate-latitude IPS-GI fields: \textit{C2}, \textit{C16}, and \textit{C37}.

    %...... Begin Sub-section ......................................................
    \subsection{Angular dispersion function} \label{subsec:Resul_ADF}

        The resulting ADFs are presented in Figure~\ref{fig:ADF}, along with the best fit (Equation~(\ref{eq:SF_tot_fit}) corrected for measurement errors, as described by Equation~(\ref{eq:ADF_corr})) in the scale bins at which we see a linear increase in dispersion. The scales above $2/3$ of the field width are not considered due to subsampling issues (see histograms in the insets in Figure~\ref{fig:ADF}, also see Section~\ref{subsec:selec_fields_SF_outlier_subsam}). Likewise, small bumps of polarization angle dispersion at small scales are not considered significant; see the discussion of Section~\ref{subsec:selec_fields_SF_outlier_subsam}.

        The ADFs of Figure~\ref{fig:ADF} do not present a steep decline or dip approaching $\ell=0$. This suggests that the turbulent correlation length $\delta$ is below the smallest scale probed by our observation, as in the case \textit{3.b} of Section~\ref{subsec:MF_cal_ADF_interpre}. Therefore, the ADFs' moderate slope is consistent with fluctuations from large-scale structures partially within the fields of view (see explanation in Section~\ref{subsec:selec_fields_SF_PA_gradient}).

        All three fields show an almost linear increasing polarization angle dispersion throughout a large range of angular scales measured, i.e.,~\textit{C16} and \textit{C37} between ${\sim}0\fdg01$ and ${\sim}0\fdg2$, which correspond to scales between ${\sim}0.03-0.6$~pc (for \textit{C16}) and ${\sim}0.09-1.7$~pc (for \textit{C37}), and \textit{C2} between ${\sim}0\fdg01$ and ${\sim}0\fdg3$, corresponding to ${\sim}0.02-0.7$~pc, assuming the dust peak distance of Table~\ref{tab:fields_proper}. In field \textit{C37}, the accuracy of the ADF holds less. The dust in \textit{C37}, which is distributed within multiple diffuse structures along $1,000$~pc, may have actual scales between $0.01-2$~pc. However, as mentioned in Section~\ref{subsec:selec_fields_cloud_thickness}, we assumed a single polarizing screen of $120$~pc thick, at an average distance of $500$~pc, for simplicity.

        The values of the turbulent parameter, $b$, obtained from fitting the ADFs, are presented in Table~\ref{tab:fields_proper}. We used $b$ to calculate the turbulent to large-scale magnetic field ratios, $\langle B_t^2 \rangle^{1/2}/B_{pos}$, as in Equation~(\ref{eq:Bt_B0_ratio}). The results presented in Table~\ref{tab:fields_proper} are around ${\sim}0.1$ for all fields studied. Therefore, highly regular magnetic fields dominate the IPS-GI intermediate-latitude observations, as suggested in \cite{Angarita_2023}. This is further confirmed by comparing optical and thermal dust polarization with diffuse dust emission at $12~\micron$ in Section~\ref{subsec:Disc_MF_structure}.
        %
        %%% --- Begin Figure -----------------------------------------------------------
        \begin{figure*}[ht!]
            \epsscale{0.377}
            \plotone{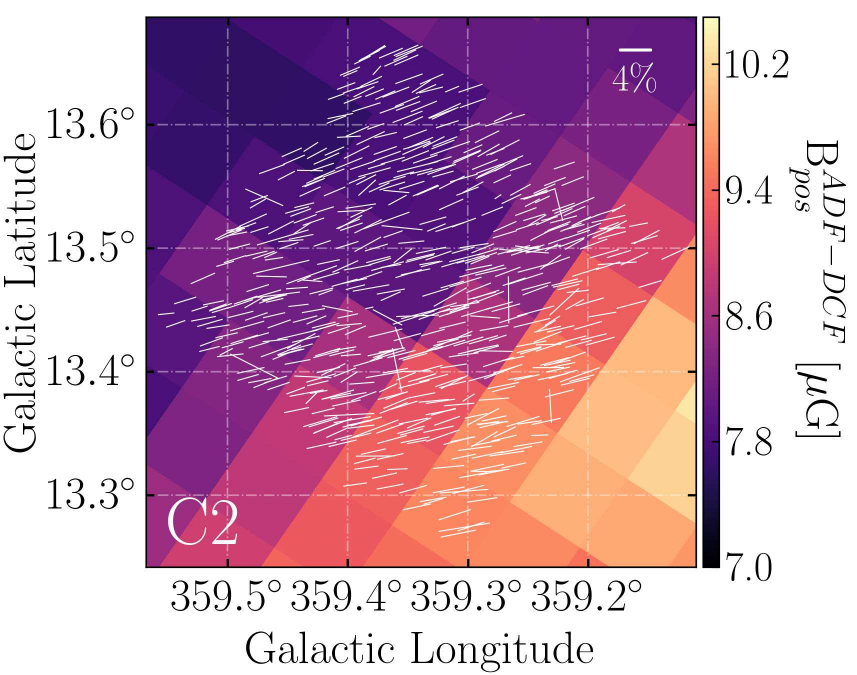}
            \epsscale{0.377}
            \plotone{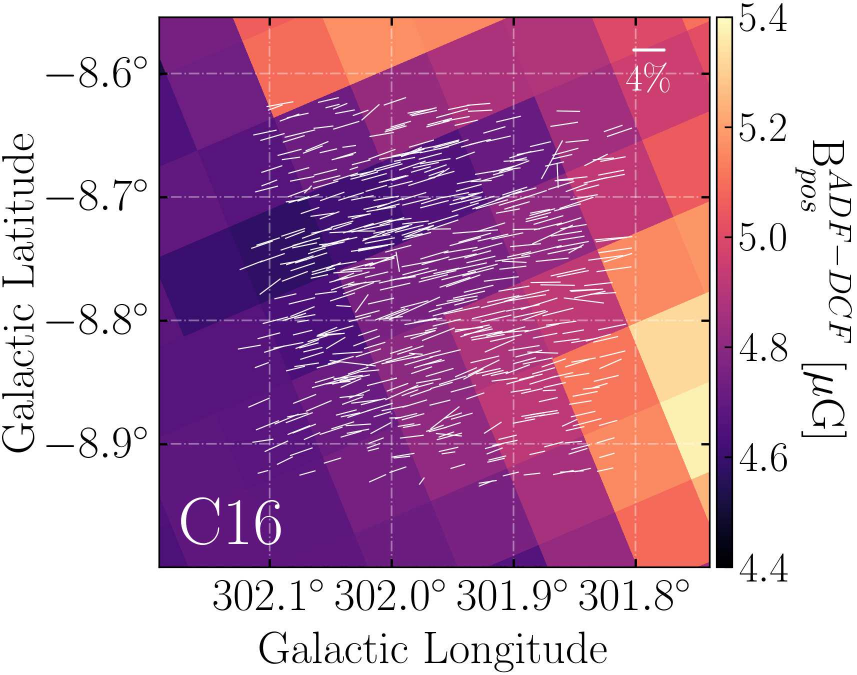}
            \epsscale{0.377}
            \plotone{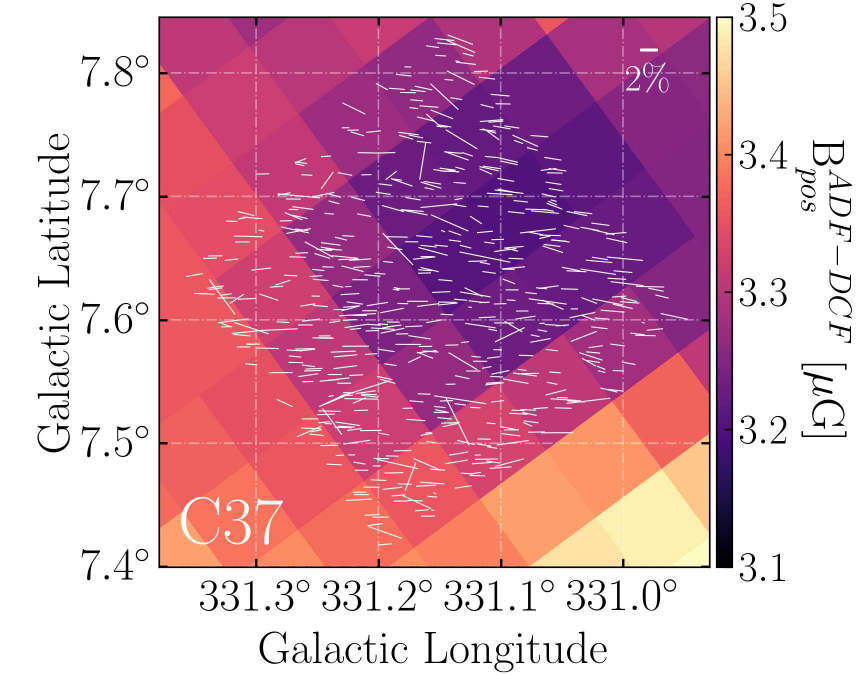}
            \caption{Magnetic field strength maps in field \textit{C2} (left), \textit{C16} (middle), and \textit{C37} (right). The white pseudo-vectors represent the IPS-GI starlight polarization.
            \label{fig:B_maps}}
        \end{figure*}
        %%% -------------------------------------------------------------------------- 

    %...... Begin Sub-section ......................................................
    \subsection{Magnetic field strength maps} \label{subsec:Resul_MF_maps}

        The average plane-of-sky magnetic field strength calculated with the DCF, ST-DCF, and ADF-DCF methods are presented in Table~\ref{tab:B0} along with the values obtained with \cite{Cho_Yoo_2016} correction to the DCF method (CY-DCF), which are discussed in Section~\ref{subsec:Disc_MF_strength}.  We used the mass density 2D map, $\rho^{LVC}$, estimated from the LVC components of the neutral medium (Figure~\ref{fig:dens_&_turb_disp}, left). Furthermore, 
        assuming a scenario in which the observed scale lengths are below the turbulent correlation length, i.e.,~$\delta>\ell_{max}$, we used the DCF and ST-DCF methods with the polarization angle dispersion corrected by the measurement errors, $\sigma_{\theta,corr}$, and the CNM turbulent velocity dispersion 2D map of the LVC components scaled down to the spatial scale of the field of view, i.e.,~$(\sigma_{v,turb})_{FoV}$. In the case where the turbulence is at small scales (see Sections~\ref{subsec:MF_cal_ADF_interpre} and \ref{subsec:selec_fields_SF_PA_gradient}), we used the ADF-DCF method with the turbulent parameter $b$ from the fit to the ADFs (Figure~\ref{fig:ADF}) and the CNM turbulent velocity dispersion 2D map scaled to the smallest scale probed by the ADFs, i.e.,~$(\sigma_{v,turb})_{ips}$ (Figure~\ref{fig:dens_&_turb_disp}, right). See the average values of the 2D maps in Table~\ref{tab:fields_proper}.  
        %
        % %%% --- Begin Table -----------------------------------------------------------
        \begin{deluxetable}{ccccc}
        % \begin{adjustwidth}{-0.4cm}{}
        % \resizebox{1.05\columnwidth}{!}{% <------ Don't forget this %
            \setlength{\tabcolsep}{3pt}
            \tablecaption{Average magnetic field strength in the intermediate-latitude IPS-GI fields. \label{tab:B0}}
            \tablehead{
            \colhead{Field} & \colhead{$\langle B_{pos}^{ADF-DCF} \rangle$} &  \colhead{$\langle B_{pos}^{DCF} \rangle$} & \colhead{$\langle B_{pos}^{ST-DCF} \rangle$} & \colhead{$\langle B_{pos}^{CY-DCF} \rangle$} \\
            \colhead{} & \colhead{($\mu$G)} & \colhead{($\mu$G)} & \colhead{($\mu$G)} & \colhead{($\mu$G)}  
            }
            \colnumbers
            \startdata
            \textit{C2}  & 9 & 16 & 7 & 3  \\
            \textit{C16} & 5 & 9 & 3 & 7  \\
            \textit{C37} & 3 & 6 & 3 & 1 \\
            \enddata
            \tablecomments{Columns: (1) IPS-GI identification; (2) ADF-DCF method, Equation~(\ref{eq:ADF_DCF}); (3) classic DCF method, Equation~(\ref{eq:DCF}) with \mbox{$f=0.5$}; (4) \cite{Skalidis_Tassis_2021} method, Equation~(\ref{eq:ST_DCF}); (5) \cite{Cho_Yoo_2016} correction to the DCF method. 
            }
        % }% <------ Don't forget this %
        % \end{adjustwidth}
        \end{deluxetable}
        %%% --------------------------------------------------------------------------
        % 
        
        The DCF and ST-DCF methods give magnetic field strengths of ${\sim}16~\mu$G and ${\sim}7~\mu$G in \textit{C2}, ${\sim}9~\mu$G and ${\sim}3~\mu$G in \textit{C16}, and ${\sim}6~\mu$G and ${\sim}3~\mu$G in \textit{C37}, respectively. The classic DCF method yields higher values than the ST-DCF method by an approximated factor of $2$ in \textit{C2} and \textit{C37}, and $3$ in \textit{C16}. However, the most probable situation is one where the dispersion is mainly attributed to large-scale field variations, and the turbulent correlation length is at small scales. 
        
        Figure~\ref{fig:B_maps} presents the plane-of-sky magnetic field strength maps calculated with the ADF-DCF method in each IPS-GI field considered. The average $B_{pos}$ obtained with the ADF-DCF method is ${\sim}9~\mu$G, ${\sim}5~\mu$G, and ${\sim}3~\mu$G, for field \textit{C2}, \textit{C16}, and \textit{C37}, respectively (Table~\ref{tab:B0}). The results of the ADF-DCF method are closer to those of the ST-DCF. Nevertheless, the direct comparison is meaningless since both methods are valid for different scenarios of the turbulent medium. 
        
        The ADF-DCF method was implemented in this research under the assumption that the observed scales are above the turbulent correlation length, i.e.,~$\delta<\ell_{min}$. This is supported by the polarization angle gradients observed in Figure~\ref{fig:heatmap_PA_gradient} (Section~\ref{subsec:selec_fields_SF_PA_gradient}), showing a smooth transition in polarization angle due to a large structure partially within the IPS-GI fields, and the profiles of the ADFs (Figure~\ref{fig:ADF}), which do not present a steep dip at the smallest scales. In particular, we highlight the similarities between the morphology of the median polarization angle gradients (Figure~\ref{fig:heatmap_PA_gradient}) and the features observed in the $B_{pos}$ map (Figure~\ref{fig:B_maps}, obtained from the combination of the $\rho$ and $\sigma_{v,turb}$ maps of Figure~\ref{fig:dens_&_turb_disp}). However, due to the limited size of the fields of view, we cannot be completely sure that the turbulence scales are below the observable scales. In the case $\delta>\ell_{max}$, we recommend considering the results of the classic DCF (Section~\ref{subsec:MF_cal_DCF}) or the ST-DCF (Section~\ref{subsec:MF_cal_ST-DCF}) methods.   
        
        Field \textit{C37} presents the lowest plane-of-sky magnetic field strength among the sample. Nonetheless, this field has the greatest uncertainties in estimations, primarily because of the numerous assumptions made given the intricate nature of very diffuse dust structures spanning a considerable pathlength. On the other hand, the average values obtained for fields \textit{C2} and \textit{C16} are in good agreement with previous estimations in the LB wall and the nearby diffuse ISM; see the discussion in Section~\ref{subsec:Disc_MF_strength}.

%...... Begin Section ......................................................
\section{Discussion} \label{sec:discu}

    We discuss below the ISM properties estimated in the intermediate-latitude IPS-GI fields and compare them with previous estimations in the local, diffuse ISM. Finally, we describe the properties of the magnetic field in the local ISM as observed by optical and thermal dust polarization.

    %...... Begin Sub-section ......................................................
    \subsection{Turbulent scales} \label{subsec:Disc_turb_scales}

        We proposed that the ADFs of Figure~\ref{fig:ADF} cannot resolve the turbulent correlation length $\delta$ in the three regions studied (Section~\ref{subsec:Resul_ADF}). However, it is possible to establish an upper limit at the smallest scale we can demonstrate, i.e.,~$~0\fdg01$. Assuming the distance to the highest dust peak along the pathlengths of \textit{C2} and \textit{C16}, i.e.,~$d_{peak}$ approximately behind or within the LB wall, the smallest scale corresponds to ${\sim}24-30$~mpc. Due to the lack of measurements of this kind in diffuse nearby structures (i.e.,~with $n_\mathrm{H}<30$~cm$^{-3}$), our results can only be compared to studies in nearby molecular clouds.

        The turbulent correlation length was measured with the ADFs method at mpc scales in various molecular clouds. For instance, \cite{Houde_2009} found a turbulent correlation length of $16$~mpc in OMC-1, assuming a distance of $450$~pc. Using starlight polarization, \cite{Franco_2010} found $\delta$ around $2-5$~mpc in different regions of the Pipe nebula assuming $d=160$~pc. \cite{Wang_JCMT_BISTRO_2019} used submillimeter polarization in the \mbox{IC 5146} molecular cloud to calculate the ADF and found a turbulent scale of ${\sim}60$~mpc adopting the distance of $813$~pc. In cases where the turbulent correlation length was not resolved \citep[e.g.,][]{Hildebrand_2009, Poidevin_2010, Chapman_2011, Karoly_2023, Hwang_2023}, the smallest scale probed by ADFs suggests upper limits of the turbulence correlation length within an approximated range of $6-120$~mpc, e.g.,~$<50$~mpc in OMC-1, OMC-2, and OMC-3 \citep{Hildebrand_2009, Poidevin_2010}, $<120$~mpc in off-cloud regions of Taurus \citep{Chapman_2011}, $<6$~mpc in L43 in Ophiuchus region \citep{Karoly_2023}, and $<19$~mpc in the Horsehead Nebula \citep{Hwang_2023}.

        Although all the above measurements have been made on dense sources, i.e.,~$n_\mathrm{H}$ between $10^3-10^5$~cm$^{-3}$ \citep{Alves_2008, Karoly_2023, Hwang_2023}, the upper limits of the turbulent correlation length in the three IPS-GI diffuse fields are on the same order of magnitude, at mpc scales. This demonstrates significant similarities in the turbulent correlation lengths at low and high densities in the local ISM ($d < 1,000$~pc).

    %...... Begin Sub-section ......................................................
    \subsection{Galactic magnetic field in the Solar neighborhood} \label{subsec:Disc_MF}

        The intermediate-latitude IPS-GI polarization measurements probe the magnetized, diffuse medium in the nearby ISM ($d<400$~pc for \textit{C2} and \textit{C16}, and $d<1,000$~pc for \textit{C37}, Figure~\ref{fig:Av_dens_vs_dist}). The following sections discuss the consistency of the magnetic field strength results with the values measured in the Solar neighborhood and the estimations made with different methods. Furthermore, we discuss the structure of the magnetic field observed in the plane of the sky with optical polarization and thermal dust polarization and the correlation between the large-scale magnetic field, the diffuse dust emission observed at $12~\micron$, and the thermal dust emission.
 
        %...... Begin Sub-section ......................................................
        \subsubsection{Magnetic field strength} \label{subsec:Disc_MF_strength}

            Regardless of the method used, our results (Table~\ref{tab:B0}) are consistent with the Galactic magnetic field strength estimations in the local diffuse ISM according to observations of starlight polarization and the Zeeman effect. For instance, \cite{Heiles_Troland_2005} found a median total magnetic field of $6\pm1.8~\mu$G in the CNM. \cite{Crutcher_2010} and \cite{Crutcher_2012} presented a maximum line-of-sight magnetic field strength of ${\sim}10~\mu$G in low-density regions of molecular clouds, with typically $n_\mathrm{H}{\sim}300$~cm$^{-3}$, using a Bayesian analysis of molecular Zeeman measurements of \ion{H}{1}, OH, and CN. Similarly, near-IR starlight polarization observations of the Taurus dark-cloud complex \citep{Chapman_2011} showed plane-of-sky field strengths of ${\sim}10~\mu$G in low-density off-cloud regions ($n_\mathrm{H}{\sim}100$~cm$^{-3}$). Later on, \cite{Medan_Andersson_2019} found an average plane-of-sky field strength in the LB wall of ${\sim}8~\mu$G, assuming a distance to the wall of $d\simeq100-315$~pc and using optical starlight polarization of OB associations at $b>30^{\circ}$. 

            One fundamental assumption of DCF-like methods is the sub-Alfv\'enic turbulence, i.e.,~$\mathcal{M}_A < 1$, where $\mathcal{M}_A$ is the Alfv\'en Mach number. We calculated $\mathcal{M}_A = \langle \sigma_{v,turb}\rangle_{FoV}/V_A$, where $V_A = B_{pos}/\sqrt{4\pi \rho}$ is the Alfv\'en velocity calculated from the plane-of-sky magnetic field strength. In regions \textit{C2}, \textit{C16}, and \textit{C37}, we obtained $\mathcal{M}_A$ values of $0.18$, $0.13$, and $0.24$, respectively, using parameters from the classic DCF method. Additionally, we calculated $\mathcal{M}_A = \sigma_{\theta}/f$, where $f$ is the classical DCF correction factor, yielding $\mathcal{M}_A < 1$ for $f$ values within the range [0.1,1]. Furthermore, $\mathcal{M}_A$ can also be estimated using the turbulent-to-total magnetic field ratio obtained from the ADF fit parameters, i.e.,~$\mathcal{M}_A = \langle B_t^2\rangle^{1/2}/B_{pos} $, which is ${\sim}0.1$ in all cases, reaffirming the sub-Alfv\'enic condition in all three regions.

            The classical DCF method does not account for the contributions of multiple turbulent cells along the line of sight. If the number of turbulent cells is large, $B_{pos}$ is likely overestimated. \cite{Cho_Yoo_2016} proposed a correction to this issue by using the relation $\delta V_c/\langle \sigma_{v,turb}\rangle \propto 1/\sqrt{N_c}$, where $\delta V_c$ is the standard deviation of the line emission centroid velocities observed towards different lines of sight, and $N_c$ is the number of turbulent cells. We computed the average $V_c$ in each pixel of our CNM model maps using the parameters of the Gaussian LVC components as in Equation~(\ref{eq:vel_M1}). Subsequently, we estimated $\delta V_c$ using Equation~(18) of \citealt{Cho_Yoo_2016} for each IPS-GI region. Then, we recalculated $B^{DCF}_{pos}$, hereinafter $B^{CY-DCF}_{pos}$, using the correction factor $\xi = 0.5$ for direct comparison with the classical DCF. We found average $B^{CY-DCF}_{pos}$ values of ${\sim}3~\mu$G, ${\sim}7~\mu$G, and ${\sim}1~\mu$G in \textit{C2}, \textit{C16}, and \textit{C37}, respectively. This suggests that DCF results are significantly overestimated in the regions of \textit{C2} and \textit{C37}, whereas for \textit{C16}, the correction is smaller.

            The magnetic field strength calculation depends considerably on knowing the ISM properties very well. In particular, the volume density assumed from 3D dust maps models (Section~\ref{subsec:selec_fields_nH}) is a key parameter in our cloud thickness calculation, which in turn affects the determination of important parameters such as $\rho$ (Section~\ref{subsec:sel_fields_density}) and $\sigma_{v,turb}$ (Section~\ref{subsec:sel_fields_sigma_v}). Finding volume densities close to $30$~cm$^{-3}$ in the IPS-GI fields implies that the density, calculated using Equation~(\ref{eq:rho}), is dominated by the cold medium \citep{Draine_2011}. However, the \ion{H}{1} line emission also shows contributions from the WNM and LNM in the three fields. As we cannot ignore the contributions from the warm medium, the turbulent velocity dispersion, as defined in Section~\ref{subsec:sel_fields_sigma_v}, characterizes the dispersion of the entire multiphase interstellar gas. Consequently, the results are a weighted mixture of ISM phases that adds uncertainty.  
            %
            %%% --- Begin Figure -----------------------------------------------------------
            \begin{figure}[ht!]
                \epsscale{2.33}
                \plottwo{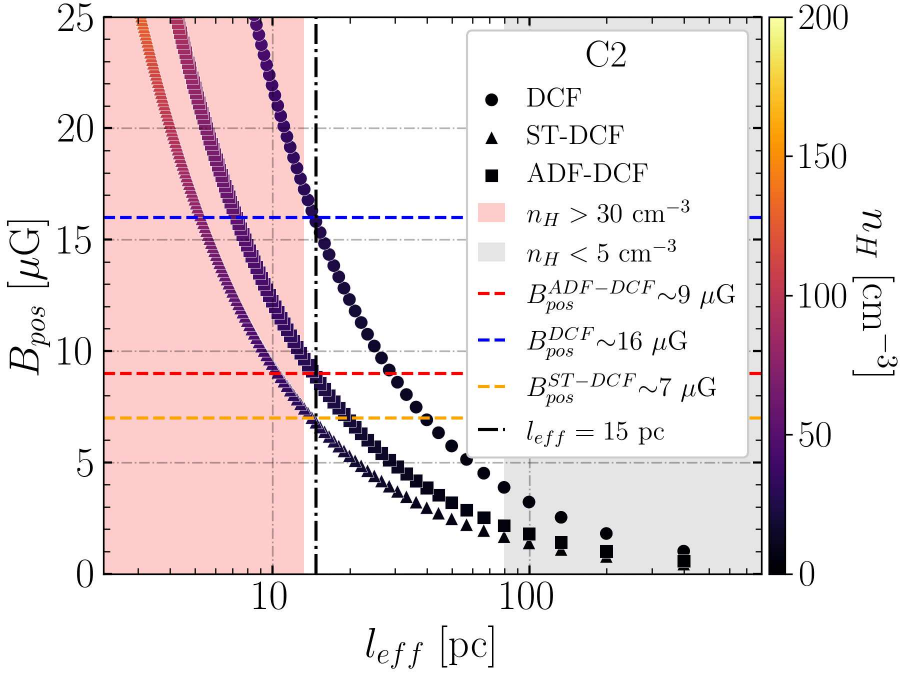}{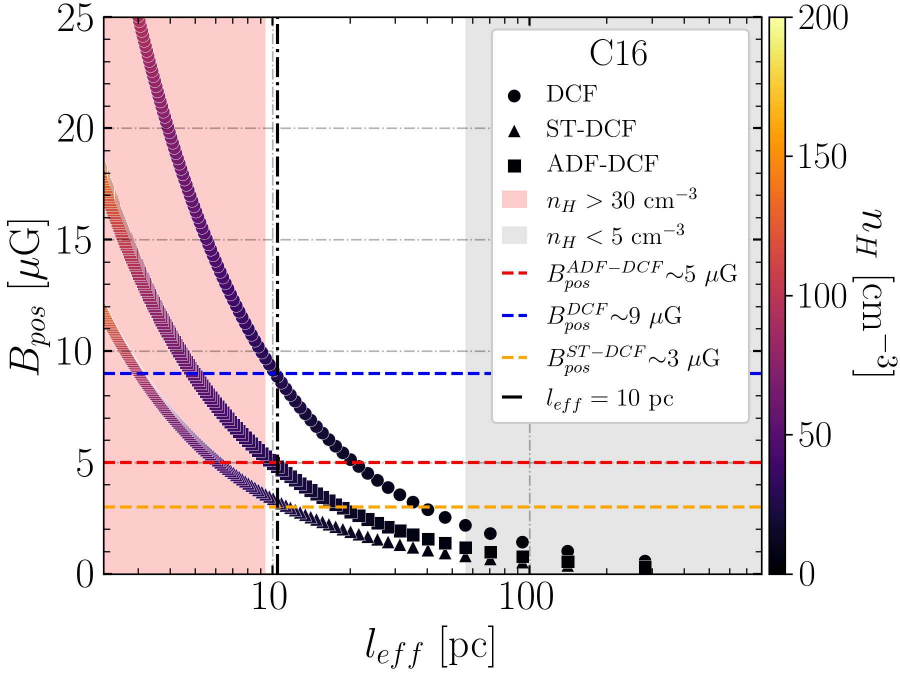}
                \epsscale{1.165}
                \plotone{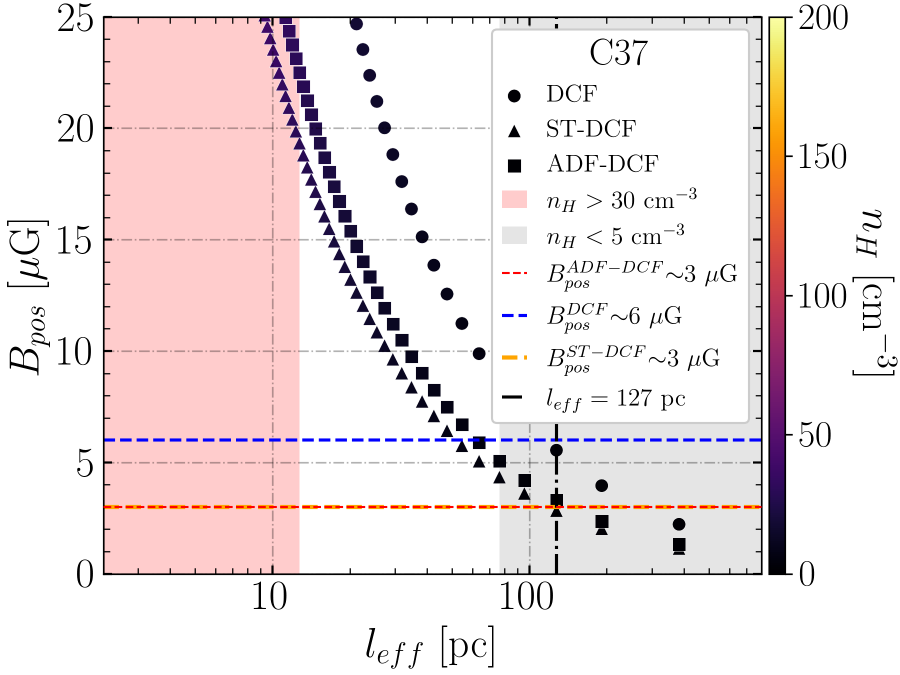}
                \caption{Magnetic field strength calculated with the DCF (the circles), ST-DCF (the triangles), and ADF-DCF (the squares) methods as a function of the effective thickness, colored by the total volume density. The red and orange horizontal dashed lines represent our $B_{pos}$ results calculated with ADF-DCF and ST-DCF, respectively. The vertical black dot-dashed line shows the effective thickness of the cloud obtained with the values of Table~\ref{tab:fields_proper}.
                \label{fig:B0_leff_nH}}
            \end{figure}
            %%% -------------------------------------------------------------------------- 

            Figure~\ref{fig:B0_leff_nH} presents the magnetic field strength as a function of the cloud thickness, colored by the total volume density, for each IPS-GI field. To construct the functions of each panel in Figure~\ref{fig:B0_leff_nH}, we used the parameters of Table~\ref{tab:fields_proper} and varied the $n_\mathrm{H}$ values between $1-200$~cm$^{-3}$ \citep{Wolfire_2003, Murray_2015}. Let us recall that we are assuming that $n_\mathrm{H}$ is mainly made up of neutral atomic hydrogen. This is supported by the good agreement between $N_\mathrm{H}^{A_V}$ and $\langle N_\mathrm{H}^{LVC} \rangle$. Furthermore, according to \cite{Murray_2015}, the $n_\mathrm{H}$ of the CNM only is expected to be within $5-120$~cm$^{-3}$. 
            
            The gray shaded area in Figure~\ref{fig:B0_leff_nH} shows low volume densities ($<5$~cm$^{-3}$) that are unlikely to represent our \textit{C2} and \textit{C16} fields. On the other hand, the volume density and, thus, the magnetic field in \textit{C37} are likely underestimated since we assumed an average peak value rather than the maximum $n_\mathrm{H}$. However, the density profiles of Figure~\ref{fig:nH} (bottom panel) shows $n_\mathrm{H}$ typically below ${\sim}5-7$~cm$^{-3}$. Furthermore, the red shaded area represents the values obtained for volume densities above $30$~cm$^{-3}$, which is a good upper limit for the diffuse cold medium \citep{Draine_2011}. \cite{Zucker_2021} argued that any estimation of $n_\mathrm{H}$ from \cite{Leike_2020} 3D dust map -- especially in dense regions -- should be considered a lower limit due to the systematic errors of the technique and the resolution limitations of the 3D map. Nevertheless, since our diffuse IPS-GI fields are far off from the center of the molecular clouds, $n_\mathrm{H}$ may not be larger than $30$~cm$^{-3}$, i.e.,~this is also the approximated maximum value of \cite{Vergely_2022} dust profiles in \textit{C2} and \textit{C16} (Figure~\ref{fig:nH}). Therefore, we would not expect our fields to be within the red area either.

            The mentioned boundaries enable us to establish approximate values for the lower and upper limits of the magnetic field strength within our fields. The maximum $B_{pos}$ are ${\sim}10~\mu$G, ${\sim}5~\mu$G, and ${\sim}7~\mu$G for \textit{C2}, \textit{C16}, and \textit{C37}, respectively. The minimum $B_{pos}$ are ${\sim}2~\mu$G and ${\sim}1~\mu$G for \textit{C2} and \textit{C16}, respectively. All the calculations were made with the ADF-DCF method.
            %
            %%% --- Begin Figure -----------------------------------------------------------
            \begin{figure*}[ht!]
                \epsscale{2.0}            \plottwo{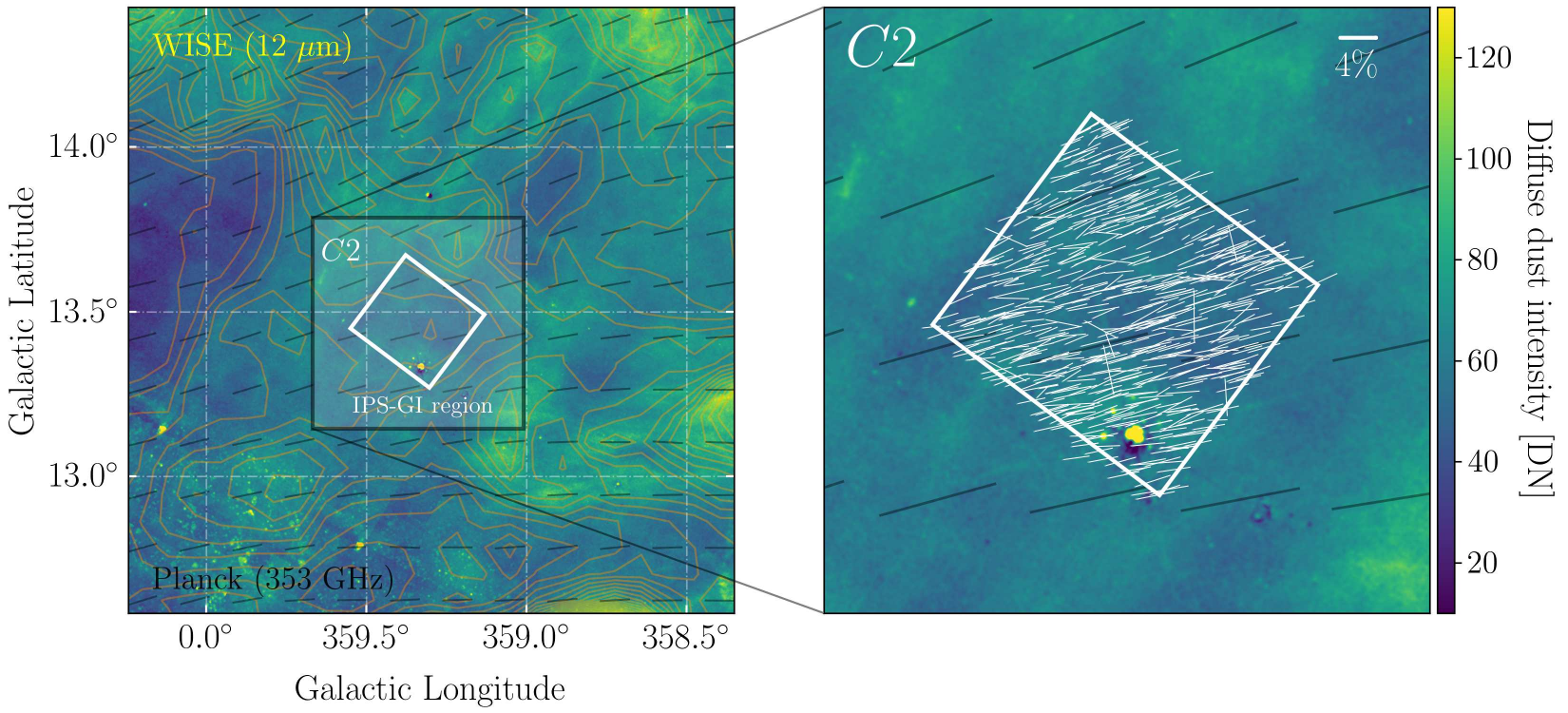}{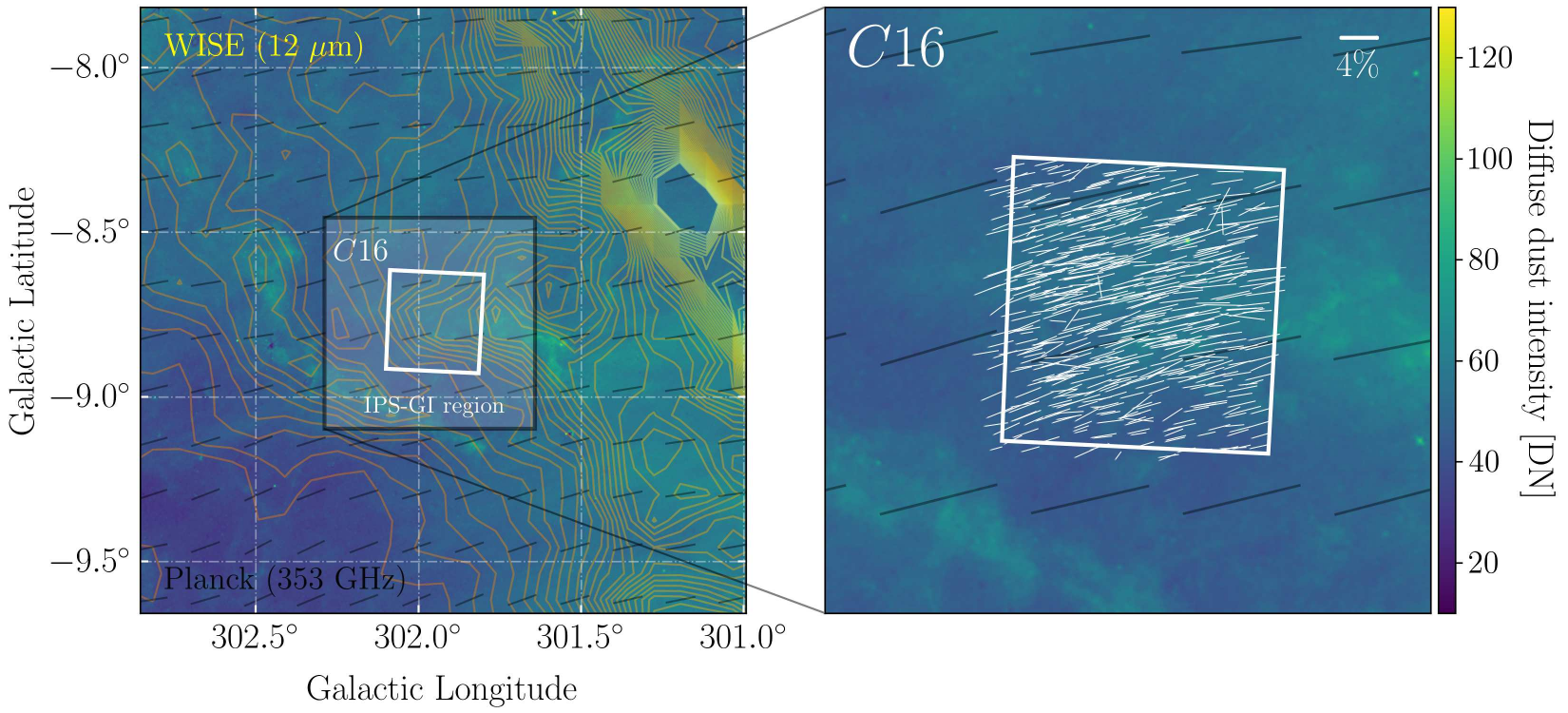}
                \epsscale{1.0}            \plotone{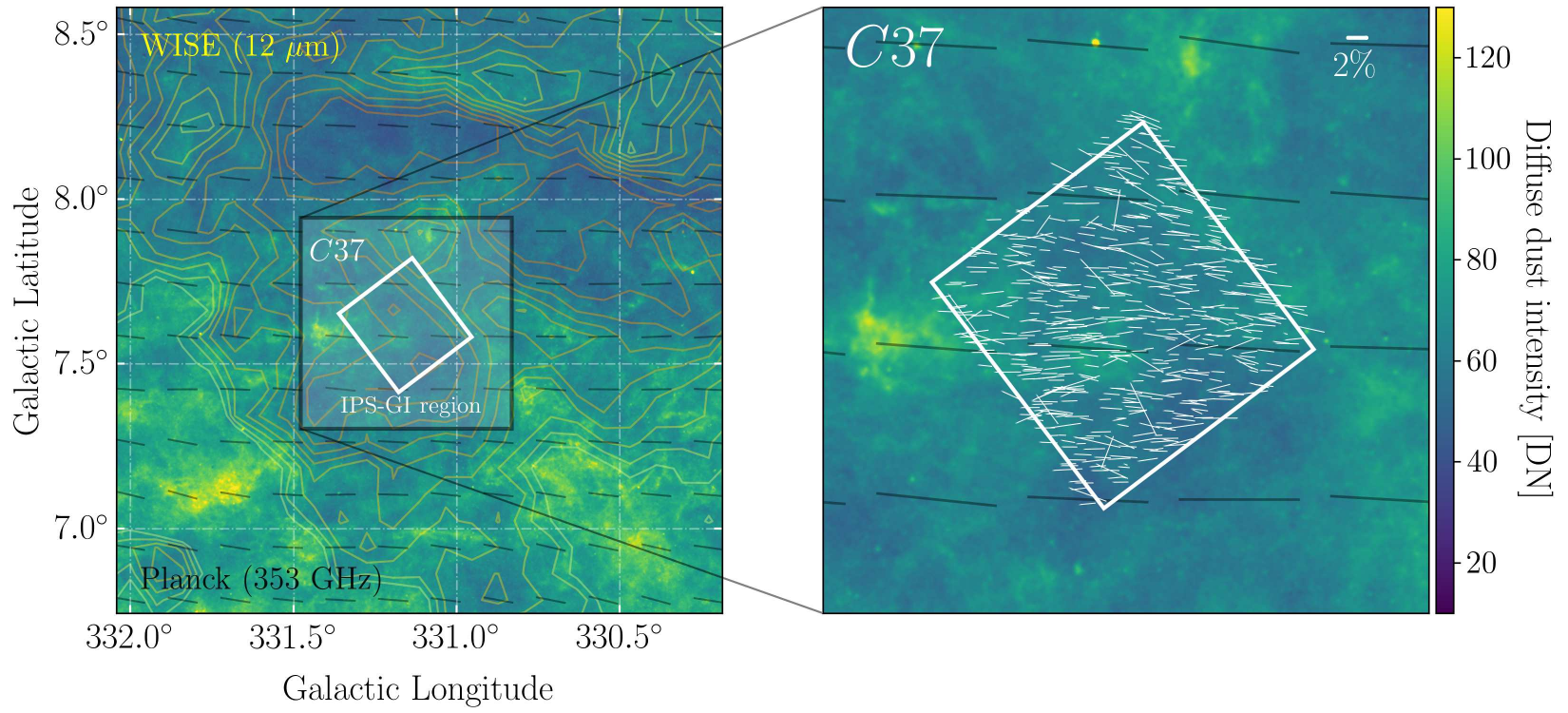}
                \caption{Left: Diffuse dust emission at $12~\micron$ from WISE (Background) compared with the thermal dust column density (the contours) and polarization (the black pseudo-vectors) at $353$~GHz from Planck in the vicinity of fields \textit{C2} (top row), \textit{C16} (middle row), and \textit{C37} (bottom row) marked by the white rectangle in the center. Right: Comparison between the optical polarization (the white pseudo-vectors) and the thermal dust polarization at 353 GHz (the black pseudo-vectors). The orientation of the thermal dust polarization was rotated by $90\degr$ so that both Planck polarization and optical polarization pseudo-vectors indicate the orientation of the plane-of-sky magnetic field.
                \label{fig:Diff_dust_12um_vs_NH353_Pol353}}
            \end{figure*}
            %%% --------------------------------------------------------------------------
 
        %...... Begin Sub-section ......................................................
        \subsubsection{Plane-of-sky magnetic field structure in the local ISM} \label{subsec:Disc_MF_structure}

            IPS-GI starlight polarization in intermediate-latitude fields probes the magnetic field coupled to nearby dust structures behind and within the LB wall ($d < 400-1,000$~pc, depending on the field, Figure~\ref{fig:Av_dens_vs_dist}). Figure~\ref{fig:Diff_dust_12um_vs_NH353_Pol353} shows the orientation of the plane-of-sky magnetic field as derived from the \cite{Planck-Collaboration_2018_20} thermal dust polarization (the black vectors) and the IPS-GI optical polarization (the white vectors) in the surroundings of the studied regions. The thermal dust polarization shows a very regular field towards our diffuse regions with an orientation in excellent agreement with starlight polarization. The consistency with optical starlight polarization also indicates that thermal dust emission in these three regions may come from the nearby ISM.

            Furthermore, the high-resolution diffuse dust at $12~\micron$ from WISE \citep{Meisner_Finkbeiner_2014} shows fine structures in the sky $1.5^{\circ}$ around the IPS-GI fields (see the background of Figure~\ref{fig:Diff_dust_12um_vs_NH353_Pol353}). Although Planck's thermal dust emission, $N_{\mathrm{H}}^{\tau_{353}}$ (the contours), is in good agreement with the diffuse dust, the extremely regular large-scale magnetic field seems to not care about small, diffuse structures and, instead, is highly ordered (Figure~\ref{fig:Diff_dust_12um_vs_NH353_Pol353}). Hence, optical starlight and thermal dust polarization probe the large-scale magnetic field, which is highly regular in the surroundings and within the intermediate-latitude IPS-GI fields, in the few hundreds of parsecs within and beyond the LB wall. 
            
            This result is consistent with our interpretation of the turbulence in the three IPS-GI fields, i.e.,~the case \textit{3.b} in Section~\ref{subsec:MF_cal_ADF_interpre}. The ADFs (Figure~\ref{fig:ADF}) are tracing the dispersion due to large-scale field fluctuations rather than pure turbulence. Consequently, the turbulent correlation lengths are below the smallest spatial scales observed in the three fields.

%...... Begin Section ......................................................
\section{Summary} \label{sec:summary}

    We used optical starlight polarization measurements from the IPS-GI catalog \citep{Versteeg_2023}, 3D dust maps from \cite{Vergely_2022} and \cite{Leike_2020}, and PPV cubes of \ion{H}{1} emission from GASS survey \citep{McClure_Griffiths_2009} to characterize the plane-of-sky magnetic field strength and the properties of the local ISM towards three intermediate Galactic latitude fields: \textit{C2}, \textit{C16}, and \textit{C37}, using the IPS-GI nomenclature. These fields sample diffuse ISM properties. To this end, we implemented the DCF \citep{Davis_1951, Chandrasekhar&Fermi_1953}, the ST-DCF \citep{Skalidis_2018}, and the ADF-DCF \citep{Hildebrand_2009} methods to estimate the plane-of-sky magnetic field strength under different regimes of the turbulence. A correction to the DCF method that accounts for the number of turbulent cells along the line of sight (i.e.,~the CY-DCF method) is also discussed.

    The polarization angle dispersion of the three fields studied exhibits a steady increase with angular scale, which may be caused by large-scale field fluctuations with angular scales not smaller than ${\sim}0\fdg2$ (i.e.,~$0.6$~pc at $d{\sim}172$~pc). Moreover, the three regions show a smooth change in the polarization orientation, probably due to structures larger than the observed region but partially within it. The above evidence suggests that the turbulent correlation length, $\delta$, is below scales of $<0\fdg01$ (i.e.,~$<0.03$~pc).

    The neutral \ion{H}{1} emission observed by GASS in the IPS-GI regions is associated with the polarizing dust structures in the nearby ISM ($\ell_c = 400-1,000$~pc). We thereby produced 2D maps of the physical properties of the nearby clouds, such as $\rho$ and $\sigma_{v, turb}$, using models derived from a Gaussian decomposition of the GASS \ion{H}{1} line emission towards the IPS-GI fields. The estimation of $n_\mathrm{H}$ is critical in our case for the accuracy of DCF-like methods. Therefore, we evaluate how the $B_{pos}$ values vary in each observed region depending on the volume density constrained by 3D dust models.    
 
    We estimated the plane-of-sky magnetic field strength between ${\sim}3~\mu$G and ${\sim}9~\mu$G with the ADF-DCF method, between ${\sim}3~\mu$G and ${\sim}7~\mu$G with the ST-DCF method, and between ${\sim}1~\mu$G and ${\sim}7~\mu$G with the CY-DCF correction to the classic DCF method in the studied regions. These values align with previous calculations using the Zeeman effect, near-IR, and optical starlight polarization in low-density regions of the local ISM. Furthermore, the turbulent to large-scale magnetic field ratios, $\langle B_t^2 \rangle^{1/2}/B_{pos}$, is ${\sim}0.1$ in all cases, indicating the dominance of the regular component of the magnetic field in the nearby ISM. 

    Finally, thermal dust polarization at 353 GHz from Planck is in good agreement with the optical starlight polarization observed by the IPS-GI in the three regions studied, which originates in the nearby ISM ($d<400-1,000$~pc, depending on the field). Therefore, Planck's polarization should be tracing the local ISM in these three regions as well. The thermal dust and optical polarization show a highly regular plane-of-sky magnetic field orientation despite the fine structures observed in the diffuse dust emission at $12~\micron$. The latter confirms our conclusion about the low turbulent to large-scale magnetic field ratios. 
    
% \newpage
% \nopagebreak
\begin{acknowledgments}
    \begin{center}
        % \textit{Acknowledgments}
        \uppercase{acknowledgments}
    \end{center}
    
    The authors are grateful to the anonymous referee, whose insightful review contributed to the improvement of this paper.       
    The authors acknowledge the Interstellar Institute's program ``II6'' and the Paris-Saclay University's Institut Pascal for hosting discussions that nourished the development of the ideas behind this work.   
    Y.A. acknowledges Dr. S. E. Clark, M. Lei (M.Sc.), and Dr. C. Zucker for their contributions to insightful discussions about the properties of the nearby ISM. 
    
    Over the years, IPS data have been gathered by a number of dedicated observers, to whom the authors are very grateful: Flaviane C. F. Benedito, Alex Carciofi, Cassia Fernandez, Tib\'erio Ferrari, Livia S. C. A. Ferreira, Viviana S. Gabriel, Aiara Lobo-Gomes, Luciana de Matos, Rocio Melgarejo, Antonio Pereyra, Nadili Ribeiro, Marcelo Rubinho, Daiane B. Seriacopi, Fernando Silva, Rodolfo Valentim, and Aline Vidotto.
    
    Y.A., M.H., and M.J.F.V. acknowledge funding from the European Research Council (ERC) under the European Union’s Horizon 2020 research and innovation programme (grant agreement No 772663).
    C.V.R. acknowledges support from {\it Conselho Nacional de Desenvolvimento Científico e Tecnológico} - CNPq (Brazil) (Proc.~310930/2021-9).    
    A.M.M.'s work and optical/NIR polarimetry at IAG have been supported over the years by several grants from S\~ao Paulo state funding agency FAPESP, especially 01/12589-1 and 10/19694-4. A.M.M. has also been partially supported by the Brazilian agency CNPq (grant 310506/2015-8). A.M.M. graduate students have been provided with grants over the years from the Brazilian agency CAPES.   
    
    This research has used data, tools, and materials developed as part of the \mbox{EXPLORE} project that has received funding from the European Union’s Horizon 2020 research and innovation programme under grant agreement No 101004214.
    Finally, this work has made use of data from the European Space Agency (ESA) mission \textit{Gaia} (\url{https://www.cosmos.esa.int/gaia}), processed by the \textit{Gaia} Data Processing and Analysis Consortium (DPAC, \url{https://www.cosmos.esa.int/web/gaia/dpac/consortium}). Funding for the DPAC has been provided by national institutions, in particular, the institutions participating in the \textit{Gaia} Multilateral Agreement.
\end{acknowledgments}

\vspace{5mm}
\facility{LNA:BC0.6m}

\software{Astropy \citep{Astropy_Collaboration_2013,Astropy_Collaboration_2018},
          Matplotlib \citep{Hunter_Matplotlib_2007}, NumPy \citep{Harris_numpy_2020}, SciPy \citep{Virtanen_SciPy_2020}}.

\newpage
%
% ...... Begin Appendix ......................................................
\appendix

\section{\texorpdfstring{\ion{H}{1}}{HI} decomposition}\label{sec:Append_HI_decomp}

    ROHSA \citep{Marchal_ROHSA_2019} is an algorithm that accounts for the spatial coherence of the emission and the multiphase nature of the gas. It performs a non-linear Gaussian regression in which $\lambda_{a}, \lambda_{\mu}$, and $\lambda_{\sigma}$ are the constant hyper-parameters for the amplitude, $\bm{a}_n > 0$, the position, $\bm{\mu}_n$, and the standard deviation, $\bm{\sigma}_n$, position-position (2D) maps of the $n$-th Gaussian component. The hyper-parameters tune the regulation term, $R(\bm{\theta})$ (with $\bm{\theta}_n$ being the set of fit parameters) that penalizes small fluctuations of the parameters in the cost function, $Q(\bm{\theta})$, i.e.,~the reduced $\bm{\chi}^2$ 2D map. A fourth term related to the width of the Gaussian components, and thus to the gas thermodynamics, is included in the cost function and parametrized by $\lambda'_{\sigma}$. This term would give a higher weight to any solution that could describe any of the Cold, Lukewarm, and Warm Neutral Medium phases (CNM, LNM, and WNM; see \citealt{Marchal_ROHSA_2019}, for more details).

\begin{figure*}[ht!]
        \gridline{
                  \fig{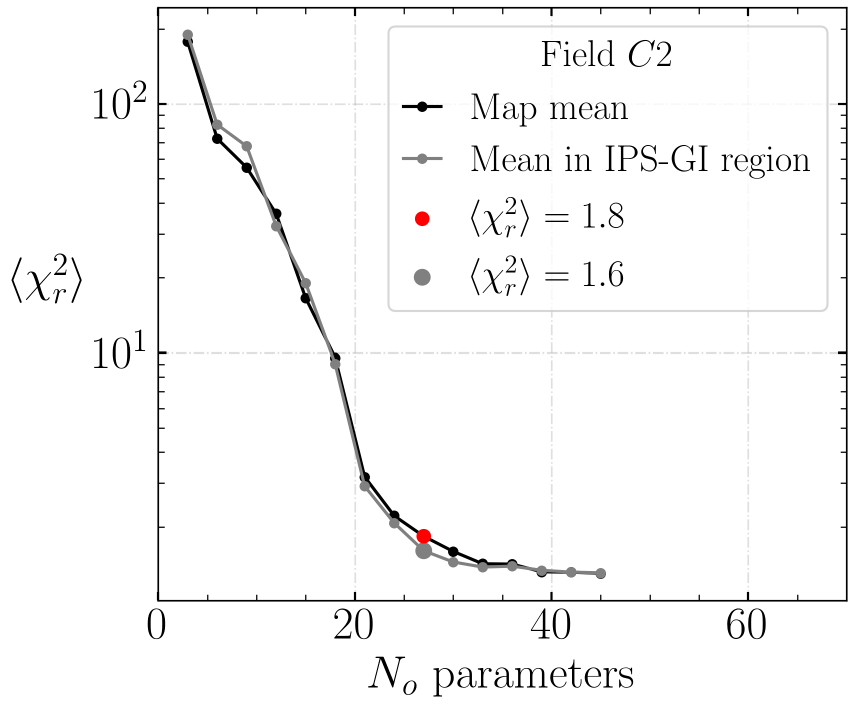}{0.32\linewidth}{}
                  \fig{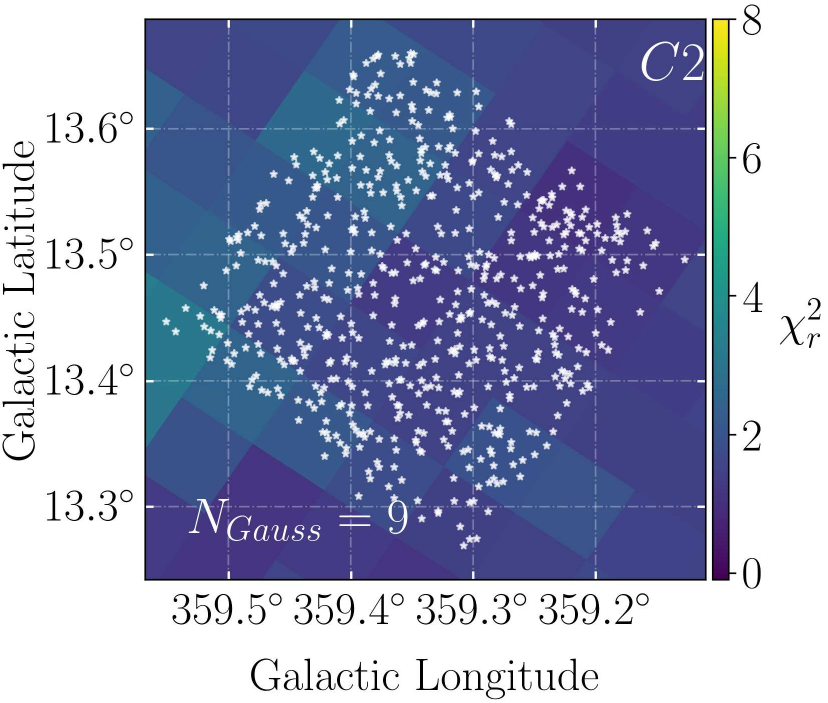}{0.32\linewidth}{}
                  \fig{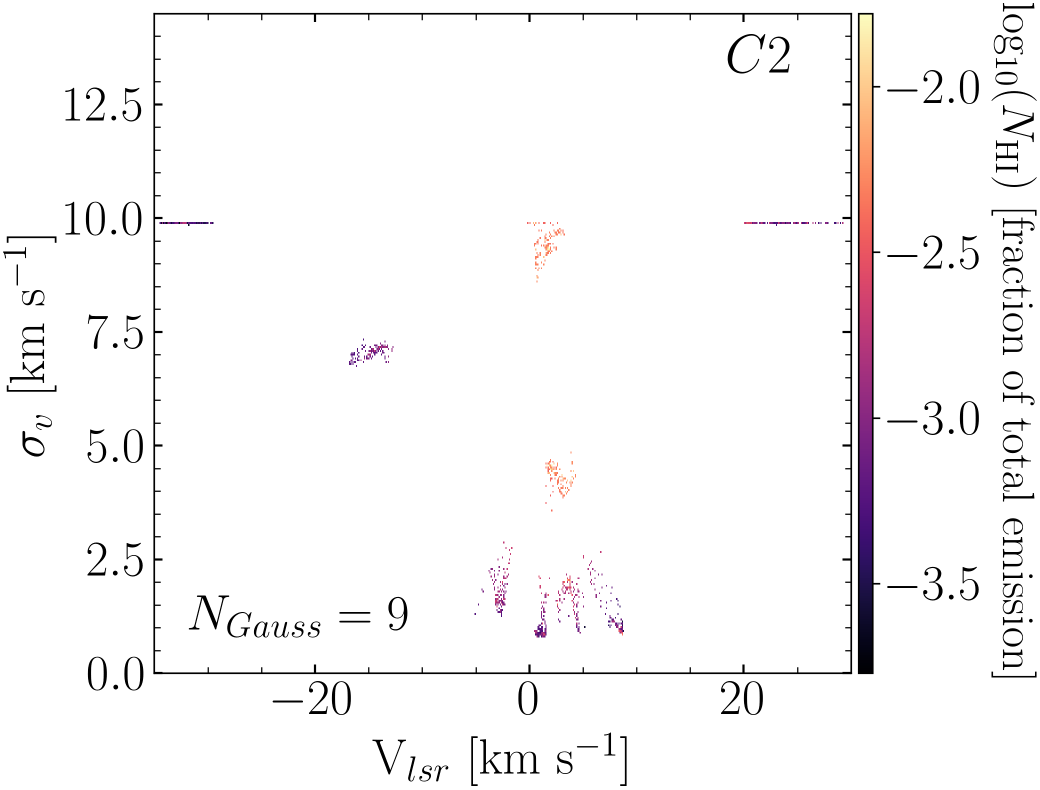}{0.35\linewidth}{}
                  }
        \gridline{
                  \fig{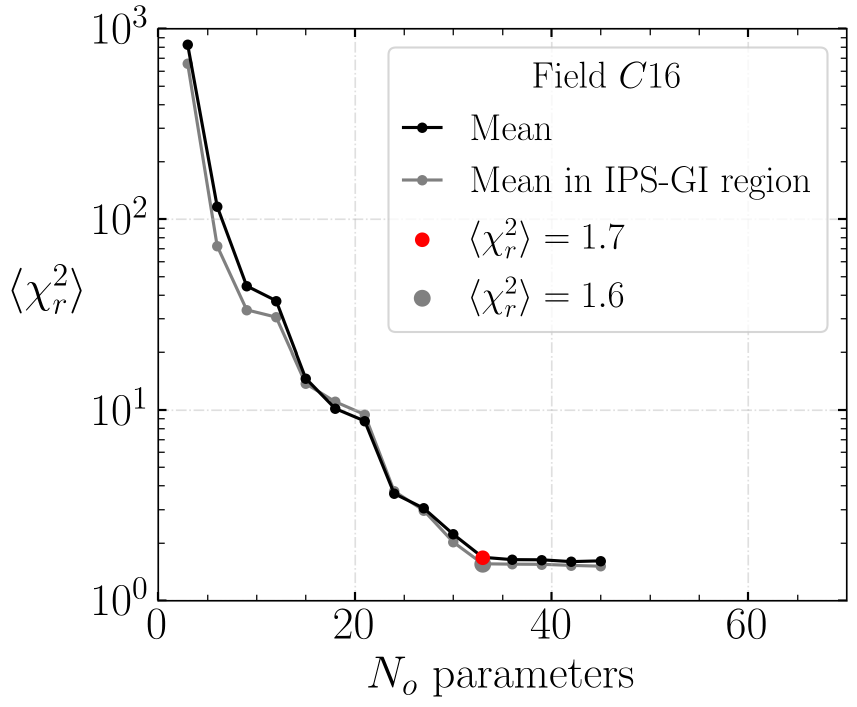}{0.32\linewidth}{}
                  \fig{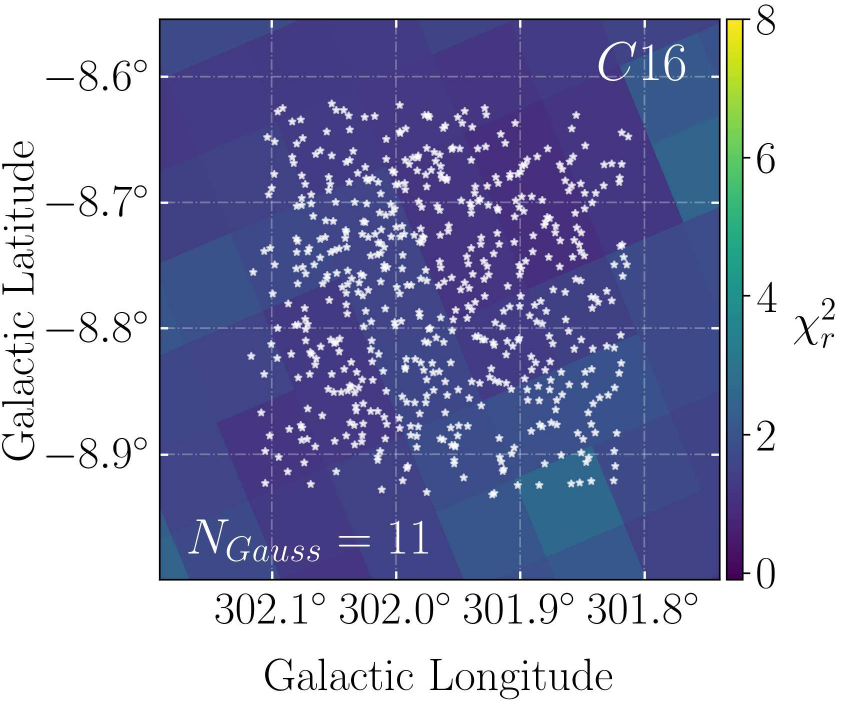}{0.32\linewidth}{}
                  \fig{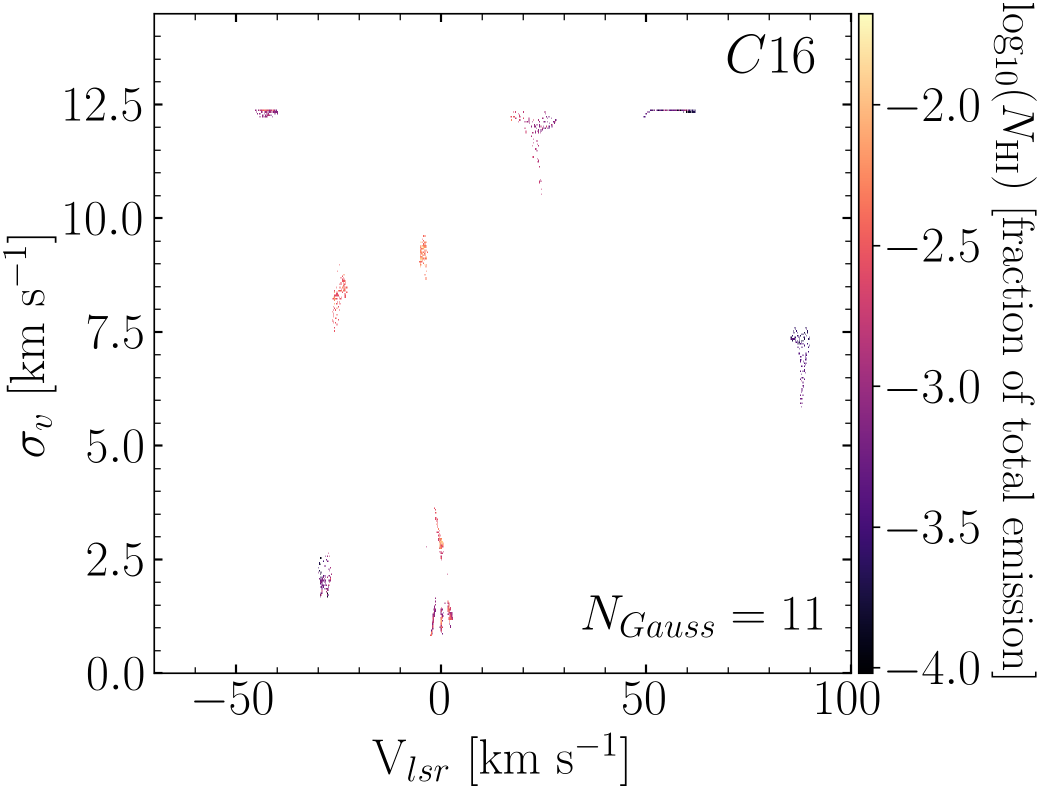}{0.35\linewidth}{}
                  }
        \gridline{
                  \fig{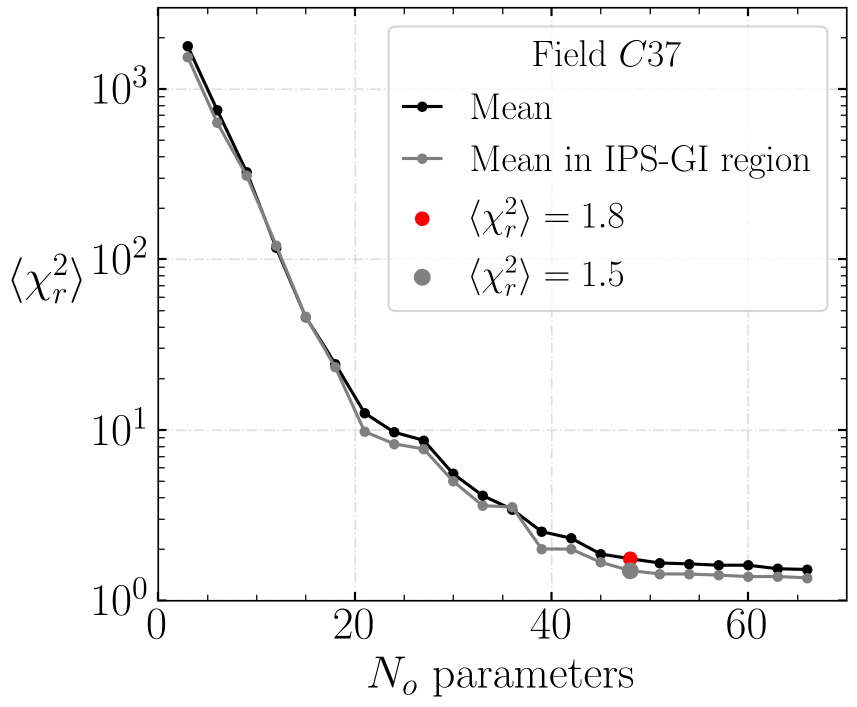}{0.32\linewidth}{}
                  \fig{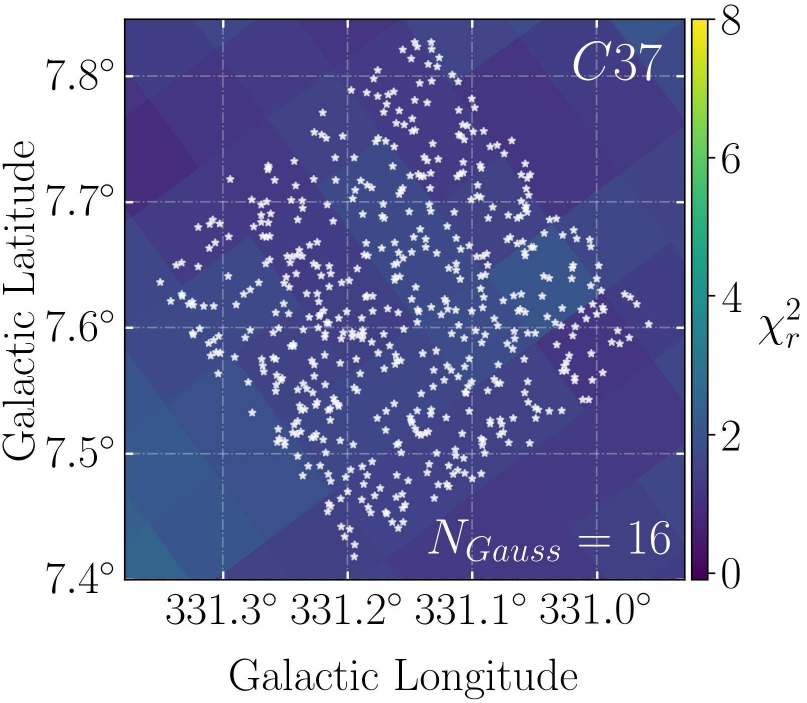}{0.32\linewidth}{}
                  \fig{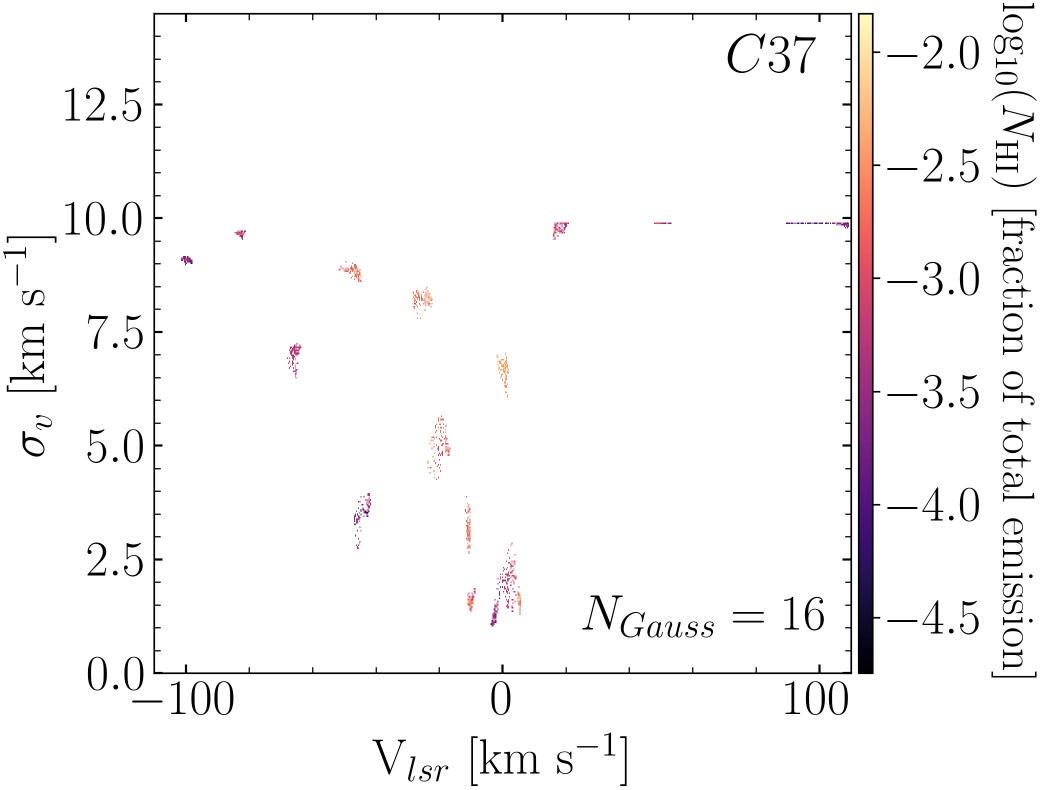}{0.35\linewidth}{}
                  }
        % \plotone{HI_maps_comp}
        \caption{Left column: average reduced $\chi^2$ as a function of the number of Gaussian parameters (Three per Gaussian component: $\bm{a}_n$, $\bm{\mu}_n$, and $\bm{\sigma}_n$). The mean $\chi^2_{r}$ as a function of the number of parameters in the IPS-GI field is shown by the gray line. The red dot represents the mean $\chi^2_{r}$ at the $N_{Gauss}$ selected. Middle column: reduced $\chi^2$ map of the total model built with Table~\ref{tab:ROHSA_param} parameters, with the position of IPS-GI measurements overlaid in white. Right column: dispersion-velocity probability distribution function weighted by the fraction of total emission of each Gaussian component (as in \citealt{Marchal_ROHSA_2019}, Figure~16).
        \label{fig:ROHSA_chi2_LVC_com}}
    \end{figure*}

    Using the \ion{H}{1} data cubes from GASS (Section~\ref{subsec:obs_HI_map}) and the set of parameters displayed in Table~\ref{tab:ROHSA_param}, we performed the Gaussian decomposition on each IPS-GI field studied (Section~\ref{sec:selec_fields}). The sets of parameters were carefully chosen (through a trial-and-error method) in each case to ensure coherent structures in the dispersion-velocity ($\sigma_{v}-v$) space weighted by the fraction of the total emission of each Gaussian, as in Figure 16 of \cite{Marchal_ROHSA_2019}; see Figure~\ref{fig:ROHSA_chi2_LVC_com}, right column. The dispersion-velocity heat map is an important metric that allows us to assess the consistency of the Gaussian components and the classification of the same into the different thermal states using the velocity dispersion thresholds ($\lesssim 3$~km~s$^{-1}$ for the CNM, $3-7$~km~s$^{-1}$ for the LNM, and $\gtrsim7$~km~s$^{-1}$ for the WNM; \citealt{Takakubo_1967, Marchal_ROHSA_2019}). So, we can distinguish between the components of the different ISM phases. 
    
    To determine the best number of Gaussian components and evaluate the quality of the modeling, we fixed the four hyper-parameters ($\lambda_{a}, \lambda_{\mu}$, $\lambda_{\sigma}$, and $\lambda'_{\sigma}$; Table~\ref{tab:ROHSA_param}) and varied the number of Gaussian components, $N_{Gauss}$, from 1 to 15 (21 in the case of \textit{C37}). The average of the reduced $\chi^2$ 2D map (e.g.,~Figure~\ref{fig:ROHSA_chi2_LVC_com}, middle column) as a function of the number of parameters (Figure~\ref{fig:ROHSA_chi2_LVC_com}, left column) helped us to find the $N_{Gauss}$ at which $\langle \chi^2_{r} \rangle$ converges towards a value close to one.  The final $N_{Gauss}$ per IPS-GI field is shown in Table~\ref{tab:ROHSA_param} and the respective $\langle \chi^2_{r} \rangle$ is displayed in Figure~\ref{fig:ROHSA_chi2_LVC_com}, left column.

    The final models of each of the ISM phases (CNM, LNM, and WNM) separately are presented in Figure~\ref{fig:ROHSA_Gauss_comp}, together with the \ion{H}{1} emission line and the total model. We also show in the insets in each panel the models made up of the LVC components, with $|V_{lsr}| < 20$~km~s$^{-1}$, which are expected to be those associated with the nearby structure under consideration. The \ion{H}{1} emission line in \textit{C2} is the simplest. All Gaussian components are centered around $V_{lsr} \approx 1.6$~km~s$^{-1}$, indicative of an LVC. Meanwhile, the \ion{H}{1} spectrum of \textit{C16} shows intermediate- and high-velocity clouds (IVC and HVC, respectively) that are unlikely related to the polarizing dust structure in the nearby ISM (Figure~\ref{fig:Av_dens_vs_dist}). Similarly, the spectrum of \textit{C37} presents IVC and HVC components that are not considered in the study of the local structure.
    
\begin{figure*}[ht!]
        \epsscale{2.4}
        \plottwo{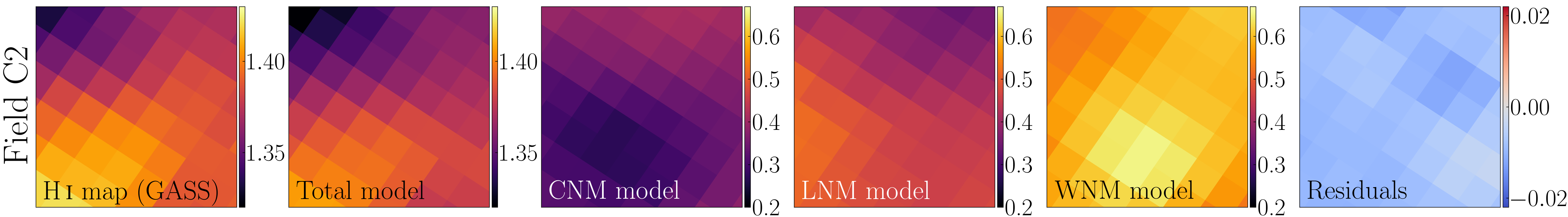}{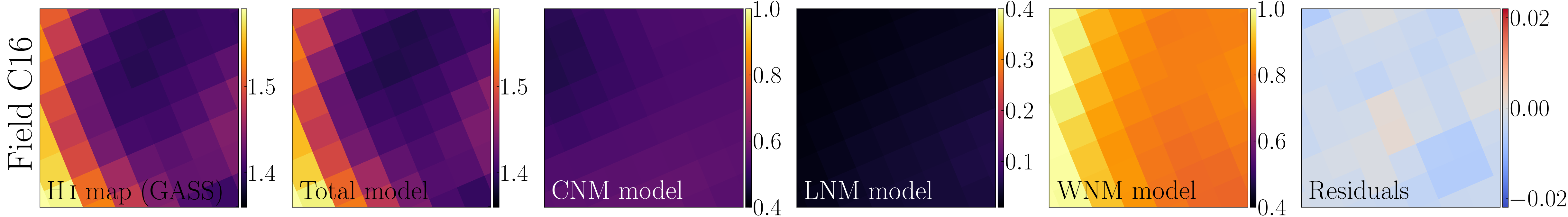}
        \epsscale{1.2}
        \plotone{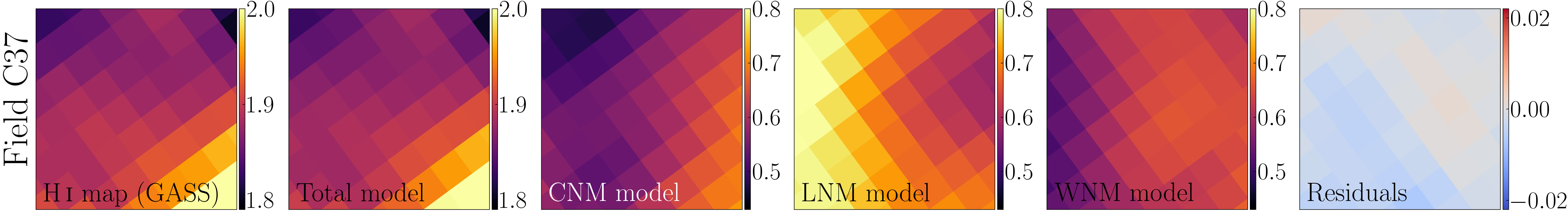}
        \caption{\ion{H}{1} column density maps per intermediate-latitude IPS-GI field in units of $10^{21}$~cm$^{-2}$. From left to right: $N_\mathrm{H}^{GASS}$ from \ion{H}{1} emission observed with GASS, $N_\mathrm{H}$ of the total model obtained with ROHSA, $N_\mathrm{H}^{cnm}$ from the CNM model, $N_\mathrm{H}^{lnm}$ from the LNM model, $N_\mathrm{H}^{wnm}$ from the WNM model, and the residuals between the observations ($N_\mathrm{H}^{GASS}$) and the total model from ROHSA ($N_\mathrm{H}$). Be aware of the different color scale of the LNM model in field \textit{C16}.
        \label{fig:ROHSA_models}}
    \end{figure*}

    Figure~\ref{fig:ROHSA_models} shows the column density of the observations compared to the modeled column density found with ROHSA and calculated with Equation~(\ref{eq:NH_integr}). The residual maps show a good agreement between the observations and the total models with a difference below ${\sim}1\%$. The CNM, LNM, and WNM models are also presented in Figure~\ref{fig:ROHSA_models} with column densities in the order of $10^{20}$~cm$^{-2}$. Besides the complexity of \textit{C37}'s \ion{H}{1} spectrum (as well as the $A_V$ density profile with distance, Figure~\ref{fig:Av_dens_vs_dist}) and the long pathlength integrated ($l_c = 1,000$~pc), the column density is comparable with the other fields and the local \ion{H}{1} structures are likely cold.
    
    %
    % %%% --- Begin Table -----------------------------------------------------------
    \begin{deluxetable}{ccccc}
        \tablecaption{Input parameters used in ROHSA per IPS-GI field. \label{tab:ROHSA_param}}
        % \tabletypesize{\scriptsize}
        % \tablewidth{5pt}
        \tablehead{
        \colhead{~~Field~~} & \colhead{~~$N_{Gauss}$~~} & \colhead{~~$\lambda_{a}, \lambda_{\mu}$, $\lambda_{\sigma}$~~} & \colhead{~~$\lambda'_{\sigma}$~~} & \colhead{~~$\langle \chi^2_{r} \rangle$~~}
        }
        \colnumbers
        \startdata
        \textit{C2} & 9 & 100 & 50 &  1.6  \\
        \textit{C16} & 11 & 100 & 50 & 1.6 \\
        \textit{C37} & 16 & 50 & 50 &  1.5 \\
        \enddata
        \tablecomments{Columns: (1) IPS-GI field name; (2) the number of Gaussian components considered for modeling; (3) ROHSA hyper-parameters for $\bm{a}_n$, $\bm{\mu}_n$, and $\bm{\sigma}_n$ 2D maps; (4) ROHSA hyper-parameter included in the cost function that accounts for the gas thermodynamics; (5) average reduced $\chi^2$ in the IPS-GI regions (Figure~\ref{fig:ROHSA_chi2_LVC_com}).
        }
    \end{deluxetable}
    %%% -------------------------------------------------------------------------- 
    
    %%% --- Begin Figure -----------------------------------------------------------
    %%% --------------------------------------------------------------------------

    %%% --- Begin Figure -----------------------------------------------------------
    %%% --------------------------------------------------------------------------

\clearpage

%...... Begin Bibliography ......................................................
%% For this sample, we use BibTeX plus aasjournals.bst to generate the
%% the bibliography. The sample631.bib file was populated from ADS. To
%% get the citations to show in the compiled file do the following:
%%
%% pdflatex sample631.tex
%% bibtext sample631
%% pdflatex sample631.tex
%% pdflatex sample631.tex

\bibliography{IPSIV_SF_gnal_ISM}{}

\newcommand{\noop}[1]{}
\begin{thebibliography}{}
\expandafter\ifx\csname natexlab\endcsname\relax\def\natexlab#1{#1}\fi
\providecommand{\url}[1]{\href{#1}{#1}}
\providecommand{\dodoi}[1]{doi:~\href{http://doi.org/#1}{\nolinkurl{#1}}}
\providecommand{\doeprint}[1]{\href{http://ascl.net/#1}{\nolinkurl{http://ascl.net/#1}}}
\providecommand{\doarXiv}[1]{\href{https://arxiv.org/abs/#1}{\nolinkurl{https://arxiv.org/abs/#1}}}

\bibitem[{{Alves} {et~al.}(2008){Alves}, {Franco}, \& {Girart}}]{Alves_2008}
{Alves}, F.~O., {Franco}, G.~A.~P., \& {Girart}, J.~M. 2008, \aap, 486, L13, \dodoi{10.1051/0004-6361:200810091}

\bibitem[{{Anders} {et~al.}(2022){Anders}, {Khalatyan}, {Queiroz}, {Chiappini}, {Ard{\`e}vol}, {Casamiquela}, {Figueras}, {Jim{\'e}nez-Arranz}, {Jordi}, {Mongui{\'o}}, {Romero-G{\'o}mez}, {Altamirano}, {Antoja}, {Assaad}, {Cantat-Gaudin}, {Castro-Ginard}, {Enke}, {Girardi}, {Guiglion}, {Khan}, {Luri}, {Miglio}, {Minchev}, {Ramos}, {Santiago}, \& {Steinmetz}}]{Anders_2022}
{Anders}, F., {Khalatyan}, A., {Queiroz}, A.~B.~A., {et~al.} 2022, \aap, 658, A91, \dodoi{10.1051/0004-6361/202142369}

\bibitem[{{Andersson} {et~al.}(2015){Andersson}, {Lazarian}, \& {Vaillancourt}}]{Andersson_2015}
{Andersson}, B.~G., {Lazarian}, A., \& {Vaillancourt}, J.~E. 2015, \araa, 53, 501, \dodoi{10.1146/annurev-astro-082214-122414}

\bibitem[{{Andersson} \& {Potter}(2005)}]{Andersson_Potter_CoalSack_2005}
{Andersson}, B.~G., \& {Potter}, S.~B. 2005, \mnras, 356, 1088, \dodoi{10.1111/j.1365-2966.2004.08538.x}

\bibitem[{{Angarita} {et~al.}(2023){Angarita}, {Versteeg}, {Haverkorn}, {Rodrigues}, {Magalh{\~a}es}, {Santos-Lima}, \& {Kawabata}}]{Angarita_2023}
{Angarita}, Y., {Versteeg}, M.~J.~F., {Haverkorn}, M., {et~al.} 2023, \aj, 166, 34, \dodoi{10.3847/1538-3881/acdc1e}

\bibitem[{{Astropy Collaboration} {et~al.}(2013){Astropy Collaboration}, {Robitaille}, {Tollerud}, {Greenfield}, {Droettboom}, {Bray}, {Aldcroft}, {Davis}, {Ginsburg}, {Price-Whelan}, {Kerzendorf}, {Conley}, {Crighton}, {Barbary}, {Muna}, {Ferguson}, {Grollier}, {Parikh}, {Nair}, {Unther}, {Deil}, {Woillez}, {Conseil}, {Kramer}, {Turner}, {Singer}, {Fox}, {Weaver}, {Zabalza}, {Edwards}, {Azalee Bostroem}, {Burke}, {Casey}, {Crawford}, {Dencheva}, {Ely}, {Jenness}, {Labrie}, {Lim}, {Pierfederici}, {Pontzen}, {Ptak}, {Refsdal}, {Servillat}, \& {Streicher}}]{Astropy_Collaboration_2013}
{Astropy Collaboration}, {Robitaille}, T.~P., {Tollerud}, E.~J., {et~al.} 2013, \aap, 558, A33, \dodoi{10.1051/0004-6361/201322068}

\bibitem[{{Astropy Collaboration} {et~al.}(2018){Astropy Collaboration}, {Price-Whelan}, {Sip{\H{o}}cz}, {G{\"u}nther}, {Lim}, {Crawford}, {Conseil}, {Shupe}, {Craig}, {Dencheva}, {Ginsburg}, {VanderPlas}, {Bradley}, {P{\'e}rez-Su{\'a}rez}, {de Val-Borro}, {Aldcroft}, {Cruz}, {Robitaille}, {Tollerud}, {Ardelean}, {Babej}, {Bach}, {Bachetti}, {Bakanov}, {Bamford}, {Barentsen}, {Barmby}, {Baumbach}, {Berry}, {Biscani}, {Boquien}, {Bostroem}, {Bouma}, {Brammer}, {Bray}, {Breytenbach}, {Buddelmeijer}, {Burke}, {Calderone}, {Cano Rodr{\'\i}guez}, {Cara}, {Cardoso}, {Cheedella}, {Copin}, {Corrales}, {Crichton}, {D'Avella}, {Deil}, {Depagne}, {Dietrich}, {Donath}, {Droettboom}, {Earl}, {Erben}, {Fabbro}, {Ferreira}, {Finethy}, {Fox}, {Garrison}, {Gibbons}, {Goldstein}, {Gommers}, {Greco}, {Greenfield}, {Groener}, {Grollier}, {Hagen}, {Hirst}, {Homeier}, {Horton}, {Hosseinzadeh}, {Hu}, {Hunkeler}, {Ivezi{\'c}}, {Jain}, {Jenness}, {Kanarek}, {Kendrew}, {Kern}, {Kerzendorf}, {Khvalko}, {King}, {Kirkby}, {Kulkarni},
  {Kumar}, {Lee}, {Lenz}, {Littlefair}, {Ma}, {Macleod}, {Mastropietro}, {McCully}, {Montagnac}, {Morris}, {Mueller}, {Mumford}, {Muna}, {Murphy}, {Nelson}, {Nguyen}, {Ninan}, {N{\"o}the}, {Ogaz}, {Oh}, {Parejko}, {Parley}, {Pascual}, {Patil}, {Patil}, {Plunkett}, {Prochaska}, {Rastogi}, {Reddy Janga}, {Sabater}, {Sakurikar}, {Seifert}, {Sherbert}, {Sherwood-Taylor}, {Shih}, {Sick}, {Silbiger}, {Singanamalla}, {Singer}, {Sladen}, {Sooley}, {Sornarajah}, {Streicher}, {Teuben}, {Thomas}, {Tremblay}, {Turner}, {Terr{\'o}n}, {van Kerkwijk}, {de la Vega}, {Watkins}, {Weaver}, {Whitmore}, {Woillez}, {Zabalza}, \& {Astropy Contributors}}]{Astropy_Collaboration_2018}
{Astropy Collaboration}, {Price-Whelan}, A.~M., {Sip{\H{o}}cz}, B.~M., {et~al.} 2018, \aj, 156, 123, \dodoi{10.3847/1538-3881/aabc4f}

\bibitem[{{Beck} \& {Wielebinski}(2013)}]{Beck_2013}
{Beck}, R., \& {Wielebinski}, R. 2013, in Planets, Stars and Stellar Systems. Volume 5: Galactic Structure and Stellar Populations, ed. T.~D. {Oswalt} \& G.~{Gilmore}, Vol.~5 (Springer Netherlands), 641, \dodoi{10.1007/978-94-007-5612-0_13}

\bibitem[{{Beresnyak} \& {Lazarian}(2019)}]{Beresnyak_2019}
{Beresnyak}, A., \& {Lazarian}, A. 2019, {Turbulence in Magnetohydrodynamics} (Berlin, Boston: De Gruyter), \dodoi{10.1515/9783110263282}

\bibitem[{{Bohlin} {et~al.}(1978){Bohlin}, {Savage}, \& {Drake}}]{bohlin_1978}
{Bohlin}, R.~C., {Savage}, B.~D., \& {Drake}, J.~F. 1978, \apj, 224, 132, \dodoi{10.1086/156357}

\bibitem[{{Boldyrev}(2002)}]{Boldyrev_2002}
{Boldyrev}, S. 2002, \apj, 569, 841, \dodoi{10.1086/339403}

\bibitem[{{Chandrasekhar} \& {Fermi}(1953)}]{Chandrasekhar&Fermi_1953}
{Chandrasekhar}, S., \& {Fermi}, E. 1953, \apj, 118, 113, \dodoi{10.1086/145731}

\bibitem[{{Chapman} {et~al.}(2011){Chapman}, {Goldsmith}, {Pineda}, {Clemens}, {Li}, \& {Kr{\v{c}}o}}]{Chapman_2011}
{Chapman}, N.~L., {Goldsmith}, P.~F., {Pineda}, J.~L., {et~al.} 2011, \apj, 741, 21, \dodoi{10.1088/0004-637X/741/1/21}

\bibitem[{{Cho} \& {Yoo}(2016)}]{Cho_Yoo_2016}
{Cho}, J., \& {Yoo}, H. 2016, \apj, 821, 21, \dodoi{10.3847/0004-637X/821/1/21}

\bibitem[{{Clark} \& {Hensley}(2019)}]{Clark_Hensley_2019}
{Clark}, S.~E., \& {Hensley}, B.~S. 2019, \apj, 887, 136, \dodoi{10.3847/1538-4357/ab5803}

\bibitem[{{Clark} {et~al.}(2015){Clark}, {Hill}, {Peek}, {Putman}, \& {Babler}}]{Clark_2015}
{Clark}, S.~E., {Hill}, J.~C., {Peek}, J.~E.~G., {Putman}, M.~E., \& {Babler}, B.~L. 2015, \prl, 115, 241302, \dodoi{10.1103/PhysRevLett.115.241302}

\bibitem[{{Clark} {et~al.}(2019){Clark}, {Peek}, \& {Miville-Desch{\^e}nes}}]{Clark_2019}
{Clark}, S.~E., {Peek}, J.~E.~G., \& {Miville-Desch{\^e}nes}, M.~A. 2019, \apj, 874, 171, \dodoi{10.3847/1538-4357/ab0b3b}

\bibitem[{{Crutcher}(2012)}]{Crutcher_2012}
{Crutcher}, R.~M. 2012, \araa, 50, 29, \dodoi{10.1146/annurev-astro-081811-125514}

\bibitem[{{Crutcher} {et~al.}(2010){Crutcher}, {Wandelt}, {Heiles}, {Falgarone}, \& {Troland}}]{Crutcher_2010}
{Crutcher}, R.~M., {Wandelt}, B., {Heiles}, C., {Falgarone}, E., \& {Troland}, T.~H. 2010, \apj, 725, 466, \dodoi{10.1088/0004-637X/725/1/466}

\bibitem[{{Davis} \& {Greenstein}(1951)}]{Davis_Greenstein_1951}
{Davis}, Leverett, J., \& {Greenstein}, J.~L. 1951, \apj, 114, 206, \dodoi{10.1086/145464}

\bibitem[{{Davis}(1951)}]{Davis_1951}
{Davis}, L. 1951, Physical Review, 81, 890, \dodoi{10.1103/PhysRev.81.890.2}

\bibitem[{{Dickey} \& {Lockman}(1990)}]{Dickey_1990}
{Dickey}, J.~M., \& {Lockman}, F.~J. 1990, \araa, 28, 215, \dodoi{10.1146/annurev.aa.28.090190.001243}

\bibitem[{{Draine}(2011)}]{Draine_2011}
{Draine}, B.~T. 2011, {Physics of the Interstellar and Intergalactic Medium} (Princeton Series in Astrophysics)

\bibitem[{{Edenhofer} {et~al.}(2023){Edenhofer}, {Zucker}, {Frank}, {Saydjari}, {Speagle}, {Finkbeiner}, \& {En{\ss}lin}}]{Edenhofer_dustmap_2023}
{Edenhofer}, G., {Zucker}, C., {Frank}, P., {et~al.} 2023, arXiv e-prints, arXiv:2308.01295, \dodoi{10.48550/arXiv.2308.01295}

\bibitem[{{Falceta-Gon{\c{c}}alves} {et~al.}(2008){Falceta-Gon{\c{c}}alves}, {Lazarian}, \& {Kowal}}]{Falceta_Gonzalves_2008}
{Falceta-Gon{\c{c}}alves}, D., {Lazarian}, A., \& {Kowal}, G. 2008, \apj, 679, 537, \dodoi{10.1086/587479}

\bibitem[{{Fitzpatrick}(2004)}]{Fitzpatrick_2004}
{Fitzpatrick}, E.~L. 2004, in Astronomical Society of the Pacific Conference Series, Vol. 309, Astrophysics of Dust, ed. A.~N. {Witt}, G.~C. {Clayton}, \& B.~T. {Draine}, 33.
\newblock \doarXiv{astro-ph/0401344}

\bibitem[{{Franco} {et~al.}(2010){Franco}, {Alves}, \& {Girart}}]{Franco_2010}
{Franco}, G.~A.~P., {Alves}, F.~O., \& {Girart}, J.~M. 2010, \apj, 723, 146, \dodoi{10.1088/0004-637X/723/1/146}

\bibitem[{{Frisch}(1995)}]{Frisch_1995}
{Frisch}, U. 1995, {Turbulence: The Legacy of A. N. Kolmogorov} (Cambridge University Press), \dodoi{10.1017/CBO9781139170666}

\bibitem[{{Gaia Collaboration} {et~al.}(2021){Gaia Collaboration}, {Brown}, {Vallenari}, {Prusti}, {de Bruijne}, {Babusiaux}, {Biermann}, {Creevey}, {Evans}, {Eyer}, {Hutton}, {Jansen}, {Jordi}, {Klioner}, {Lammers}, {Lindegren}, {Luri}, {Mignard}, {Panem}, {Pourbaix}, {Randich}, {Sartoretti}, {Soubiran}, {Walton}, {Arenou}, {Bailer-Jones}, {Bastian}, {Cropper}, {Drimmel}, {Katz}, {Lattanzi}, {van Leeuwen}, {Bakker}, {Cacciari}, {Casta{\~n}eda}, {De Angeli}, {Ducourant}, {Fabricius}, {Fouesneau}, {Fr{\'e}mat}, {Guerra}, {Guerrier}, {Guiraud}, {Jean-Antoine Piccolo}, {Masana}, {Messineo}, {Mowlavi}, {Nicolas}, {Nienartowicz}, {Pailler}, {Panuzzo}, {Riclet}, {Roux}, {Seabroke}, {Sordo}, {Tanga}, {Th{\'e}venin}, {Gracia-Abril}, {Portell}, {Teyssier}, {Altmann}, {Andrae}, {Bellas-Velidis}, {Benson}, {Berthier}, {Blomme}, {Brugaletta}, {Burgess}, {Busso}, {Carry}, {Cellino}, {Cheek}, {Clementini}, {Damerdji}, {Davidson}, {Delchambre}, {Dell'Oro}, {Fern{\'a}ndez-Hern{\'a}ndez}, {Galluccio}, {Garc{\'\i}a-Lario},
  {Garcia-Reinaldos}, {Gonz{\'a}lez-N{\'u}{\~n}ez}, {Gosset}, {Haigron}, {Halbwachs}, {Hambly}, {Harrison}, {Hatzidimitriou}, {Heiter}, {Hern{\'a}ndez}, {Hestroffer}, {Hodgkin}, {Holl}, {Jan{\ss}en}, {Jevardat de Fombelle}, {Jordan}, {Krone-Martins}, {Lanzafame}, {L{\"o}ffler}, {Lorca}, {Manteiga}, {Marchal}, {Marrese}, {Moitinho}, {Mora}, {Muinonen}, {Osborne}, {Pancino}, {Pauwels}, {Petit}, {Recio-Blanco}, {Richards}, {Riello}, {Rimoldini}, {Robin}, {Roegiers}, {Rybizki}, {Sarro}, {Siopis}, {Smith}, {Sozzetti}, {Ulla}, {Utrilla}, {van Leeuwen}, {van Reeven}, {Abbas}, {Abreu Aramburu}, {Accart}, {Aerts}, {Aguado}, {Ajaj}, {Altavilla}, {{\'A}lvarez}, {{\'A}lvarez Cid-Fuentes}, {Alves}, {Anderson}, {Anglada Varela}, {Antoja}, {Audard}, {Baines}, {Baker}, {Balaguer-N{\'u}{\~n}ez}, {Balbinot}, {Balog}, {Barache}, {Barbato}, {Barros}, {Barstow}, {Bartolom{\'e}}, {Bassilana}, {Bauchet}, {Baudesson-Stella}, {Becciani}, {Bellazzini}, {Bernet}, {Bertone}, {Bianchi}, {Blanco-Cuaresma}, {Boch}, {Bombrun}, {Bossini},
  {Bouquillon}, {Bragaglia}, {Bramante}, {Breedt}, {Bressan}, {Brouillet}, {Bucciarelli}, {Burlacu}, {Busonero}, {Butkevich}, {Buzzi}, {Caffau}, {Cancelliere}, {C{\'a}novas}, {Cantat-Gaudin}, {Carballo}, {Carlucci}, {Carnerero}, {Carrasco}, {Casamiquela}, {Castellani}, {Castro-Ginard}, {Castro Sampol}, {Chaoul}, {Charlot}, {Chemin}, {Chiavassa}, {Cioni}, {Comoretto}, {Cooper}, {Cornez}, {Cowell}, {Crifo}, {Crosta}, {Crowley}, {Dafonte}, {Dapergolas}, {David}, {David}, {de Laverny}, {De Luise}, {De March}, {De Ridder}, {de Souza}, {de Teodoro}, {de Torres}, {del Peloso}, {del Pozo}, {Delbo}, {Delgado}, {Delgado}, {Delisle}, {Di Matteo}, {Diakite}, {Diener}, {Distefano}, {Dolding}, {Eappachen}, {Edvardsson}, {Enke}, {Esquej}, {Fabre}, {Fabrizio}, {Faigler}, {Fedorets}, {Fernique}, {Fienga}, {Figueras}, {Fouron}, {Fragkoudi}, {Fraile}, {Franke}, {Gai}, {Garabato}, {Garcia-Gutierrez}, {Garc{\'\i}a-Torres}, {Garofalo}, {Gavras}, {Gerlach}, {Geyer}, {Giacobbe}, {Gilmore}, {Girona}, {Giuffrida}, {Gomel}, {Gomez},
  {Gonzalez-Santamaria}, {Gonz{\'a}lez-Vidal}, {Granvik}, {Guti{\'e}rrez-S{\'a}nchez}, {Guy}, {Hauser}, {Haywood}, {Helmi}, {Hidalgo}, {Hilger}, {H{\l}adczuk}, {Hobbs}, {Holland}, {Huckle}, {Jasniewicz}, {Jonker}, {Juaristi Campillo}, {Julbe}, {Karbevska}, {Kervella}, {Khanna}, {Kochoska}, {Kontizas}, {Kordopatis}, {Korn}, {Kostrzewa-Rutkowska}, {Kruszy{\'n}ska}, {Lambert}, {Lanza}, {Lasne}, {Le Campion}, {Le Fustec}, {Lebreton}, {Lebzelter}, {Leccia}, {Leclerc}, {Lecoeur-Taibi}, {Liao}, {Licata}, {Lindstr{\o}m}, {Lister}, {Livanou}, {Lobel}, {Madrero Pardo}, {Managau}, {Mann}, {Marchant}, {Marconi}, {Marcos Santos}, {Marinoni}, {Marocco}, {Marshall}, {Martin Polo}, {Mart{\'\i}n-Fleitas}, {Masip}, {Massari}, {Mastrobuono-Battisti}, {Mazeh}, {McMillan}, {Messina}, {Michalik}, {Millar}, {Mints}, {Molina}, {Molinaro}, {Moln{\'a}r}, {Montegriffo}, {Mor}, {Morbidelli}, {Morel}, {Morris}, {Mulone}, {Munoz}, {Muraveva}, {Murphy}, {Musella}, {Noval}, {Ord{\'e}novic}, {Orr{\`u}}, {Osinde}, {Pagani}, {Pagano},
  {Palaversa}, {Palicio}, {Panahi}, {Pawlak}, {Pe{\~n}alosa Esteller}, {Penttil{\"a}}, {Piersimoni}, {Pineau}, {Plachy}, {Plum}, {Poggio}, {Poretti}, {Poujoulet}, {Pr{\v{s}}a}, {Pulone}, {Racero}, {Ragaini}, {Rainer}, {Raiteri}, {Rambaux}, {Ramos}, {Ramos-Lerate}, {Re Fiorentin}, {Regibo}, {Reyl{\'e}}, {Ripepi}, {Riva}, {Rixon}, {Robichon}, {Robin}, {Roelens}, {Rohrbasser}, {Romero-G{\'o}mez}, {Rowell}, {Royer}, {Rybicki}, {Sadowski}, {Sagrist{\`a} Sell{\'e}s}, {Sahlmann}, {Salgado}, {Salguero}, {Samaras}, {Sanchez Gimenez}, {Sanna}, {Santove{\~n}a}, {Sarasso}, {Schultheis}, {Sciacca}, {Segol}, {Segovia}, {S{\'e}gransan}, {Semeux}, {Shahaf}, {Siddiqui}, {Siebert}, {Siltala}, {Slezak}, {Smart}, {Solano}, {Solitro}, {Souami}, {Souchay}, {Spagna}, {Spoto}, {Steele}, {Steidelm{\"u}ller}, {Stephenson}, {S{\"u}veges}, {Szabados}, {Szegedi-Elek}, {Taris}, {Tauran}, {Taylor}, {Teixeira}, {Thuillot}, {Tonello}, {Torra}, {Torra}, {Turon}, {Unger}, {Vaillant}, {van Dillen}, {Vanel}, {Vecchiato}, {Viala}, {Vicente},
  {Voutsinas}, {Weiler}, {Wevers}, {Wyrzykowski}, {Yoldas}, {Yvard}, {Zhao}, {Zorec}, {Zucker}, {Zurbach}, \& {Zwitter}}]{Gaia_Collaboration_2021b}
{Gaia Collaboration}, {Brown}, A.~G.~A., {Vallenari}, A., {et~al.} 2021, \aap, 649, A1, \dodoi{10.1051/0004-6361/202039657}

\bibitem[{{Goldreich} \& {Sridhar}(1995)}]{Goldreich_Sridhar_1995}
{Goldreich}, P., \& {Sridhar}, S. 1995, \apj, 438, 763, \dodoi{10.1086/175121}

\bibitem[{{Green}(2018)}]{Green_dustmaps_2018}
{Green}, G. 2018, The Journal of Open Source Software, 3, 695, \dodoi{10.21105/joss.00695}

\bibitem[{{Harris} {et~al.}(2020){Harris}, Millman, van~der Walt, Gommers, Virtanen, Cournapeau, Wieser, Taylor, Berg, Smith, Kern, Picus, Hoyer, van Kerkwijk, Brett, Haldane, del R{\'{i}}o, Wiebe, Peterson, G{\'{e}}rard-Marchant, Sheppard, Reddy, Weckesser, Abbasi, Gohlke, \& Oliphant}]{Harris_numpy_2020}
{Harris}, C.~R., Millman, K.~J., van~der Walt, S.~J., {et~al.} 2020, Nature, 585, 357, \dodoi{10.1038/s41586-020-2649-2}

\bibitem[{{Haverkorn}(2015)}]{Haverkorn_2015}
{Haverkorn}, M. 2015, in Astrophysics and Space Science Library, Vol. 407, Magnetic Fields in Diffuse Media, ed. A.~{Lazarian}, E.~M. {de Gouveia Dal Pino}, \& C.~{Melioli}, 483, \dodoi{10.1007/978-3-662-44625-6_17}

\bibitem[{{Haverkorn} {et~al.}(2008){Haverkorn}, {Brown}, {Gaensler}, \& {McClure-Griffiths}}]{Haverkorn_2008}
{Haverkorn}, M., {Brown}, J.~C., {Gaensler}, B.~M., \& {McClure-Griffiths}, N.~M. 2008, \apj, 680, 362, \dodoi{10.1086/587165}

\bibitem[{{Haverkorn} {et~al.}(2004){Haverkorn}, {Gaensler}, {McClure-Griffiths}, {Dickey}, \& {Green}}]{Haverkorn_2004}
{Haverkorn}, M., {Gaensler}, B.~M., {McClure-Griffiths}, N.~M., {Dickey}, J.~M., \& {Green}, A.~J. 2004, \apj, 609, 776, \dodoi{10.1086/421341}

\bibitem[{{Heiles}(2000)}]{Heiles_2000}
{Heiles}, C. 2000, \aj, 119, 923, \dodoi{10.1086/301236}

\bibitem[{{Heiles} \& {Troland}(2003)}]{Heiles_Troland_2003}
{Heiles}, C., \& {Troland}, T.~H. 2003, \apj, 586, 1067, \dodoi{10.1086/367828}

\bibitem[{{Heiles} \& {Troland}(2005)}]{Heiles_Troland_2005}
---. 2005, \apj, 624, 773, \dodoi{10.1086/428896}

\bibitem[{{Hensley} {et~al.}(2022){Hensley}, {Murray}, \& {Dodici}}]{Hensley_2022}
{Hensley}, B.~S., {Murray}, C.~E., \& {Dodici}, M. 2022, \apj, 929, 23, \dodoi{10.3847/1538-4357/ac5cbd}

\bibitem[{{Hildebrand} {et~al.}(2009){Hildebrand}, {Kirby}, {Dotson}, {Houde}, \& {Vaillancourt}}]{Hildebrand_2009}
{Hildebrand}, R.~H., {Kirby}, L., {Dotson}, J.~L., {Houde}, M., \& {Vaillancourt}, J.~E. 2009, \apj, 696, 567, \dodoi{10.1088/0004-637X/696/1/567}

\bibitem[{{Hiltner}(1949)}]{Hiltner_1949}
{Hiltner}, W.~A. 1949, Science, 109, 165, \dodoi{10.1126/science.109.2825.165}

\bibitem[{{Houde} {et~al.}(2009){Houde}, {Vaillancourt}, {Hildebrand}, {Chitsazzadeh}, \& {Kirby}}]{Houde_2009}
{Houde}, M., {Vaillancourt}, J.~E., {Hildebrand}, R.~H., {Chitsazzadeh}, S., \& {Kirby}, L. 2009, \apj, 706, 1504, \dodoi{10.1088/0004-637X/706/2/1504}

\bibitem[{{Hunter}(2007)}]{Hunter_Matplotlib_2007}
{Hunter}, J.~D. 2007, Computing in Science \& Engineering, 9, 90, \dodoi{10.1109/MCSE.2007.55}

\bibitem[{{Hwang} {et~al.}(2023){Hwang}, {Pattle}, {Parsons}, {Go}, \& {Kim}}]{Hwang_2023}
{Hwang}, J., {Pattle}, K., {Parsons}, H., {Go}, M., \& {Kim}, J. 2023, \aj, 165, 198, \dodoi{10.3847/1538-3881/acc460}

\bibitem[{{Jones}(2015)}]{Jones_bookChap_2015}
{Jones}, T.~J. 2015, in Astrophysics and Space Science Library, Vol. 407, Magnetic Fields in Diffuse Media, ed. A.~{Lazarian}, E.~M. {de Gouveia Dal Pino}, \& C.~{Melioli}, 153, \dodoi{10.1007/978-3-662-44625-6_7}

\bibitem[{{Karoly} {et~al.}(2023){Karoly}, {Ward-Thompson}, {Pattle}, {Berry}, {Whitworth}, {Kirk}, {Bastien}, {Ching}, {Coud{\'e}}, {Hwang}, {Kwon}, {Soam}, {Wang}, {Hasegawa}, {Lai}, {Qiu}, {Arzoumanian}, {Bourke}, {Byun}, {Chen}, {Chen}, {Chen}, {Chen}, {Cho}, {Choi}, {Choi}, {Choi}, {Chrysostomou}, {Chung}, {Dai}, {Debattista}, {Di Francesco}, {Diep}, {Doi}, {Duan}, {Duan}, {Eswaraiah}, {Fanciullo}, {Fiege}, {Fissel}, {Franzmann}, {Friberg}, {Friesen}, {Fuller}, {Furuya}, {Gledhill}, {Graves}, {Greaves}, {Griffin}, {Gu}, {Han}, {Hoang}, {Houde}, {Hull}, {Inoue}, {Inutsuka}, {Iwasaki}, {Jeong}, {Johnstone}, {K{\"o}nyves}, {Kang}, {Kang}, {Kataoka}, {Kawabata}, {Kemper}, {Kim}, {Kim}, {Kim}, {Kim}, {Kim}, {Kim}, {Kim}, {Kirchschlager}, {Kobayashi}, {Koch}, {Kusune}, {Kwon}, {Lacaille}, {Law}, {Lee}, {Lee}, {Lee}, {Lee}, {Lee}, {Lee}, {Li}, {Li}, {Li}, {Li}, {Lin}, {Liu}, {Liu}, {Liu}, {Liu}, {Longmore}, {Lu}, {Lyo}, {Mairs}, {Matsumura}, {Matthews}, {Moriarty-Schieven}, {Nagata}, {Nakamura}, {Nakanishi},
  {Ngoc}, {Ohashi}, {Onaka}, {Park}, {Parsons}, {Peretto}, {Priestley}, {Pyo}, {Qian}, {Rao}, {Rawlings}, {Rawlings}, {Retter}, {Richer}, {Rigby}, {Sadavoy}, {Saito}, {Savini}, {Seta}, {Sharma}, {Shimajiri}, {Shinnaga}, {Tahani}, {Tamura}, {Tang}, {Tang}, {Tomisaka}, {Tram}, {Tsukamoto}, {Viti}, {Wang}, {Wu}, {Xie}, {Yang}, {Yen}, {Yoo}, {Yuan}, {Yun}, {Zenko}, {Zhang}, {Zhang}, {Zhang}, {Zhou}, {Zhu}, {de Looze}, {Andr{\'e}}, {Dowell}, {Eden}, {Eyres}, {Falle}, {Le Gouellec}, {Poidevin}, {Robitaille}, \& {van Loo}}]{Karoly_2023}
{Karoly}, J., {Ward-Thompson}, D., {Pattle}, K., {et~al.} 2023, \apj, 952, 29, \dodoi{10.3847/1538-4357/acd6f2}

\bibitem[{{Kobulnicky} {et~al.}(1994){Kobulnicky}, {Molnar}, \& {Jones}}]{Kobulnicky_1994}
{Kobulnicky}, H.~A., {Molnar}, L.~A., \& {Jones}, T.~J. 1994, \aj, 107, 1433, \dodoi{10.1086/116956}

\bibitem[{{Kolmogorov}(1941)}]{Kolmogorov_1941}
{Kolmogorov}, A. 1941, Akademiia Nauk SSSR Doklady, 30, 301

\bibitem[{{Larson}(1979)}]{Larson_1979}
{Larson}, R.~B. 1979, \mnras, 186, 479, \dodoi{10.1093/mnras/186.3.479}

\bibitem[{{Lei} \& {Clark}(2023)}]{Lei&Clark_2023}
{Lei}, M., \& {Clark}, S.~E. 2023, arXiv e-prints, arXiv:2312.03846, \dodoi{10.48550/arXiv.2312.03846}

\bibitem[{{Leike} {et~al.}(2020){Leike}, {Glatzle}, \& {En{\ss}lin}}]{Leike_2020}
{Leike}, R.~H., {Glatzle}, M., \& {En{\ss}lin}, T.~A. 2020, \aap, 639, A138, \dodoi{10.1051/0004-6361/202038169}

\bibitem[{{Lenz} {et~al.}(2017){Lenz}, {Hensley}, \& {Dor{\'e}}}]{Lenz_2017}
{Lenz}, D., {Hensley}, B.~S., \& {Dor{\'e}}, O. 2017, \apj, 846, 38, \dodoi{10.3847/1538-4357/aa84af}

\bibitem[{{Liu} {et~al.}(2022){Liu}, {Qiu}, \& {Zhang}}]{Liu_2022}
{Liu}, J., {Qiu}, K., \& {Zhang}, Q. 2022, \apj, 925, 30, \dodoi{10.3847/1538-4357/ac3911}

\bibitem[{{Liu} {et~al.}(2021){Liu}, {Zhang}, {Commer{\c{c}}on}, {Valdivia}, {Maury}, \& {Qiu}}]{Liu_2021}
{Liu}, J., {Zhang}, Q., {Commer{\c{c}}on}, B., {et~al.} 2021, \apj, 919, 79, \dodoi{10.3847/1538-4357/ac0cec}

\bibitem[{{Magalh{\~a}es} {et~al.}(2005){Magalh{\~a}es}, {Pereyra}, {Melgarejo}, {de Matos}, {Carciofi}, {Benedito}, {Valentim}, {Vidotto}, {da Silva}, {de Souza}, {Faria}, \& {Gabriel}}]{Magalhaes_2005}
{Magalh{\~a}es}, A.~M., {Pereyra}, A., {Melgarejo}, R., {et~al.} 2005, in Astronomical Society of the Pacific Conference Series, Vol. 343, Astronomical Polarimetry: Current Status and Future Directions, ed. A.~{Adamson}, C.~{Aspin}, C.~{Davis}, \& T.~{Fujiyoshi}, 305

\bibitem[{{Marchal} \& {Miville-Desch{\^e}nes}(2021)}]{Marchal_2021}
{Marchal}, A., \& {Miville-Desch{\^e}nes}, M.-A. 2021, \apj, 908, 186, \dodoi{10.3847/1538-4357/abd108}

\bibitem[{{Marchal} {et~al.}(2019){Marchal}, {Miville-Desch{\^e}nes}, {Orieux}, {Gac}, {Soussen}, {Lesot}, {d'Allonnes}, \& {Salom{\'e}}}]{Marchal_ROHSA_2019}
{Marchal}, A., {Miville-Desch{\^e}nes}, M.-A., {Orieux}, F., {et~al.} 2019, \aap, 626, A101, \dodoi{10.1051/0004-6361/201935335}

\bibitem[{{McClure-Griffiths} {et~al.}(2009){McClure-Griffiths}, {Pisano}, {Calabretta}, {Ford}, {Lockman}, {Staveley-Smith}, {Kalberla}, {Bailin}, {Dedes}, {Janowiecki}, {Gibson}, {Murphy}, {Nakanishi}, \& {Newton-McGee}}]{McClure_Griffiths_2009}
{McClure-Griffiths}, N.~M., {Pisano}, D.~J., {Calabretta}, M.~R., {et~al.} 2009, \apjs, 181, 398, \dodoi{10.1088/0067-0049/181/2/398}

\bibitem[{{Medan} \& {Andersson}(2019)}]{Medan_Andersson_2019}
{Medan}, I., \& {Andersson}, B.~G. 2019, \apj, 873, 87, \dodoi{10.3847/1538-4357/ab063c}

\bibitem[{{Meisner} \& {Finkbeiner}(2014)}]{Meisner_Finkbeiner_2014}
{Meisner}, A.~M., \& {Finkbeiner}, D.~P. 2014, \apj, 781, 5, \dodoi{10.1088/0004-637X/781/1/5}

\bibitem[{{Miville-Desch{\^e}nes} {et~al.}(2017){Miville-Desch{\^e}nes}, {Murray}, \& {Lee}}]{Miville-Deschenes_2017}
{Miville-Desch{\^e}nes}, M.-A., {Murray}, N., \& {Lee}, E.~J. 2017, \apj, 834, 57, \dodoi{10.3847/1538-4357/834/1/57}

\bibitem[{{Murray} {et~al.}(2015){Murray}, {Stanimirovi{\'c}}, {Goss}, {Dickey}, {Heiles}, {Lindner}, {Babler}, {Pingel}, {Lawrence}, {Jencson}, \& {Hennebelle}}]{Murray_2015}
{Murray}, C.~E., {Stanimirovi{\'c}}, S., {Goss}, W.~M., {et~al.} 2015, \apj, 804, 89, \dodoi{10.1088/0004-637X/804/2/89}

\bibitem[{{Panopoulou} {et~al.}(2023){Panopoulou}, {Markopoulioti}, {Bouzelou}, {Millar-Blanchaer}, {Tinyanont}, {Blinov}, {Pelgrims}, {Johnson}, {Skalidis}, \& {Soam}}]{Panopoulou_catalog_2023}
{Panopoulou}, G.~V., {Markopoulioti}, L., {Bouzelou}, F., {et~al.} 2023, arXiv e-prints, arXiv:2307.05752, \dodoi{10.48550/arXiv.2307.05752}

\bibitem[{{Planck Collaboration XII} {et~al.}(2020){Planck Collaboration XII}, {Aghanim}, {Akrami}, {Alves}, {Ashdown}, {Aumont}, {Baccigalupi}, {Ballardini}, {Banday}, {Barreiro}, {Bartolo}, {Basak}, {Benabed}, {Bernard}, {Bersanelli}, {Bielewicz}, {Bock}, {Bond}, {Borrill}, {Bouchet}, {Boulanger}, {Bracco}, {Bucher}, {Burigana}, {Calabrese}, {Cardoso}, {Carron}, {Chary}, {Chiang}, {Colombo}, {Combet}, {Crill}, {Cuttaia}, {de Bernardis}, {de Zotti}, {Delabrouille}, {Delouis}, {Di Valentino}, {Dickinson}, {Diego}, {Dor{\'e}}, {Douspis}, {Ducout}, {Dupac}, {Efstathiou}, {Elsner}, {En{\ss}lin}, {Eriksen}, {Falgarone}, {Fantaye}, {Fernandez-Cobos}, {Ferri{\`e}re}, {Finelli}, {Forastieri}, {Frailis}, {Fraisse}, {Franceschi}, {Frolov}, {Galeotta}, {Galli}, {Ganga}, {G{\'e}nova-Santos}, {Gerbino}, {Ghosh}, {Gonz{\'a}lez-Nuevo}, {G{\'o}rski}, {Gratton}, {Green}, {Gruppuso}, {Gudmundsson}, {Guillet}, {Handley}, {Hansen}, {Helou}, {Herranz}, {Hivon}, {Huang}, {Jaffe}, {Jones}, {Keih{\"a}nen}, {Keskitalo}, {Kiiveri},
  {Kim}, {Krachmalnicoff}, {Kunz}, {Kurki-Suonio}, {Lagache}, {Lamarre}, {Lasenby}, {Lattanzi}, {Lawrence}, {Le Jeune}, {Levrier}, {Liguori}, {Lilje}, {Lindholm}, {L{\'o}pez-Caniego}, {Lubin}, {Ma}, {Mac{\'\i}as-P{\'e}rez}, {Maggio}, {Maino}, {Mandolesi}, {Mangilli}, {Marcos-Caballero}, {Maris}, {Martin}, {Mart{\'\i}nez-Gonz{\'a}lez}, {Matarrese}, {Mauri}, {McEwen}, {Melchiorri}, {Mennella}, {Migliaccio}, {Miville-Desch{\^e}nes}, {Molinari}, {Moneti}, {Montier}, {Morgante}, {Moss}, {Natoli}, {Pagano}, {Paoletti}, {Patanchon}, {Perrotta}, {Pettorino}, {Piacentini}, {Polastri}, {Polenta}, {Puget}, {Rachen}, {Reinecke}, {Remazeilles}, {Renzi}, {Ristorcelli}, {Rocha}, {Rosset}, {Roudier}, {Rubi{\~n}o-Mart{\'\i}n}, {Ruiz-Granados}, {Salvati}, {Sandri}, {Savelainen}, {Scott}, {Sirignano}, {Sunyaev}, {Suur-Uski}, {Tauber}, {Tavagnacco}, {Tenti}, {Toffolatti}, {Tomasi}, {Trombetti}, {Valiviita}, {Vansyngel}, {Van Tent}, {Vielva}, {Villa}, {Vittorio}, {Wandelt}, {Wehus}, {Zacchei}, \&
  {Zonca}}]{Planck-Collaboration_2018_20}
{Planck Collaboration XII}, {Aghanim}, N., {Akrami}, Y., {et~al.} 2020, \aap, 641, A12, \dodoi{10.1051/0004-6361/201833885}

\bibitem[{{Poidevin} {et~al.}(2010){Poidevin}, {Bastien}, \& {Matthews}}]{Poidevin_2010}
{Poidevin}, F., {Bastien}, P., \& {Matthews}, B.~C. 2010, \apj, 716, 893, \dodoi{10.1088/0004-637X/716/2/893}

\bibitem[{{Ram{\'\i}rez} {et~al.}(2017){Ram{\'\i}rez}, {Magalh{\~a}es}, {Davidson}, {Pereyra}, \& {Rubinho}}]{Ramirez_2017}
{Ram{\'\i}rez}, E.~A., {Magalh{\~a}es}, A.~M., {Davidson}, James~W., J., {Pereyra}, A., \& {Rubinho}, M. 2017, \pasp, 129, 055001, \dodoi{10.1088/1538-3873/aa54a7}

\bibitem[{{Saury} {et~al.}(2014){Saury}, {Miville-Desch{\^e}nes}, {Hennebelle}, {Audit}, \& {Schmidt}}]{Saury_2014}
{Saury}, E., {Miville-Desch{\^e}nes}, M.~A., {Hennebelle}, P., {Audit}, E., \& {Schmidt}, W. 2014, \aap, 567, A16, \dodoi{10.1051/0004-6361/201321113}

\bibitem[{{Savage} \& {Mathis}(1979)}]{Savage_1979}
{Savage}, B.~D., \& {Mathis}, J.~S. 1979, \araa, 17, 73, \dodoi{10.1146/annurev.aa.17.090179.000445}

\bibitem[{{Serkowski}(1958)}]{Serkowski_1958}
{Serkowski}, K. 1958, \actaa, 8, 135

\bibitem[{{Serkowski}(1962)}]{Serkowski_1962}
---. 1962, Advances in Astronomy and Astrophysics, 1, 289, \dodoi{10.1016/B978-1-4831-9919-1.50009-1}

\bibitem[{{Skalidis} {et~al.}(2018){Skalidis}, {Panopoulou}, {Tassis}, {Pavlidou}, {Blinov}, {Komis}, \& {Liodakis}}]{Skalidis_2018}
{Skalidis}, R., {Panopoulou}, G.~V., {Tassis}, K., {et~al.} 2018, \aap, 616, A52, \dodoi{10.1051/0004-6361/201832827}

\bibitem[{{Skalidis} \& {Tassis}(2021)}]{Skalidis_Tassis_2021}
{Skalidis}, R., \& {Tassis}, K. 2021, \aap, 647, A186, \dodoi{10.1051/0004-6361/202039779}

\bibitem[{{Soler} {et~al.}(2016){Soler}, {Alves}, {Boulanger}, {Bracco}, {Falgarone}, {Franco}, {Guillet}, {Hennebelle}, {Levrier}, {Martin}, \& {Miville-Desch{\^e}nes}}]{Soler_2016}
{Soler}, J.~D., {Alves}, F., {Boulanger}, F., {et~al.} 2016, \aap, 596, A93, \dodoi{10.1051/0004-6361/201628996}

\bibitem[{{Sun} \& {Han}(2004)}]{Sun_Han_SF_2004}
{Sun}, X.~H., \& {Han}, J.~L. 2004, in The Magnetized Interstellar Medium, ed. B.~{Uyaniker}, W.~{Reich}, \& R.~{Wielebinski}, 25--30, \dodoi{10.48550/arXiv.astro-ph/0402180}

\bibitem[{{Takakubo}(1967)}]{Takakubo_1967}
{Takakubo}, K. 1967, \bain, 19, 125

\bibitem[{{V{\'a}zquez-Semadeni}(2015)}]{Vazquez_Semadeni_2015}
{V{\'a}zquez-Semadeni}, E. 2015, in Astrophysics and Space Science Library, Vol. 407, Magnetic Fields in Diffuse Media, ed. A.~{Lazarian}, E.~M. {de Gouveia Dal Pino}, \& C.~{Melioli}, 401, \dodoi{10.1007/978-3-662-44625-6_14}

\bibitem[{{Vergely} {et~al.}(2022){Vergely}, {Lallement}, \& {Cox}}]{Vergely_2022}
{Vergely}, J.~L., {Lallement}, R., \& {Cox}, N.~L.~J. 2022, \aap, 664, A174, \dodoi{10.1051/0004-6361/202243319}

\bibitem[{{Versteeg} {et~al.}(2024){Versteeg}, {Angarita}, {Magalh{\~a}es}, {Haverkorn}, {Rodrigues}, {Santos-Lima}, \& {Kawabata}}]{Versteeg_2024}
{Versteeg}, M.~J.~F., {Angarita}, Y., {Magalh{\~a}es}, A.~M., {et~al.} 2024, \aj, 167, 177, \dodoi{10.3847/1538-3881/ad2e08}

\bibitem[{{Versteeg} {et~al.}(2023){Versteeg}, {Magalh{\~a}es}, {Haverkorn}, {Angarita}, {Rodrigues}, {Santos-Lima}, \& {Kawabata}}]{Versteeg_2023}
{Versteeg}, M.~J.~F., {Magalh{\~a}es}, A.~M., {Haverkorn}, M., {et~al.} 2023, \aj, 165, 87, \dodoi{10.3847/1538-3881/aca8fd}

\bibitem[{{Virtanen} {et~al.}(2020){Virtanen}, Gommers, Oliphant, Haberland, Reddy, Cournapeau, Burovski, Peterson, Weckesser, Bright, {van der Walt}, Brett, Wilson, Millman, Mayorov, Nelson, Jones, Kern, Larson, Carey, Polat, Feng, Moore, {VanderPlas}, Laxalde, Perktold, Cimrman, Henriksen, Quintero, Harris, Archibald, Ribeiro, Pedregosa, {van Mulbregt}, \& {SciPy 1.0 Contributors}}]{Virtanen_SciPy_2020}
{Virtanen}, P., Gommers, R., Oliphant, T.~E., {et~al.} 2020, Nature Methods, 17, 261, \dodoi{10.1038/s41592-019-0686-2}

\bibitem[{{Wakker}(2001)}]{Wakker_2001}
{Wakker}, B.~P. 2001, \apjs, 136, 463, \dodoi{10.1086/321783}

\bibitem[{{Wang} {et~al.}(2019){Wang}, {Lai}, {Eswaraiah}, {Pattle}, {Di Francesco}, {Johnstone}, {Koch}, {Liu}, {Tamura}, {Furuya}, {Onaka}, {Ward-Thompson}, {Soam}, {Kim}, {Lee}, {Lee}, {Mairs}, {Arzoumanian}, {Kim}, {Hoang}, {Hwang}, {Liu}, {Berry}, {Bastien}, {Hasegawa}, {Kwon}, {Qiu}, {Andr{\'e}}, {Aso}, {Byun}, {Chen}, {Chen}, {Chen}, {Ching}, {Cho}, {Choi}, {Chrysostomou}, {Chung}, {Coud{\'e}}, {Doi}, {Dowell}, {Drabek-Maunder}, {Duan}, {Eyres}, {Falle}, {Fanciullo}, {Fiege}, {Franzmann}, {Friberg}, {Friesen}, {Fuller}, {Gledhill}, {Graves}, {Greaves}, {Griffin}, {Gu}, {Han}, {Hatchell}, {Hayashi}, {Holland}, {Houde}, {Inoue}, {Inutsuka}, {Iwasaki}, {Jeong}, {Kanamori}, {Kang}, {Kang}, {Kang}, {Kataoka}, {Kawabata}, {Kemper}, {Kim}, {Kim}, {Kim}, {Kim}, {Kirk}, {Kobayashi}, {Konyves}, {Kwon}, {Lacaille}, {Lee}, {Lee}, {Lee}, {Lee}, {Li}, {Li}, {Li}, {Liu}, {Liu}, {Lyo}, {Matsumura}, {Matthews}, {Moriarty-Schieven}, {Nagata}, {Nakamura}, {Nakanishi}, {Ohashi}, {Park}, {Parsons}, {Pascale}, {Peretto},
  {Pon}, {Pyo}, {Qian}, {Rao}, {Rawlings}, {Retter}, {Richer}, {Rigby}, {Robitaille}, {Sadavoy}, {Saito}, {Savini}, {Scaife}, {Seta}, {Shinnaga}, {Tang}, {Tomisaka}, {Tsukamoto}, {van Loo}, {Wang}, {Whitworth}, {Yen}, {Yoo}, {Yuan}, {Yun}, {Zenko}, {Zhang}, {Zhang}, {Zhang}, {Zhou}, \& {Zhu}}]{Wang_JCMT_BISTRO_2019}
{Wang}, J.-W., {Lai}, S.-P., {Eswaraiah}, C., {et~al.} 2019, \apj, 876, 42, \dodoi{10.3847/1538-4357/ab13a2}

\bibitem[{{Wolfire} {et~al.}(2003){Wolfire}, {McKee}, {Hollenbach}, \& {Tielens}}]{Wolfire_2003}
{Wolfire}, M.~G., {McKee}, C.~F., {Hollenbach}, D., \& {Tielens}, A.~G.~G.~M. 2003, \apj, 587, 278, \dodoi{10.1086/368016}

\bibitem[{{Zhang} {et~al.}(2023){Zhang}, {Green}, \& {Rix}}]{Zhang_2023}
{Zhang}, X., {Green}, G.~M., \& {Rix}, H.-W. 2023, \mnras, 524, 1855, \dodoi{10.1093/mnras/stad1941}

\bibitem[{{Zucker} {et~al.}(2021){Zucker}, {Goodman}, {Alves}, {Bialy}, {Koch}, {Speagle}, {Foley}, {Finkbeiner}, {Leike}, {En{\ss}lin}, {Peek}, \& {Edenhofer}}]{Zucker_2021}
{Zucker}, C., {Goodman}, A., {Alves}, J., {et~al.} 2021, \apj, 919, 35, \dodoi{10.3847/1538-4357/ac1f96}

\end{thebibliography}
\bibliographystyle{aasjournal}

%% Include this line if you are using the \added, \replaced, \deleted
%% commands to see a summary list of all changes at the end of the article.
% \listofchanges

\end{document}